\documentclass[a4paper,aps,prd,twocolumn,eqsecnum,amsmath,amssymb,nofootinbib,
  superscriptaddress]{revtex4-2}

\usepackage{graphicx}
\usepackage{dcolumn}
\usepackage{bm}
\usepackage{natbib}
\usepackage{multirow}
\usepackage{color}
\usepackage{graphics}
\usepackage{epsfig}
\usepackage{booktabs,multirow}
\usepackage{hhline}
\usepackage{slashed}
\usepackage{rccol}
\usepackage{mathrsfs}
\usepackage{enumitem}
\usepackage[dvipsnames]{xcolor,colortbl}
\usepackage{subfigure}
\usepackage{xcolor}\colorlet{linkequation}{blue}
\usepackage[colorlinks=true, 
            linkcolor=red,   
            filecolor=red,   
            citecolor=green, 
            urlcolor=blue,
            hyperfootnotes=true]{hyperref}
\usepackage[normalem]{ulem}

\begin{document}


\def\nuebar{{\rm \bar{\nu}_e}}
\def\nuebare{{\rm \bar{\nu}_{e}-e}}
\def\nue{{\rm \nu_e}}
\def\nuee{{\rm \nu_{e}-e}}
\def\s2tw{{\rm \sin ^2 \theta_{W}}}
\def\munu{{\rm \mu_{\nu}}}
\def\cpkkd{\rm{kg^{-1} keV^{-1} day^{-1}}}
\def\beqa{\begin{eqnarray}}
\def\eeqa{\end{eqnarray}}

\newcommand{\be}{\ensuremath{\beta}}
\newcommand{\al}{\ensuremath{\alpha}}
\newcommand{\sa}{\ensuremath{\sin\alpha}}
\newcommand{\ca}{\ensuremath{\cos\alpha}}
\newcommand{\ta}{\ensuremath{\tan\alpha}}
\newcommand{\sbt}{\ensuremath{\sin\beta}}
\newcommand{\cbt}{\ensuremath{\cos\beta}}
\newcommand{\cba}{\ensuremath{c_{\beta-\alpha}}}
\newcommand{\ma}{\ensuremath{m_{A}}}
\newcommand{\mh}{\ensuremath{m_{h^0}}}
\newcommand{\mH}{\ensuremath{m_{H^0}}}
\newcommand{\mev}{\mbox{~MeV}}
\newcommand{\gev}{\mbox{~GeV}}
\newcommand{\tev}{\mbox{~TeV}}
\newcommand{\ben}{\begin{enumerate}}
\newcommand{\een}{\end{enumerate}}
\newcommand{\bc}{\begin{center}}
\newcommand{\ec}{\end{center}}
\newcommand{\mb}{\mbox{\ }}
\newcommand{\vs}{\vspace}
\newcommand{\ra}{\rightarrow}
\newcommand{\la}{\leftarrow}
\newcommand{\ul}{\underline}
\newcommand{\ds}{\displaystyle}
\definecolor{LightCyan}{rgb}{0.0, 1, 0.94}

\newcommand{\as}{Institute of Physics, Academia Sinica, Taipei 11529, Taiwan}
\newcommand{\deu}{Department of Physics,
Dokuz Eyl\"{u}l University, Buca, \.{I}zmir TR35160, T\"{u}rkiye}
\newcommand{\itu}{Department of Physics Engineering,
  Istanbul Technical University, Sarıyer, İstanbul TR34467, T\"{u}rkiye} 
\newcommand{\ktu}{Department of Physics,
  Karadeniz Technical University, TR61080 Trabzon, T\"{u}rkiye}
\newcommand{\scu}{College of Physics, Sichuan University, Chengdu 610065, China}
\newcommand{\sbcu}{Department of Physics, School of Physical and Chemical Sciences, Central University of South Bihar, Gaya 824236, India}
\newcommand{\gla}{Department of Physics, Institute of Applied Sciences and Humanities, GLA University, Mathura 281406, India}
\newcommand{\garw}{Department of Physics, H.N.B. Garhwal University, Srinagar 246174, India}
\newcommand{\bhu}{Department of Physics, Institute of Science, Banaras Hindu University, Varanasi 221005, India}

\newcommand{\corrmd}{muhammed.deniz@deu.edu.tr}
\newcommand{\corrhtw}{htwong@gate.sinica.edu.tw}
\newcommand{\corrsk}{karadags@itu.edu.tr}

\title{Constraints on new physics with light mediators and
  generalized neutrino interactions via coherent elastic
  neutrino nucleus scattering}

\author{S.~Karada\u{g}} \altaffiliation[Corresponding Author: ]
       { \corrsk } \affiliation{\itu} \affiliation{\as}
\author{ M.~Deniz } \altaffiliation[Corresponding Author: ]
       { \corrmd } \affiliation{ \deu }
\author{ S.~Karmakar } \affiliation{\as} \affiliation{\gla}
\author{ M.~K.~Singh } \affiliation{\as} \affiliation{\bhu}
\author{M.~Demirci} \affiliation{\ktu}
\author{Greeshma ~C.} \affiliation{\as} \affiliation{\sbcu}
\author{H.~B.~ Li} \affiliation{\as}
\author{ S.~T.~Lin } \affiliation{\scu}
\author{M.~F.~Mustamin} \affiliation{\ktu}
\author{V.~Sharma}\affiliation{\garw}
\author{L.~Singh} \affiliation{\sbcu}
\author{ M.~K.~Singh } \affiliation{\gla}
\author{V.~Singh} \affiliation{\sbcu}
\author{ H.~T.~Wong } \affiliation{\as}

\collaboration{The TEXONO Collaboration}

\date{\today}

\begin{abstract} 

We investigate new physics effects on coherent elastic neutrino nucleus scattering within the framework of nonstandard interactions and generalized neutrino interactions. Additionally, we examine the possibility of light mediators from a simplified model that includes all possible Lorentz-invariant interactions of vector, axialvector, scalar, pseudoscalar, and tensor types. Constraints and allowed regions at the $90\%$ CL for masses and couplings in each new physics scenario have been obtained through the analysis of TEXONO data, which includes two datasets from a high-purity $n$-type point contact germanium detector in 2016 and an advanced $p$-type point contact Ge detector in 2025. The results are presented in comparison with other reactor and accelerator-based neutrino experiments for complementarity.


\end{abstract}


\maketitle

\section{Introduction}\label{sec:intro}
The interactions of neutrinos with matter~\cite{pdg2024} have long been considered promising avenues for revealing new physical phenomena. The neutrino-nucleus elastic scattering process~\cite{Freedman:1973yd, Freedman:1977},

\begin{equation}
  \nu_\alpha(\bar{\nu}_\alpha) + N(\mathcal{Z},\mathcal{N}) \longrightarrow
  \nu_\alpha(\bar{\nu}_\alpha) + N(\mathcal{Z},\mathcal{N}),
\end{equation}
where $\alpha$ stands for lepton flavor ($e$, $\mu$ or $\tau$) for incoming or outgoing neutrinos and $N(\mathcal{Z}$, $\mathcal{N})$ denotes the atomic nucleus, is characterized by its atomic number ($\mathcal{Z}$) and neutron number ($\mathcal{N}$). This process is also referred to as coherent elastic neutrino-nucleus scattering (CE$\nu$NS) and serves as an important channel for testing the Standard Model (SM) of electroweak theory, especially in the low energy regime.

\subsection{Coherent elastic neutrino-nucleus scattering (CE$\nu$NS)} \label{subsec:vN}
CE$\nu$NS, which involves neutral current (NC) interactions between neutrinos and quarks in an atomic nucleus, is essentially a tree-level process defined within the SM. However, radiative corrections such as those resulting from loop diagrams also contribute to this interaction and must also be taken into account in precise calculations~\cite{Erler:2005}. In this process, neutrinos interact coherently with a nucleus, whether protons or neutrons, via neutral boson exchange at low energies. Coherent means that neutrinos can reach the interior of the nucleus without breaking it, i.e. the interaction occurs with the nucleus as a whole. Therefore, the wavelength of the process is larger than the nuclear radius and results in a momentum exchange on the order of a few hundred MeV. This means that the complexity of the strong interaction can be encapsulated in a nuclear form factor. The process yields the largest cross section compared to other neutrino interactions in the low energy regime so far. The cross section depends on the number of nucleons involved. In particular, the observed cross section increases with the square of the number of neutrons, since the weak mixing angle affects and limits the contribution from proton interactions. However, this process is difficult to detect because the nuclear recoil energy is limited to the low keV scale. This criterion is necessary to ensure that the nucleus interacts coherently with the neutrino. This challenging goal was finally achieved in 2017 with the advancement of accelerator neutrino technology by the COHERENT Collaboration~\cite{Akimov:2017ade,Akimov:2022csi}, several decades after its initial formulation~\cite{Freedman:1973yd}.

Since the first detection of CE$\nu$NS, numerous efforts have been made to better understand the phenomena. The COHERENT Collaboration updated their experiment using liquid Argon (LAr)~\cite{Akimov:2020pdx}. They also announced the first detection of CE$\nu$NS with a $p$-type point-contact germanium (pPCGe) detector, reinforcing the importance of Ge detectors in precision studies of low-energy neutrino interactions~\cite{COHERENT-Ge:arXiv2024} and plan to further upgrade their experimental facilities by adapting a liquid Xenon detector as a target~\cite{COHERENT:2015}. The CENNS team at Fermilab also aims to detect the CE$\nu$NS process using the booster neutrino beam at the facility~\cite{Brice:2013fwa}. Furthermore, reactor neutrino experiments are also contributing to the detection of the process in order to achieve lower nuclear threshold energy. The CONUS Collaboration provided strong limits~\cite{Bonet:2020awv} with Ge detectors at the Brokdorf Nuclear Power Plant in Germany. The follow-up CONUS+ project recently reported preliminary results~\cite{CONUSpl:2025} on a positive observation of CE$\nu$NS at a significance of 3.7$\sigma$ at the Leibstadt nuclear power plant in Switzerland, marking a significant step in reactor neutrino CE$\nu$NS searches. Alongside CONUS, other reactor-based efforts include TEXONO~\cite{Wong:2006} with a Ge detector located at the Kuo-Sheng Nuclear Power Plant in Taiwan; CONNIE~\cite{Aguilar-Arevalo:2019zme} with a charge coupled device (CCD) detector array at the Angra Power Plant in Rio de Janeiro, Brazil; RICOCHET~\cite{Chooz:2017} with both Ge-based and metallic Zn-based smaller detector at the Chooz Nuclear Power Plant in France; $\nu$GeN~\cite{Nugen:2022} with a high purity Ge detector at the Kalinin Nuclear Power Plant in Russia; Dresden-II~\cite{Dresden:2021} with a p-type point contact Ge detector in the USA,  MINER~\cite{Miner:2017} with low threshold cryogenic germanium and silicon detectors at the Nuclear Science Center at Texas in the USA; $\nu$CLEUS~\cite{nucleus:2017} with CaWO$_4$ and Al$_2$O$_3$ calorimeter arrays and several cryogenic veto detectors operated at millikelvin temperatures; and the RED-100~\cite{Red:2022} with two-phase Xenon emission detector at the Kalinin Nuclear Power Plant.

Nuclear power plant experiments have high potential for observing the CE$\nu$NS process, as they meet the coherency criteria ($>95\%$) due to their high rate of low-energy neutrino flux~\cite{Kerman:2016}. Dark matter (DM) experiments, such as XENONnT~\cite{XENON:2020kmp}, DARWIN~\cite{DARWIN:2016hyl}, LZ~\cite{LZ:2019sgr}, and PandaX-4T~\cite{PandaX:2018wtu}, are also influenced by this channel, as CE$\nu$NS plays a crucial role in the neutrino-floor phenomenon, providing an unavoidable background for DM experiments.  Recently, the PandaX-4T~\cite{Panda:2024} and XENONnT~\cite{Xenon:2024} experiments have reported the first measurements of nuclear recoils from solar $^8$B neutrinos via CE$\nu$NS with the corresponding statistical significance of 2.64$\sigma$ and 2.73$\sigma$, respectively, representing the first experimental step toward the neutrino fog.

\subsection{Nonstandard interactions in CE$\nu$NS} \label{subsec:NSIcevns}
The CE$\nu$NS process opens up numerous opportunities to expand our understanding of physics. For example, it allows the investigation of neutrino and nuclear properties, such as study of the neutrino magnetic moment~\cite{Vogel:1989, Billard:2018jnl}. Recent research has also explored potential explanations of sterile neutrinos~\cite{Miranda:2020syh}. Additionally, nuclear properties may be examined through this process~\cite{Cadeddu:2017etk}. CE$\nu$NS plays an important role in astrophysical processes~\cite{Freedman:1973yd, Wilson:1974} and has been proposed for detecting supernova neutrinos~\cite{Horowitz:2003}, providing a compact and transportable neutrino detector for real-time monitoring nuclear reactors~\cite{Learned:2005}, and studying neutron density distributions~\cite{Patton:2012}. Moreover, the potential appearance of new fermions linked to neutrino mass and their role in explaining DM has been recently investigated~\cite{Brdar:2018qqj}. The study of weakly interacting massive particles~\cite{Liu:2017drf, Schumann:2019eaa} as a candidate of DM is also related to CE$\nu$NS, as both share a similar nuclear recoil energy range in the few keV scale. CE$\nu$NS events from solar and atmospheric neutrinos represent an irreducible background~\cite{Monroe:2007, Gutlein:2010} for future DM direct detection experiments~\cite{Drees:2014,DarkSide-20k}. Additionally, measuring of the CE$\nu$NS scattering cross section serves as a sensitive probe for new physics~\cite{Krauss:1991, Barranco:2005, Scholberg:2006}, potentially enabling the detection of new light bosons~\cite{Farzan:2018gtr,Pablo:2024}, which is crucial for advancing our understanding of novel interactions in nature.

Probing nonstandard interactions (NSI) through neutrino-nucleus scattering has been a subject of intense interest in recent years~\cite{Farzan:2018gtr, Pablo:2024, Barranco:2005, Barranco:2011wx, CONUSpl:2025-BSM, Lindner:2017, Miranda:2015, Amir:2019, Amir:2021, Venegas:CONUSpl2025, Chattaraj:CONUSpl2025}. These studies have investigated various NSI scenarios, such as those based on simplified models focusing on light mediators~\cite{Kosmas:2017tsq,CONUS:2022, CONNIE:2020, DresdenII:2022, Corona:2022, Miranda:2020nsi,Coloma:2022nsi,Lindner:2024nsi,Cadeddu:2021nsi, DemirciM:2024nsi,Majumdar:prd2022,Romeri-exo3:jhep2023,Romeri:DM-LM2024}, model-independent approaches effective in the high mediator mass regime~\cite{Kosmas:2017tsq,Giunti:2019xpr, Mustamin:2021mtq, NSItensor1:prd2024, Chatterjee:2023nsi,Canas:2020nsi,Romeri-exo3:jhep2023,Coherent-Ge:PRD2024}, and exotic neutral current interactions~\cite{Flores-exo2:prd2022, Romeri-exo3:jhep2023, AristizabalSierra:2018eqm,Majumdar:DresdenExo}, along with experimental observables and implications for neutrino flavor physics~\cite{flavor:ICube2021,Miranda:2015,atmospheric:2021} and astrophysical processes~\cite{nsi:status2019, DM:Sierra2015,DM:Park2023}.  
In addition, the studies combining complementary sources such as reactor and accelerator neutrinos~\cite{CONUSpl:2025-BSM,Romeri:CONUSpl2025,Coloma:2022nsi} have helped to improve sensitivity to multiple interaction types. Collectively, these studies provide a strong foundation for further investigation of a wide range of NSI scenarios and underscore the need for ongoing research using both current and future neutrino experimental data.

\subsection{Scope and structure of this work} \label{subsec:structure}
In this work,  we provide a comprehensive analysis of the TEXONO reactor neutrino experiment's sensitivity to all five possible Lorentz-invariant NSI interaction types, both within a simplified model and model-independent frameworks. We explore a simplified model as an alternative approach to investigating physics beyond the SM (BSM). Specifically, we consider generalized neutrino interactions (GNI) as well as the possibility of light mediators from simplified model, which includes vector (V), axialvector (A), scalar (S), pseudoscalar (P), and tensor (T) types. Each is assumed to involve light mediators that couple to neutrinos and the quark constituents of the nucleus. Notably, recent studies have employed an effective neutrino-quark approximation that includes these interaction types~\cite{AristizabalSierra:2018eqm}. The connections between this effective approximation and NSI have also been examined~\cite{Bischer:2019ttk}. Simplified model provides an alternative framework in particle physics to bridge theoretical prediction with experimental measurements. The experimental parameters considered in this model are limited to cross sections, new particle masses, and couplings. It is important to emphasize that, unlike comprehensive BSM extensions involving numerous particles and decay chains, this model does not fully represent the complexities of actual physics~\cite{Mccoy:2018}. Nevertheless, explanatory models remain valuable for consideration, including those that can be embedded within potential SM extensions, either dependently or independently.

In this work, in particular, we present the first NSI analysis based on TEXONO experiments, using both $n$-type point-contact germanium (nPCGe) (2016) data~\cite{Soma:2016} and recently released pPCGe (2025) data~\cite{Texono:2024}. In the TEXONO experiment, which studies neutrino interactions in a reactor environment dominated by low-energy electron antineutrinos, we aimed to provide a systematic assessment of how advances in detector technology have increased sensitivity to NSI parameters. We also intend to highlight the important role of the TEXONO experiment in CE$\nu$NS studies by comparing our results with the existing bounds from other experiments in the literature, both reactor-based and accelerator-based. To this end in this work, we cover all possible interaction types in both the simplified model and model-independent approach, as well as exotic neutral current interactions.

We also note that the $\nu$–e scattering channel, while also sensitive to NSI effects and widely studied in the literature~\cite{nsiastrophys, reactornubsm, nsiboundlsnd, nsiboundcombined, nsimuon}, is beyond the scope of this work due to its distinct kinematic features and systematic considerations. A proper treatment of this channel requires a dedicated and comprehensive analysis. Specifically, TEXONO Collaboration has previously investigated new physics BSM with this channel given in Refs.~\cite{nueNSI:MD2010, nueNSI:SB2015, nueNSI:MD2017, nueNSI:BS2017}. This channel can also be studied specifically for generalized NSI using both TEXONO-CsI(Tl)~\cite{nue:MD2010} and TEXONO-Ge\cite{Texono:2024} data and eventually combines the results with CE$\nu$NS constraints within a more comprehensive NSI framework. 
For additional context, we refer readers to Refs.\cite{Majumdar:prd2022, Candela:JHEP2024, Romeri-exo3:jhep2023, CONUSpl:2025-BSM, nue:Xu2019, Bonet:2020awv}, which examine NSI sensitivity through $\nu$–e scattering across different experimental settings.

In the following sections, we first provide a brief discussion of the theoretical formulation of CE$\nu$NS, covering differential cross sections and the effect of the form factor in Sec.~\ref{sec:formalism}. In Sec.~\ref{sec:gni}, we delve into GNI derived from the simplified model. The formalism and structure of the model-independent NSI of neutrinos and the exotic neutral current interactions are given in Secs.~\ref{sec:nsi} and ~\ref{sec:nsi-exo}, respectively. The observation of CE$\nu$NS by analysis of the TEXONO experiment and the calculation of expected event rates based on both SM and BSM NSI theories are discussed in Sec.~\ref{sec:cevns}. Our analysis methods and final results are introduced in Sec.~\ref{sec:resdis}, starting from the relevant Lagrangian for the new interactions considered in the model and using $\chi^2$-analysis methods. This section also presents our findings, including new constraints on coupling and mass parameter spaces, as well as on four-Fermi pointlike interactions of $\varepsilon_{\alpha \beta}^{fX}$ parameter space and exotic neutral-current interactions of $\xi_X$ parameter space, along with discussions of potential implications. Finally, we summarize and conclude our work in Sec.~\ref{sec:conc}.

\begin{figure}[!htb]
	\centering
	\includegraphics[scale=0.18]{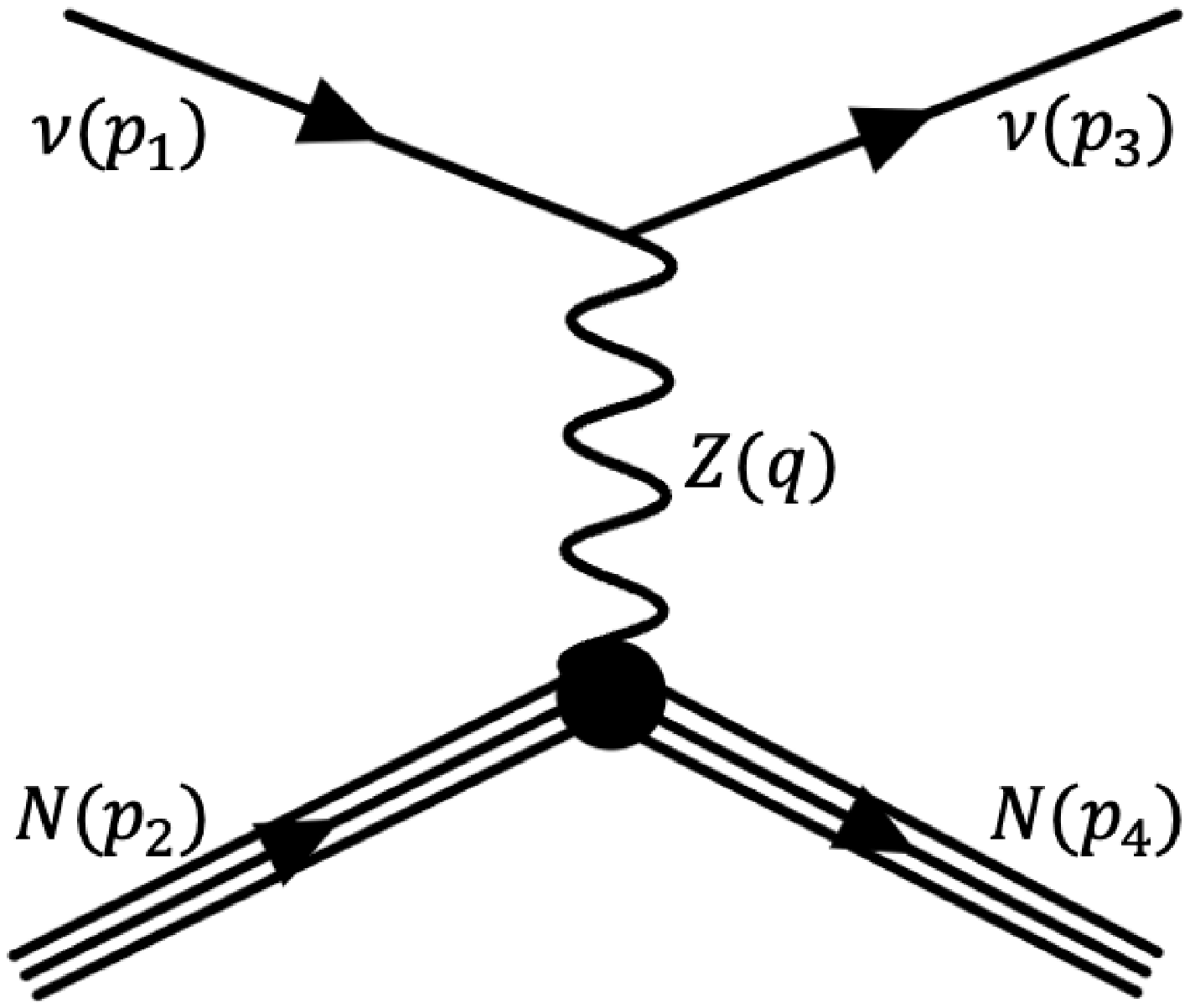}
	\caption{Feynman diagram for the CE$\nu$NS process.}
	\label{fig:cevns}
\end{figure}

\section{Formalism} \label{sec:formalism}
The scattering amplitude for the CE$\nu$NS process, as shown in Fig.~\ref{fig:cevns}, can be expressed as ($q^2 \ll m_Z^2$)
\begin{align}
   \mathscr{M} &= \frac{G_F}{\sqrt{2}} \bar{\nu}(p_3) \gamma^\mu(1-\gamma^5) \nu(p_1)  \notag \\
   & \times {\overline{N}}(p_4)\gamma_\mu \left(Q_{\text{SM}_V}-\gamma^5Q_{\text{SM}_A}\right)
   N(p_2),
\end{align}
where $G_F$ is the Fermi constant. We have used $G_F/\sqrt{2}=g_W^2/8m_W^2$ and $m_W = m_Z \cos\theta_W$, where $g_W$ and $m_W(m_Z)$ are the weak coupling constant and $W(Z)$ boson mass, respectively. 

The SM differential cross section, incorporating both vector and axialvector weak charge 
components can be written as,
 \begin{align} 
\left[{\frac{d\sigma}{dT_N}} \right]_{\text{SM}} &=\frac{G_{F}^{2} m_N}{4 \pi}
\Big( Q_{\text{SM}_V}^2 \mathcal{K}_V +  Q_{\text{SM}_A}^2 \mathcal{K}_A  \Big), 
\label{eq:smcevns}
\end{align}
where $m_N$ is the nucleus mass, and the kinetic terms of vector and
axialvector, $\mathcal{K}_V$ and $\mathcal{K}_A$, can be described respectively as
\begin{align}
  \mathcal{K}_V &= 1 - \frac{m_N T_N}{2 E_{\nu}^2}
  - \frac{T_N}{E_{\nu}} + \frac{T_N^2}{2E^2_{\nu}},  \nonumber \\
  \mathcal{K}_A &= 1 + \frac{m_N T_N}{2 E_{\nu}^2}  
  - \frac{T_N}{E_{\nu}}+ \frac{T_N^2}{2E^2_{\nu}}.
\end{align}
The vector part of the SM weak charge $Q_{\text{SM}_V}$ can be written as
\begin{align}
Q_{\text{SM}_V} &= g_p^V \mathcal{Z} F_{\mathcal{Z}}(q^2) + g_n^V \mathcal{N}
F_{\mathcal{N}}(q^2) \notag \\ 
&= \left[\mathcal{Z} \left(1-4\sin^2\theta_W \right)
  - \mathcal{N} \right] F_V(q^2),
\label{eq:qsmv}
\end{align} 
where $\sin^2\theta_W$ is the weak mixing angle with a value of 0.23867~\cite{pdg2024} at $q^2 \simeq 0$~\cite{Erler:2005}, and $g_n^V = -1 $ and $g_p^V = 1 - 4 \sin^2\theta_W \simeq 0.0453$ are tree-level vector neutrino-neutron and neutrino-proton couplings, respectively. Including radiative corrections, these couplings are modified to $g_n^V \simeq -1.0234$ and $g_p^V(\nu_e)\simeq0.0762$,  reflecting flavor dependence of proton coupling for electron neutrinos~\cite{Corona:2024}. The cross section in Eq.~\eqref{eq:smcevns} is enhanced by $\mathcal{N}^2$ due to the multiplication factor of $g_p^V$, making the contribution from the $\mathcal{Z}$ boson relatively small. Despite the smallness of this factor, it remains important because it includes the weak mixing angle, $\sin^2\theta_W$.

Nuclear physics effects in CE$\nu$NS are incorporated via the weak nuclear form factors $F_\mathcal{N}$ for neutrons and $F_\mathcal{Z}$ for protons given in Eq.~\eqref{eq:qsmv}.
In this work, these are assumed to be equal such that ${F_{\mathcal{N}}(q^2) = F_{\mathcal{Z}}(q^2) \equiv F_V(q^2)}.$ The three-momentum transfer $(q \equiv \vert \vec{q} \vert)$ in terms of the measurable nuclear recoil energy of the target is given by $ q^2 = 2 m_N T + T^2 \simeq 2 m_N T $. On the theoretical side, the CE$\nu$NS cross section depends on the nuclear matrix element models and the nuclear structure details, which can be introduced through the nuclear form factors for protons and neutrons. In this study, we consider the Helm parametrization for the form factor~\cite{Helm:1956zz},
\begin{equation}
F_{\text{Helm}}(q^2) = 3\frac{j_1(qR)}{qR} e^{-\frac{1}{2}q^2 s^2}. \label{eq:ffhelm}
\end{equation}
Here, $j_1(x)=\sin x/x^2-\cos x/x$ is the  first order spherical Bessel
function. Nuclear radius is given by $R=\sqrt{c^2 - 5s^2}$,
with the nuclear parameters $c=1.2 \mathcal{A}^{1/3} $ fm, $s=0.5$ fm~\cite{Duda:2006uk}.

Although the SM axialvector term in the CE$\nu$NS cross section in Eq.\eqref{eq:smcevns} plays a minor role due to its strong dependence on the nuclear spin, we have nonetheless accounted for it in our analyses. The axial contribution to the SM weak charge, $Q_{SM_A}$, is determined through the axialvector form factor $F_A(q^2)$, derived from nuclear spin structure function calculations, which describe the momentum dependence of the nuclear spin response in scattering processes, as detailed in Ref.~\cite{SpSn:prd2020},
\begin{align}
F_A(q^2) =&\frac{8 \pi}{2J+1} \Big[ (g_A^0)^2 \ S_{00}^{\mathcal{T}}(q^2) + g_A^0 g_A^1 \ S_{01}^{\mathcal{T}}(q^2) \notag \\
& + (g_A^1)^2 \ S_{11}^{\mathcal{T}}(q^2) \Big], 
\label{eq:ffA}
\end{align}
where $J$ is the total angular momentum of the nucleus in its ground state, $S_{ij}^{\mathcal{T}}(q^2)$ is the transverse spin structure function,  and $g_A^0$ and $g_A^1$ are the isoscalar and isovector axial coupling coefficients, respectively.  In the SM case, only the $S_{11}^{\mathcal{T}}(q^2)$ term contributes. The isovector coefficients is given by,
\begin{align}
 g_A^1=\frac{g_p^A-g_n^A}{2} = \Delta_u^p - \Delta_d^p = \Delta_d^n - \Delta_u^n,
\label{eq:isovec}
\end{align}
where $\Delta_q^{\mathfrak{n}=n,p}$ represents the contribution of the quark-spin content of the nucleon $\mathfrak{n}$. The axial coupling coefficients are defined as $g_n^A=\Delta_u^n-\Delta_d^n$ with $g_p^A=-g_n^A$. Neglecting strangeness and two-body currents, the axial vector contribution to the SM weak charge can be expressed as,
\begin{equation}
 Q_{SM_A}^2 = \frac{8 \pi}{2J+1}  (\Delta_u^p - \Delta_d^p)^2 \ S_{11}^{\mathcal{T}}(q^2).
\end{equation}
The numerical values of coefficients are $ \Delta_u^p=\Delta_d^n  = 0.842$,  and $ \Delta_d^p=\Delta_u^n  = -0.427$~\cite{SpSn:prd2020,Belanger-ffval:2014}. The most common stable Ge isotopes considered, along with their natural abundances, are $^{70}$Ge (20.8\%), $^{72}$Ge (27.5\%), $^{73}$Ge (7.8\%), $^{74}$Ge (36.5\%), and $^{76}$Ge (6.2\%).  Among these, only $^{73}$Ge has a nonzero nuclear spin ($J = 9/2$); the others have $J = 0$ due to their even numbers of protons and neutrons. The axialvector contribution to the cross section vanishes for spin-zero nuclei.

\subsection{New light mediators from simplified model} \label{sec:gni}
In this study we focused on new neutrino physics caused by new light mediators. Light new physics is motivated by the lack of signals in the collider experiments. Consequently, the coherent neutrino nucleus scattering at low energies has been a good candidate channel to test new physics scenarios with light mediator interactions~\cite{Farzan:2018gtr}. We also consider explicit future realizations of reactor antineutrinos to probe coherent scattering and its contribution to light new physics BSM. Therefore, in our work, we extend the scope beyond this standard framework to also consider light mediator models.

\begin{figure}[!htb]
	\centering
	\includegraphics[scale=0.18]{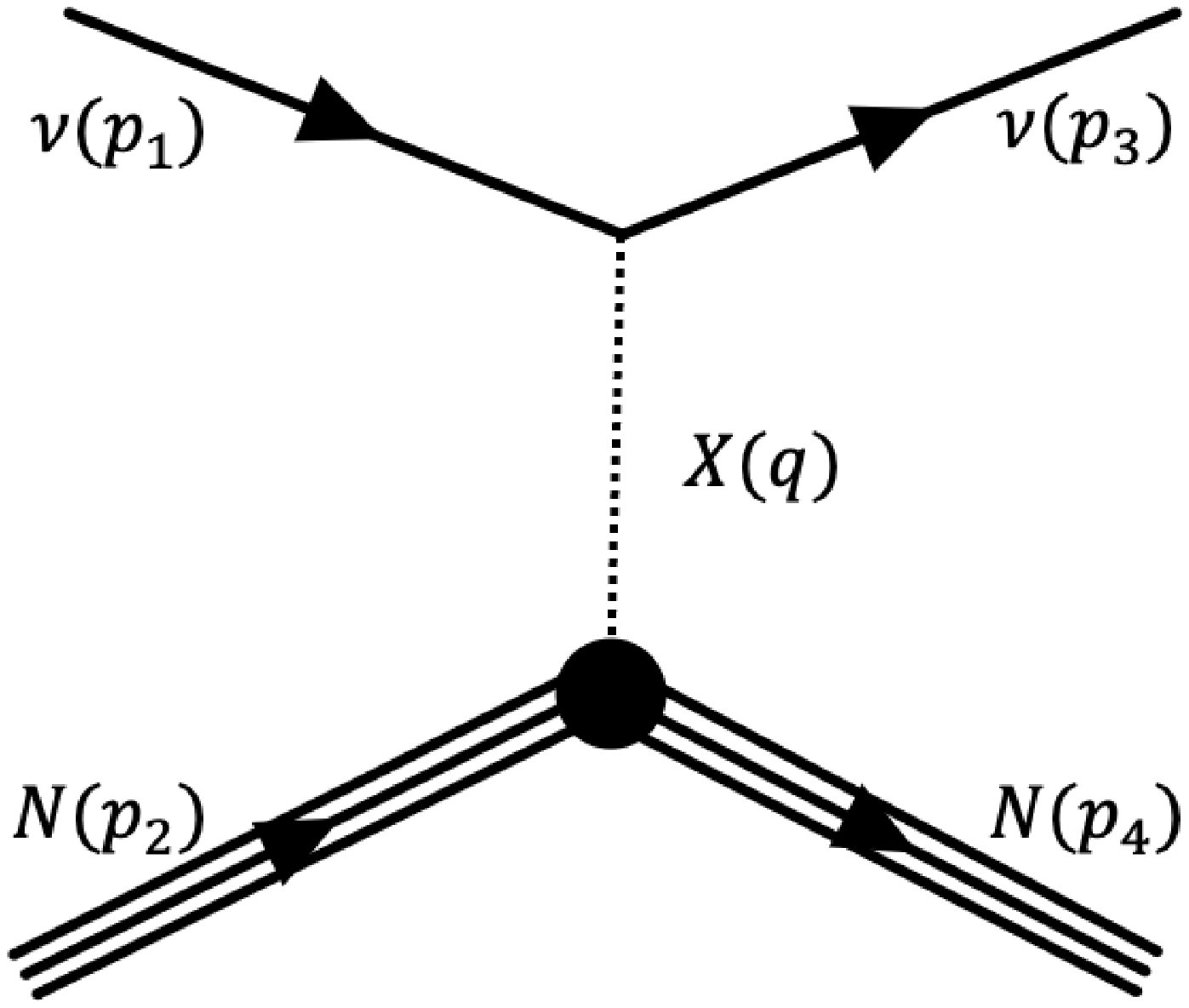}
	\caption{Representative Feynman diagram for CE$\nu$NS process with new mediator X.}
	\label{fig:cevns_newmed}
\end{figure}

To parametrize new physics at a very low scale, the light mediator models are constructed to include only a few new particles and interactions in general vector, axialvector, scalar, pseudoscalar and tensor form~\cite{Cerdeno:2016sfi, Barranco:2011wx}. The appearance of new interactions in the CE$\nu$NS process is illustrated in Fig.~\ref{fig:cevns_newmed}. These are so-called simplified models that have been also intensively studied in the dark matter searches at the LHC~\cite{Abdallah:2015}. For measurements of neutrino-electron scattering or neutrino-nucleus scattering, such models are of interest because the mediators can have a pronounced effect on the recorded recoil spectra, most notably for mediator masses smaller than the maximum momentum transfer. The Lagrangians responsible for these interactions can be written as
\begin{align}
  \mathscr{L}_V \supset [g_{V\nu} \bar{\nu}_L \gamma^\mu \nu_L
    + g_{Vq} \bar{q} \gamma^\mu q] V_\mu, \notag \\
  \mathscr{L}_A \supset [g_{A\nu} \bar{\nu}_L \gamma^\mu \nu_L
    - g_{Aq} \bar{q} \gamma^\mu \gamma^5 q] A_\mu, \notag \\
  \mathscr{L}_S \supset [(g_{S\nu} \bar{\nu}_R \nu_L + H.c.)
    + g_{Sq} \bar{q} q] S, \notag \\
  \mathscr{L}_P \supset [(g_{P\nu} \bar{\nu}_R \nu_L + H.c.)
    - ig_{Pq}\gamma^5 \bar{q} q] P, \notag \\
  \mathscr{L}_T \supset [g_{T\nu} \bar{\nu}_R \sigma^{\mu\nu} \nu_L
    - g_{Tq} \bar{q} \sigma^{\mu\nu} q] T_{\mu\nu},
  \label{eq:lagrange}
\end{align}
where $\sigma_{\mu\nu} = \frac{i}{2}[\gamma_\mu,\gamma_\nu] = \frac{i}{2}
(\gamma_\mu\gamma_\nu-\gamma_\nu\gamma_\mu)$. 

With these Lagrangians, we can match the new mediator interactions from the quark level to the nuclear scale. We find the following for each interaction type:
\begin{equation}
	\mathscr{L}_{XN} = g_{XN} \overline{N}N X,
  \label{eq:nuc-scale}
\end{equation}
where $X=V,~A,~S,~P,~T$. 

The coupling of the new mediators with the nucleus for vector, scalar, and pseudoscalar interactions can simply be written as,
\begin{align}
  g_{VN} =& \Big[\mathcal{Z}\sum_q Q'^p_qg_{Vq}
  + \mathcal{N}\sum_q Q'^n_qg_{Vq}\Big] F_V(q^2), \notag \\
  g_{SN} =& \Big[\mathcal{Z}\sum_{q} g_{Sq}\frac{m_p}{m_q}f_{Tq}^p
  + \mathcal{N}\sum_{q} g_{Sq} \frac{m_n}{m_q} f_{Tq}^n \Big] F_S(q^2),  \notag \\
g_{PN} =&  \Big[\mathcal{Z} \sum_{q} g_{Pq} \frac{m_p}{m_q} \Delta_q^p + \mathcal{N} \sum_{q} g_{Pq} \frac{m_n}{m_q} \Delta_q^n  \Big] \notag \\
&\times \Big(1-  \sum_{q'}\frac{\bar{m}}{m_{q'}}\Big)  \  F_P(q^2),
  \label{eq:couplingVSP}
\end{align}
where $\bar{m}$ is constant defined as $\bar{m}=(1/m_u + 1/m_d + 1/m_s)^{-1}$ \cite{Cirelli:JCAP2013,Cheng:JHEP2012}. Here, the spin dependence of the pseudoscalar interaction is neglected, given that its cross section is suppressed by the recoil energy, and the associated spin contribution is subdominant~\cite{Candela:JHEP2024}.

For axialvector and tensor interactions the definitions of the coupling constants can be expressed in terms of individual form factors, which include spin structure functions. The relevant spin structure function is defined as $S_{00}^{\mathcal{T}}(q^2)$. In the case of tensor interactions, the longitudinal component of the spin structure function, $S_{00}^{\mathcal{L}}(q^2)$, also contributes and should be included in the analysis.
The axialvector and tensor quark couplings satisfy the relations $g_{Au}=g_{Ad} \equiv g_{Aq}$ and $g_{Tu}=g_{Td} \equiv g_{Tq}$, respectively.  Following the definition in Eq~\eqref{eq:ffA},  the effective couplings at the nucleus level can be expressed as,
\begin{align}
  g_{AN} &= \sqrt{\frac{8 \pi}{2J+1}}  \Big[ g_{Aq}( \Delta_u^p +  \Delta_d^p) \Big], \notag \\
  g_{TN} &= \sqrt{\frac{8 \pi}{2J+1}} \Big[  g_{Tq} (\delta_u^p +\delta_d^p) \Big].  
  \label{eq:couplingAT}
\end{align}
The parameters in Eqs.~\eqref{eq:couplingVSP} and ~\eqref{eq:couplingAT} are listed in Table~\ref{tab:simparval}. It is worth noting that values for parameters involving strange quarks can be found in the literature~\cite{Belanger-ffval:2014,SpSn:prd2020}; however, we have observed that their impact on our calculations is negligible. Therefore, we only consider the first lepton family quarks, namely the up and down quarks. In addition, we emphasize that the pseudoscalar interaction with the nucleus was considered negligible in previous works~\cite{Cerdeno:2016sfi, Dent:2016wcr,Chattaraj:2025}, whereas in this study, we incorporate this interaction into our analysis to provide a more comprehensive treatment. Furthermore, while tensor and axialvector interactions are evaluated using the form factors derived from nuclear spin structure functions, we retain the Helm form factor parametrization from Eq.~\eqref{eq:ffhelm} for vector, scalar, and pseudoscalar interactions.
\begin{table}[h!]
  \begin{center}
    \caption{Values of the parameters for simplified models.}
    \begin{tabular}{|c|c|c|}    
      \hline
      \textbf{Parameter} & \textbf{Value} &
      \textbf{Reference} \\
      \hline
      $m_u$ & 2.16 & \multirow{5}{*}{~\cite{pdg2024}} \\ 
      $m_d$ & 4.67 & \\
      $m_s$ & 93.4 & \\
      $m_n$ & 939.6 & \\
      $m_p$ & 938.3 &\\
      \hline
	  $Q'^p_u = Q'^n_d$ & 2 & \multirow{2}{*}{~\cite{Corona:2022}} \\ 
	  $Q'^n_u = Q'^p_d$ & 1 &  \\
	  \hline
      $f_{T_u}^p$ & 0.0208 & \multirow{4}{*}
          {~\cite{SpSn:prd2020,Hoferichter:prl2015}} \\
	  $f_{T_u}^n$ & 0.0411 &  \\
	  $f_{T_d}^p$ & 0.0189 &  \\
	  $f_{T_d}^n$ & 0.0451 &  \\
	  \hline
	  $\Delta_u^p = \Delta_d^n$ & 0.842 & \multirow{2}{*}
              {~\cite{SpSn:prd2020,Belanger-ffval:2014}} \\
	      $\Delta_u^n = \Delta_d^p$ & $-$0.427&  \\
	  \hline
	  $\delta_u^p = \delta_d^n$ & 0.784 & \multirow{2}{*}
              {~\cite{SpSn:prd2020,Gupta:prd2018}} \\
	      $\delta_u^n = \delta_d^p$ & $-$0.204&  \\
  	      \hline 
    \end{tabular}
    \label{tab:simparval}
  \end{center}
\end{table}

Based on these considerations, we obtain the differential cross sections for the case of,
 
\begin{enumerate}[label=(\roman*)]
  
\item Vector interaction, 

Vector differential cross section without interference can be written as  
\begin{equation}
\left[ {\frac{d\sigma }{dT_{N}}}\right]_{V} = \frac{G_{F}^{2}m_{N}}{4\pi }
\frac{2Q_{V}^{2}}{{\left( m_{V}^{2}+2m_{N}T_{N} \right)^{2}}}~\mathcal{K}_V.
\label{eq:dSdT_V}
\end{equation}
The interference term of the vector differential cross section can be written as 
\begin{align}
  \left[ {\frac{d\sigma }{dT_{N}}}\right]_{V_{\text{INT}}} &= \frac{G_{F}^{2}m_{N}}{4\pi}
  \frac{2\sqrt{2}Q_{V} Q_{\text{SM}_{V}}}{m_{V}^{2}+2m_{N}T_{N}}~\mathcal{K}_V .
 \label{eq:dSdT_Vint}
\end{align}

\item Axialvector interaction, 
 
\begin{equation}
\left[ {\frac{d\sigma }{dT_{N}}}\right]_{A} = \frac{G_{F}^{2}m_{N}}{4\pi}
\frac{2 Q_{A}^{2}}{\left( m_{A}^{2}+2m_{N}T_{N}\right)^{2}} S_{00}^{\mathcal{T}}(q^2)~\mathcal{K}_A.
  \label{eq:dSdT_A}
\end{equation}

Although the SM CE$\nu$NS cross section given in Eq.~\eqref{eq:smcevns} includes an axialvector component, the interference between the SM and axialvector contributions are not
considered here since the relevant spin-structure function for the SM case is $S_{11}^{\mathcal{T}}(q^2)$, while for the axialvector case it is $S_{00}^{\mathcal{T}}(q^2)$~\cite{SpSn:prd2020,Chattaraj:2025}.

\item Scalar interaction, 
\begin{equation}
\left[{\frac{d\sigma}{dT_N}} \right]_{S} = \frac{G_{F}^{2} m_N}{4 \pi}
\frac{Q_{S}^{2}}{\left( m_{S}^2 + 2 m_N T_N \right)^2}~\mathcal{K}_S,
  \label{eq:dSdT_S}
\end{equation}
where $\mathcal{K}_S$ can be defined as
\begin{equation}
\mathcal{K}_S = \frac{m_N T_N}{ E_{\nu}^2} + \frac{T_N^2}{2E_\nu^2}.
\end{equation}

\item Pseudoscalar interaction, 
\begin{equation}
\left[{\frac{d\sigma}{dT_N}} \right]_{P}= \frac{G_{F}^{2} m_N}{4 \pi}
\frac{Q_{P}^{2}}{\left( m_{P}^2 + 2 m_N T_N \right)^2}~\mathcal{K}_P,
  \label{eq:dSdT_P}
\end{equation}
where $\mathcal{K}_P$ can be defined as
\begin{equation}
\mathcal{K}_P = \frac{T_N^2}{2E_\nu^2}.
\end{equation}

\item Tensor interaction,
\begin{align}
\left[{\frac{d\sigma}{dT_N}} \right]_{T} =& \frac{G_{F}^{2} m_N}{4 \pi}
 \frac{Q_{T}^2}{\left( m_{T}^2 + 2 m_N T_N \right)^2} \notag \\
 \times & \bigg[ 8 S_{00}^{\mathcal{T}}(q^2)~\mathcal{K}_V  + 16 S_{00}^{\mathcal{L}}(q^2)  \Big( 1 - \frac{T_N}{E_{\nu}}\Big) \bigg].
  \label{eq:dSdT_T}
\end{align}

\end{enumerate}
In all these equations the new charges $Q_X$ can be defined as, 
\begin{equation}
Q_X = \frac{g_{X\nu} g_{XN} }{G_F},
\end{equation}
where $X$ stands for the different interaction channels of V, A, S, P, and T.

\subsection{Model independent nonstandard interactions of neutrino} \label{sec:nsi}
\begin{figure}[!htb]
	\centering
	\includegraphics[scale=0.18]{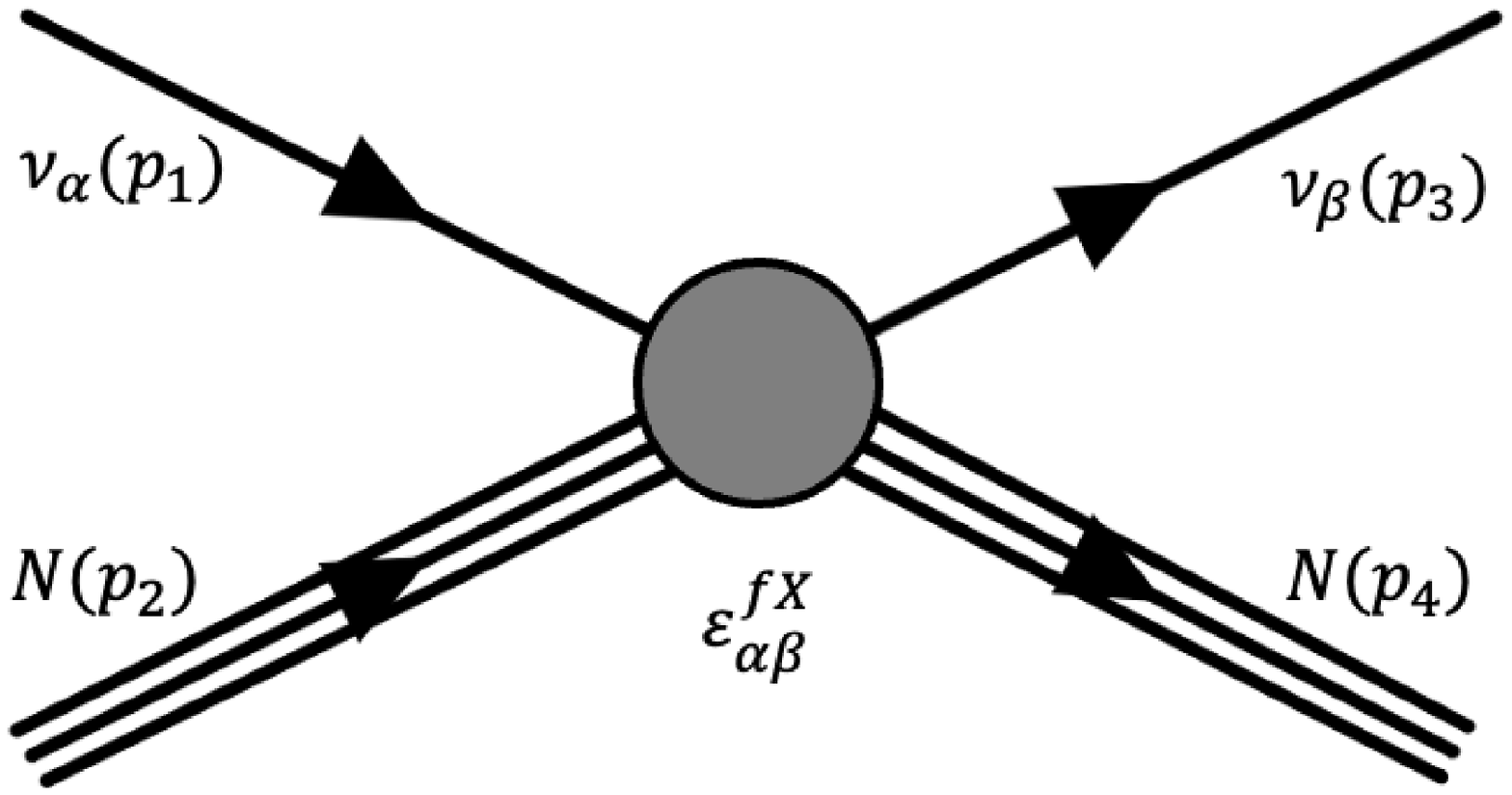}
	\caption{Feynman diagram for four-Fermi pointlike interaction.}
	\label{fig:feyn_point}
\end{figure}

One of the most extensively studied models in the search of new physics BSM is the NSI of neutrinos~\cite{Davidson:2003}. NSI of neutrinos, in a model-independent approach, can be described as a four-Fermi pointlike interaction, as shown in Fig.~\ref{fig:feyn_point}. 
This model introduces a new type of contact interaction between neutrinos and matter that differs from the SM theory~\cite{Miranda:2015}. In general, NSI can induce a NC and charge current (CC) processes, each implying the presence of a new mediator with heavier mass~\cite{Berezhiani:2002} or in the same order as the electroweak (EW) theory~\cite{Farzan:2016}. For CE$\nu$NS, only the NC case is relevant. Neutrinos with NSI influence quarks inside a nucleus, interacting either as a nonuniversal flavor-conserving (FC) or flavor-violating (FV) process. 
In general, the Lagrangians given in terms of the model-independent coupling constant in Eq.~\eqref{eq:lagrange} can be modified to define the model-independent NSI by replacing the coupling $g_X$ with $\varepsilon$  using the relation given in Eq.~\eqref{eq:g-epsilon}. Specifically, the V-A type for model-independent NSI can be formulated as
\begin{equation}
  -\mathscr{L}_{NSI}^{eff} = \varepsilon_{\alpha \beta}^{fP} 2 \sqrt{2} G_F
  (\bar{\nu}_\alpha \gamma_\mu L \nu_\beta) (\bar{f'}\gamma^\mu P f),
\label{eq:NSI}
\end{equation}
with $\varepsilon_{\alpha\beta}^{fP}$ as the coupling that describes the strength of the NSI, $f$ is a first generation SM fermion (e, u, or d) in which $ f = f' $ refers to NC NSI, whereas CC NSI are obtained for $f' \neq f$, $P$ denotes the chiral projector
$ L, \, R = (1\pm \gamma^5)/2 $, and $\alpha$ and $\beta$
denote the neutrino flavor $(e,\, \mu, \, \text{or}\, \tau)$.

The relevant Lagrangian for the CE$\nu$NS NC process can be written as
\begin{equation}
  -\mathscr{L}_{NSI}^{eff} = \frac{G_F}{\sqrt{2}}
  [\bar{\nu}_\alpha \gamma_\mu (1-\gamma^5)\nu_\beta] [\bar{f}\gamma^\mu
  (\varepsilon_{\alpha \beta}^{fV}-\varepsilon_{\alpha \beta}^{fA}\gamma^5 )f].
\label{eq:NSI-VA}
\end{equation}

The NSI coupling $\varepsilon_{\alpha \beta}^{fX}$ can be expressed as
\begin{equation}
  \varepsilon_{\alpha \beta}^{fX} \simeq \frac{g_{X\nu}g_{XN}}{\sqrt{2}G_F M_X^2},
  \label{eq:g-epsilon}
\end{equation}
where $g_X$ and $m_X$ stand for the coupling constant and mass of the mediator, respectively.

Hence, the differential cross section of NSI can be written as,

\begin{enumerate}[label=(\roman*)]

\item for the vector case 
\begin{align}
  \left[ {\frac{d\sigma }{dT_{N}}}\right]_{V}^{NSI} =&
  \frac{G_{F}^{2}m_{N}}{\pi } |F_V(q^2)|^2
  \Big[ {\varepsilon_{\alpha \beta}^{uV}}
    \left( \mathcal{N}+2\mathcal{Z} \right) \notag \\
    &  + {\varepsilon_{\alpha \beta}^{dV}}
    \left( 2\mathcal{N}+\mathcal{Z} \right) \Big]^2~\mathcal{K}_V,
  \label{eq:dSdT_V_eps}
\end{align}
where $\alpha = \beta = e, ~\mu, ~\tau$ is corresponding to the FC type of interaction and $\alpha \neq \beta = e, ~\mu, ~\tau$ is corresponding to the FV case.

The interference term for the FC case only can be written as 
\begin{align}
  \left[ {\frac{d\sigma }{dT_{N}}}\right]_{V_{\text{INT}}}^{FC}  =&
  \frac{G_{F}^{2}m_{N}}{\pi} |F_V(q^2)|^2 \notag \\
  & \times \Big[ \varepsilon_{\alpha \alpha}^{uV}
    \left( \mathcal{N}+2\mathcal{Z} \right) +
     \varepsilon_{\alpha \alpha}^{dV}
    \left( 2\mathcal{N}+\mathcal{Z} \right) \Big] \notag \\
  & \times \Big[ \left(1 - 4 \sin^2\theta_W \right)\mathcal{Z}
    -\mathcal{N} \Big]~\mathcal{K}_V, 
  \label{eq:dSdT_Vint_eps}
\end{align}

\item for the axialvector case; 
\begin{align}
  \left[ {\frac{d\sigma }{dT_{N}}}\right]_{A}^{NSI} =&
  \frac{G_{F}^{2}m_{N}}{\pi } \Big(\frac{8 \pi}{2J+1}\Big) \notag \\
  \times &\frac{1}{4}\Big[\varepsilon_{\alpha \beta}^{uA} \Big(\Delta_u^p \sqrt{S_p^{\mathcal{T}}(q^2)} + \Delta_u^n  \sqrt{S_n^{\mathcal{T}}(q^2)} \Big) \notag \\
&+ \varepsilon_{\alpha \beta}^{dA} \Big(\Delta_d^p \sqrt{S_p^{\mathcal{T}}(q^2)}+ \Delta_d^n \sqrt{S_n^{\mathcal{T}}(q^2)}\Big) \Big]^2 \  \mathcal{K}_A,
  \label{eq:dSdT_A_eps}
\end{align}
where $S_{\mathfrak{n}}^{\mathcal{T}}(q^2)$ denotes the transverse spin structure functions in a proton/neutron basis, with $\mathfrak{n}=p,n$.
In this section, we adopt the proton/neutron basis, which enables a clearer identification and evaluation of the individual spin-dependent contributions from protons and neutrons, and provides a more direct connection to the associated quark-level NSI couplings. For a detailed analysis and discussion of this formulation, see Ref.~\cite{SpSn:prd2020}.

In the standard NSI framework, the effective parameters $\varepsilon_{\alpha \beta}^{fX}$  are typically defined for vector and axial-vector interactions only. In this work, however, we adopt a phenomenological extension of this notation to scalar, pseudoscalar, and tensor interactions, for consistency with the generalized Lorentz-invariant S, P, V, A, T operator basis~\cite{NSItensor1:prd2024}. The differential cross section of S, P, and T type NSI can be expressed using relevant Lagrangian given in Eqs.~\eqref{eq:lagrange} and ~\eqref{eq:nuc-scale} with the definition of couplings in Eqs.~\eqref{eq:couplingVSP} and ~\eqref{eq:couplingAT} as

\item for the scalar case;
\begin{align}
  \left[ {\frac{d\sigma }{dT_{N}}}\right]_{S}^{NSI}
  =& \frac{G_{F}^{2}m_{N}}{2\pi } |F_S(q^2)|^2 \notag \\
   \times& \left[{\varepsilon_{\alpha \beta}^{uS}}
    \Big(\mathcal{N}\frac{m_n}{m_u}f^n_{Tu}+\mathcal{Z}
    \frac{m_p}{m_u}f^p_{Tu} \Big) \right. \notag \\
    & \left. + {\varepsilon_{\alpha \beta}^{dS}}
   \Big (\mathcal{N}\frac{m_n}{m_d}f^n_{Td} +
    \mathcal{Z}\frac{m_p}{m_d}f^p_{Td} \Big)  \right]^2~\mathcal{K}_S,
  \label{eq:dSdT_S_eps}
\end{align}

\item for the pseudoscalar case;
\begin{align}
  \left[ {\frac{d\sigma }{dT_{N}}}\right]_{P}^{NSI} =&
  \frac{G_{F}^{2}m_{N}}{2\pi } |F_P(q^2)|^2 \notag \\
   \times& \bigg[{\varepsilon_{\alpha \beta}^{uP}}
    \Big(\mathcal{N}\frac{m_n}{m_u}\Delta^n_u+\mathcal{Z}
    \frac{m_p}{m_u}\Delta^p_u \Big)  \notag \\
    & + {\varepsilon_{\alpha \beta}^{dP}}
    \Big(\mathcal{N}\frac{m_n}{m_d}\Delta^n_d+\mathcal{Z}\frac{m_p}{m_d}\Delta^p_d \Big)
    \bigg]^2 \notag \\
 \times &  \Big(1-  \sum_{q'}\frac{\bar{m}}{m_{q'}}\Big)^2~\mathcal{K}_P,
  \label{eq:dSdT_P_eps}
\end{align}

\item for the tensor case;
\begin{align}
  \left[ {\frac{d\sigma }{dT_{N}}}\right]_{T}^{NSI}& =
 \frac{G_{F}^{2}m_{N}}{\pi } \Big(\frac{8 \pi}{2J+1}\Big) \notag \\
\times \bigg\{ \Big[&\varepsilon_{\alpha \beta}^{uT} \Big(\delta_u^p \sqrt{S_p^{\mathcal{T}}(q^2)} + \delta_u^n  \sqrt{S_n^{\mathcal{T}}(q^2)}\Big) \notag \\
+& \varepsilon_{\alpha \beta}^{dT} \Big(\delta_d^p \sqrt{S_p^{\mathcal{T}}(q^2)}+ \delta_d^n \sqrt{S_n^{\mathcal{T}}(q^2)}\Big) \Big]^2~\mathcal{K}_V \notag \\
+ \Big[&\varepsilon_{\alpha \beta}^{uT} \Big(\delta_u^p \sqrt{S_p^{\mathcal{L}}(q^2)} + \delta_u^n  \sqrt{S_n^{\mathcal{L}}(q^2)}\Big) \notag \\
+& \varepsilon_{\alpha \beta}^{dT} \Big(\delta_d^p \sqrt{S_p^{\mathcal{L}}(q^2)}+ \delta_d^n \sqrt{S_n^{\mathcal{L}}(q^2)}\Big) \Big]^2   \Big( 2 - \frac{2T_N}{E_{\nu}}\Big) \bigg\},
  \label{eq:dSdT_T_eps}
\end{align}
where $S_{\mathfrak{n}}^{\mathcal{T}}(q^2)$ and $S_{\mathfrak{n}}^{\mathcal{L}}(q^2)$  denote the transverse and longitudinal spin structure functions in proton/neutron basis, respectively.
We follow a similar notation as in the axialvector case. For tensor NSI, both transverse and longitudinal spin structure functions contribute to the cross section~\cite{SpSn:prd2020}. 

\end{enumerate}

\subsection{``Exotic'' neutral currents in coherent $\nu$-N scattering} \label{sec:nsi-exo}
Apart from the model-independent NSI in epsilon-space $\varepsilon_{\alpha \beta}^{qX}$ additional, more ``exotic'' new interactions could be presented. 
The $\varepsilon_{\alpha \beta}^{qX}$ coefficients are defined as dimensionless four-Fermi operators normalized to $G_F$, in the context of standard or generalized NSI.
On the other hand, the ``exotic'' interaction parameters appear in the GNI formulation based on a quark-level and nucleus scale effective Lagrangian and include all Lorentz-invariant couplings. These parametrizations differ in interpretation and origin but can be related via matching procedures in the heavy mediator limit. Five possible types of such ``exotic'' interactions can be introduced namely vector (V), axialvector (A), scalar (S), pseudoscalar (P), and tensor (T). 
The Lagrangian at the quark level for these interactions, considering all Lorentz-invariant interactions of neutrinos with first generation quarks, can be formulated as
\begin{equation}
  \mathscr{L} = \frac{G_F}{\sqrt{2}} \sum_{X} \bar{\nu}\Gamma^X\nu
  \Big[\bar{q}\Gamma^X(C_X^q + i \gamma^5 D_X^q )q \Big].
  \label{eq:lag-exo}
\end{equation}
Here $X$ represents V, A, S, P, and T. We only consider the first generation of quarks, where q stands for u and d quarks, and $\Gamma^X$ can be described as 
\begin{equation}
  \Gamma^X = \{\gamma^\mu,\gamma^\mu\gamma^5, I, i\gamma^5, \sigma^{\mu\nu}\}.
  \label{eq:gamma-exoq}
\end{equation}

Analogous to Eq.~\eqref{eq:g-epsilon}, the coefficients $C_X^q $ and $D_X^q$ are expected to be of the order $(g_X^2/g^2) (m_W^2 /m_X^2 )$, where $m_X$ represents the mass of the new exchange particles and $g_X$ denotes their coupling constants. The coefficients $C_X^q $ and $D_X^q$ in 
Eq.~\eqref{eq:lag-exo} are dimensionless and, in principle, can be complex numbers. However for simplicity, we assume these parameters $C_X^q $ and $D_X^q$ to be real~\cite{Lindner:2017}.

The effective Lagrangian of neutrino-nucleus interaction, transition from the quark level to the nucleus scale, can be expressed as,
\begin{align}
  \mathscr{L}_{\nu-N} & \supset
  \sum_{X=S,V,T} \bar{\nu}\Gamma^X\nu \Big[\overline{N} C_X \Gamma^X N\Big]
  \nonumber \\
  & + \sum_{(X,Y)=(S,P),(V,A)}\bar{\nu}\Gamma^X\nu
  \Big[\overline{N}i D_X \Gamma^Y N \Big].
  \label{eq:lag-exoN}
\end{align}
Here the expression for $C_X$ and $D_X$ can be derived from Eqs.~\eqref{eq:couplingVSP} and ~\eqref{eq:couplingAT}. It is worth noting that different notation for the couplings ($g_{XN}$) was used in Sec.~\ref{sec:gni}. The notation adopted here follows the same conventions in Ref.~\cite{AristizabalSierra:2018eqm}.

Since the effective couplings for nuclear spin-dependent interactions are determined by the sum over spin-up and spin-down nucleons, axial and pseudoscalar currents are suppressed in heavy nuclei. Consequently, the most significant contributions are expected from scalar, vector and tensor quark currents. As such, we can constrain the quark currents to be considered as $C^q_S$,  $D^q_P$, $C^q_V$, $D^q_A$, and $C^q_T$. The expression for $D_P$ can be obtained from that of $C_S$ by replacing $C^q_S$ with $D^q_P$, while $D_A$ can be derived from $C_V$ by substituting $C^q_V$ with $D^q_A$~\cite{AristizabalSierra:2018eqm, Flores-exo2:prd2022}.
\begin{align}
  \left[{\frac{d\sigma}{dT_N}} \right]_{EXO} =
  &\frac{G_{F}^{2} m_N}{4 \pi} |F(q^2)|^2 ~\mathcal{N}^2 
\Big[ (\xi_V + \xi_{\text{SM}})^2~\mathcal{K}_V \nonumber \\ 
& +  \xi_A^2~\mathcal{K}_A - 2 \xi_V \xi_A~ \mathcal{K}_{VA} + \xi_S^2~\mathcal{K}_S 
\nonumber \\ & + \xi_T^2~\mathcal{K}_T - R~\mathcal{K}_{VA} \Big], 
\label{eq:exo}
\end{align}
where $\mathcal{K}_T$ and $\mathcal{K}_{VA}$ are
\begin{align}
  \mathcal{K}_T &= 1 - \frac{m_N T_N}{4E_{\nu}^2}
  - \frac{T_N}{E_{\nu}} + \frac{T_N^2}{4E^2_{\nu}}, \nonumber \\
  \mathcal{K}_{VA} &= \frac{T_N^2}{2E_{\nu}^2} -\frac{T_N}{E_{\nu}},
\label{eq:KT}
\end{align}
and the new exotic couplings can be defined in terms of the coefficients
$C_X $ and $D_X$ as
\begin{align}
  & \xi_V^2 = \frac{C_V^2+D_A^2}{\mathcal{N}^2} \nonumber \\ 
  & \xi_A^2 = \frac{D_V^2+C_A^2}{\mathcal{N}^2} \nonumber \\ 
  & \xi_V \xi_A = \frac{C_V D_V + C_A D_A} {\mathcal{N}^2}\nonumber \\  
  & \xi_S^2 = \frac{C_S^2+D_P^2} {\mathcal{N}^2}\nonumber \\ 
  & \xi_T^2 = 8\frac{C_T^2+D_T^2} {\mathcal{N}^2}\nonumber \\ 
  & R = 2\frac{C_PC_T - C_SC_T + D_TD_P - D_TD_S} {\mathcal{N}^2}.
\label{eq:exo-coup}
\end{align}
While different notations exist in the literature to represent the SM contribution~\cite{AristizabalSierra:2018eqm,Flores-exo2:prd2022}, we adopt the convention used in Refs.~\cite{NSItensor1:prd2024,Romeri-exo3:jhep2023}.  Specifically,  $\xi_{\text{SM}} =[\mathcal{Z} \left(1-4\sin^2\theta_W \right)-\mathcal{N}]/ \mathcal{N}$ is used to represent the SM contribution in Eq.~\eqref{eq:exo}. The first two lines of Eq.~\eqref{eq:exo} encompass the vector interaction including the SM contribution, the axialvector interaction, and the scalar interaction, which involves interference with pseudoscalar terms. The last line includes the tensor interaction, while the R term accounts for the interference 
between the (pseudo) scalar and tensor interactions.

\section{Experimental Data and Event Rates} \label{sec:cevns}
\begin{figure}[!htb]
	\centering
	\includegraphics[scale=0.4]{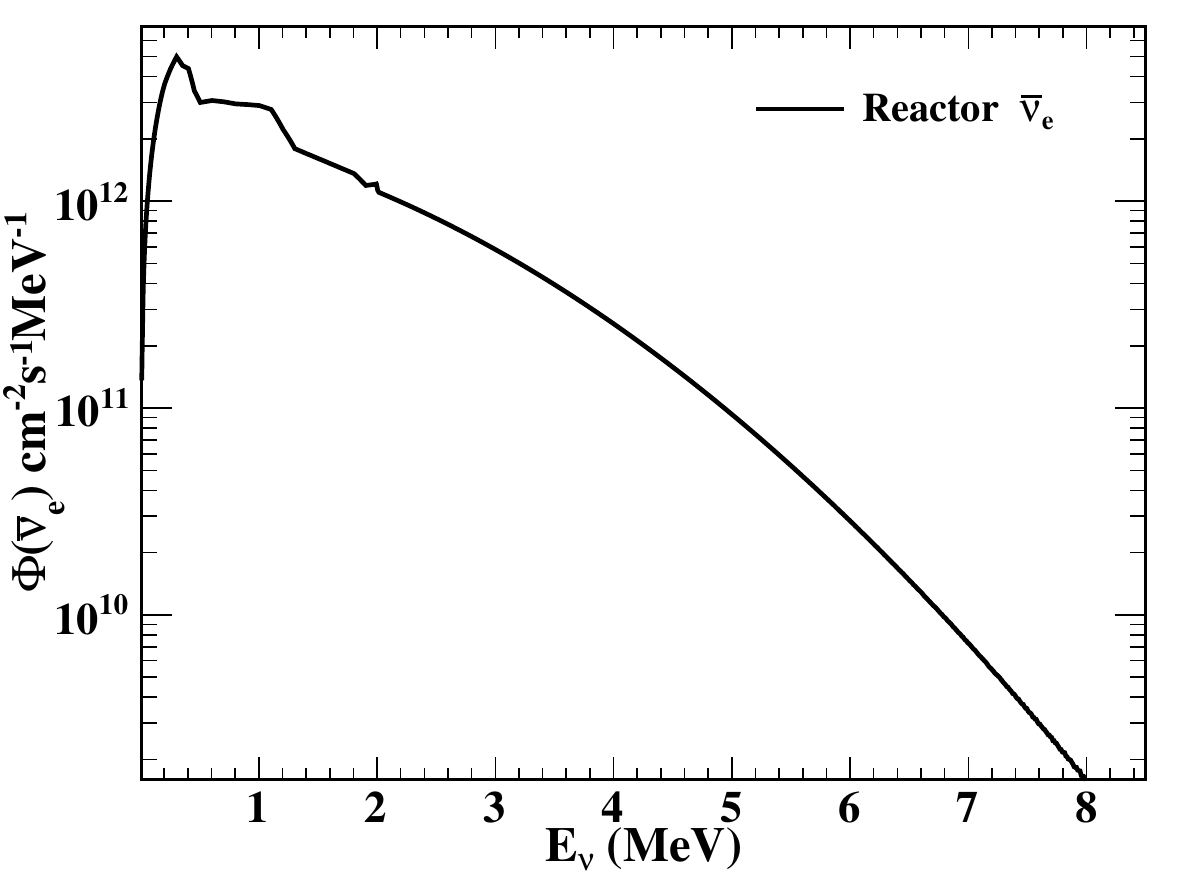}
	\caption{Typical neutrino spectrum $(\Phi_{\nu})$ for reactor $\bar{\nu}_e$ corresponding to a total flux of $6.35 \times 10^{12}~\mathrm{cm}^{-2}\mathrm{s}^{-1}$ at the detector site (see Fig.~5b in Ref.~\cite{Wong:flux2007}.)}
	\label{fig:nuspec}
\end{figure}

The number of events can be calculated from
\begin{equation}
  R_X = N_{tar} \int_{T_{N_{min}}}^{T_{N_{max}}} dT_N  \int_{E_{\nu_{min}}}^{E_{\nu_{max}}} dE_\nu
  \frac{d\Phi}{dE_\nu} \Big[\frac{d\sigma}{dT_N}\Big]_X,
 \label{eq:event-rate}
\end{equation} 
where $X$ represents different interaction channels such as the SM, NSI, etc., $d\Phi/dE_\nu$ is the neutrino spectra shown in Fig.~\ref{fig:nuspec}, $d\sigma/dT_N $ is a differential cross section of the process, and $N_{tar}$ is the total number of target nuclei.

A nuclear recoil threshold of $T_{N_{min}} \simeq 2.96$~eV is the minimum required average ionization energy in Ge to detect neutrinos with energy $E_\nu$. From kinematics, the maximum recoil energy is defined as
$T_{N_{max}} = 2E_\nu^2/(m_N + 2E_\nu) \simeq 2E_\nu^2/m_N $. Here $E_{\nu}$ is running from 0 to $\sim$ 8~MeV with 10~keV bin size.

Scattering of neutrinos off nuclei provides a valuable opportunity to test the SM and to explore potential new physics BSM through various experimental programs. In particular, nuclear reactors and SNS  serve as optimal facilities for studying CE$\nu$NS processes~\cite{Wong:2006, COHERENT:2015, Avignone:2015}. SNS-based neutrino sources, generated from pion and muon decays at rest (DAR$-\pi$), offer a complementary flux in the intermediate-energy range~\cite{COHERENT:2015}. Meanwhile, nuclear reactors produce significantly larger electron antineutrino fluxes (${\bar\nu}_e$) in the low-energy regime compared to other neutrino sources, making them well suited for precision measurements. The TEXONO experiment, utilizing low-threshold detectors, plays a crucial role in probing neutrino interactions and searching for potential NSI effects through reactor-based neutrino experiments. The reactor $\bar{\nu}_e$ spectrum used in our analysis shown in Fig.~\ref{fig:nuspec} is derived from the total spectrum at the typical reactor operation conditions, as given in Fig.~5b in Ref.~\cite{Wong:flux2007}. The total $\bar{\nu}_e$ flux is determined by the reactor thermal power, i.e. $\Phi(\bar{\nu}_e) = 6.35 \times 10^{12}\mathrm{cm}^{-2}\mathrm{s}^{-1}$ at the experimental site.  This value carries a conservative uncertainty of 5\%, consistent with those used in similar reactor-based experiments~\cite{Texono:2024, Fernandez:flux, Colaresi:2022}. This spectrum is used to calculate the expected recoil spectra throughout our analysis.

As part of this effort, the TEXONO Collaboration is conducting an intense research program on low-energy neutrino physics ($<$10 MeV) at the Kuo-Sheng Nuclear Power Plant in Taiwan. The collaboration conducted an experiment in 2016 using a high-purity nPCGe detector cooled by liquid nitrogen and upgraded its facilities in 2025 with an advanced pPCGe detector. In the present study,  124.2(70.3) kg-days reactor ON(OFF) data collected at Kuo-Sheng Neutrino Laboratory with a threshold energy of 300 eV~\cite{Soma:2016},  and 242(357) kg-days reactor ON(OFF) data with an improved threshold energy of 200 eV~\cite{Texono:2024} are analyzed.

In order to predict the number of observable events in the detectors, it is essential to account for detector quenching, resolution, and detection efficiency functions. First, the nuclear recoil energy $T_N$ must be converted into the electron-equivalent energy $E_{ee}$ using the quenching function, $Q_F$, such that
\begin{equation}
E_{ee} = Q_F(T_N)\times T_N.
\end{equation}
Then the differential cross section of $d\sigma/dT_N$ can be converted into
$d\sigma/dE_{ee}$ by
\begin{align}
 \frac{d\sigma}{dE_{ee}} &= \left( \frac{d\sigma}{dT_N} \right)/\left( \frac{dE_{ee}}{dT_N} \right) \notag \\ 
  &= \left( \frac{d\sigma}{dT_N} \right) / \left( \frac{dQ_F}{dT_N} T_N + Q_F \right).
\end{align}

Subsequently, the predicted electron-equivalent energy $E_{ee}$ must be smeared by an energy resolution function in order to transform the calculated event rate into the observable event rate within specific energy bins. Finally, the predicted event rate must be processed through the detector efficiency function to align the predicted and observed rates, ensuring they are directly comparable.

The $Q_F$ for the TEXONO high-purity germanium (HPGe) detector is provided using the standard Lindhard model, expressed as
\begin{equation}
Q_F = \frac{k \  g(\epsilon)}{1+k \ g(\epsilon)},
\end{equation}
where $g(\epsilon)$ is from fitting~\cite{Lewin:1996} with the function of
\begin{equation}
g(\epsilon) = 3\epsilon^{0.15} + 0.7\epsilon^{0.6} + \epsilon ,
\end{equation}
where $\epsilon = 11.5~Z^{-7/3}~T_N$, with $T_N$ in keV, and $k$ is a measure of the electronic energy loss. There is some intense effort on quantifying the impact of quenching factor uncertainties on CE$\nu$NS cross section measurement with a germanium detector~\cite{Bonhomme:2022, Yulun:2025}. In the latest study with a pPCGe(2025) detector, the TEXONO collaboration chose the Lindhard model with parameter $k \approx 0.162$~\cite{Texono:2024}.

The energy resolution  function is described as a Gaussian energy resolution function given as
\begin{equation}
\sigma^2 = \sigma_o^2 + E_{ee} \eta F,
\end{equation}
where $\sigma_o$, obtained by an rms of the test pulser, is $49 \  eV_{ee}$ for the nPCGe (2016) detector and $31.37 \ eV_{ee}$ for the pPCGe (2025) detector,
$\eta = 2.96 \ eV_{ee}$ is the average energy required for photons to create an electron-hole pair in Ge, and $F \approx 0.105$ is the Fano factor taken from Ref.~\cite{Colaresi:2022}.

\section{Analysis and Results}\label{sec:resdis}
\subsection{Event rate predictions} \label{subsec:dRdT}
In this section, we present our analysis results obtained from TEXONO nPCGe(2016) and pPCGe (2025) data for each model considered.

\begin{figure*}[htb!]
	\centering
	\subfigure[]{\includegraphics[scale=0.4]{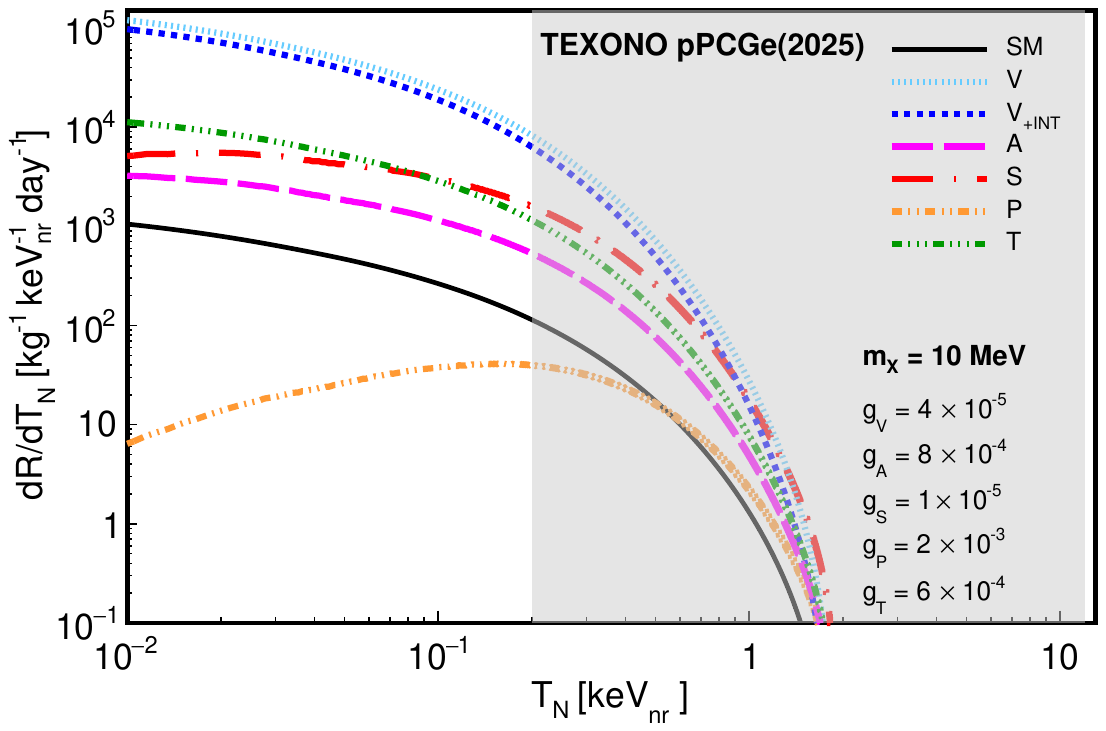}}    
	\subfigure[]{\includegraphics[scale=0.4]{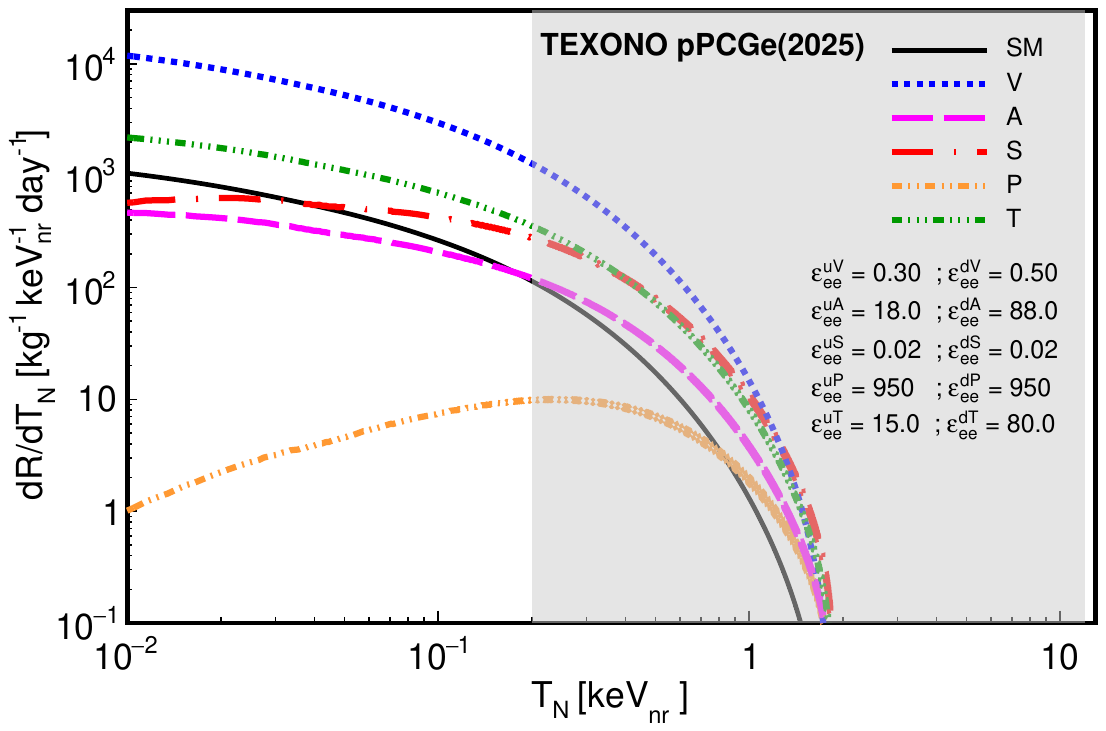}}     
\caption{Event rates as a function of the recoil energy $T_N$ for V, A, S, P, and T type interactions with typical neutrino spectra for (a) new light mediators with specific values of mass and coupling, and for (b) model independent NSI with specific values of $\varepsilon_{ee}^{uX}-\varepsilon_{ee}^{dX}$ parameters.
These values are presented for the TEXONO pPCGe (2025)~\cite{Texono:2024} data. V$_{\text{+INT}}$ represents a V type NSI with interference with the SM. The gray-shaded regions indicate the region of interest for each experiment.}
	\label{fig:dRdT}
\end{figure*}

\begin{figure*}[htb!]
	\centering
	\subfigure[]{\includegraphics[scale=0.4]{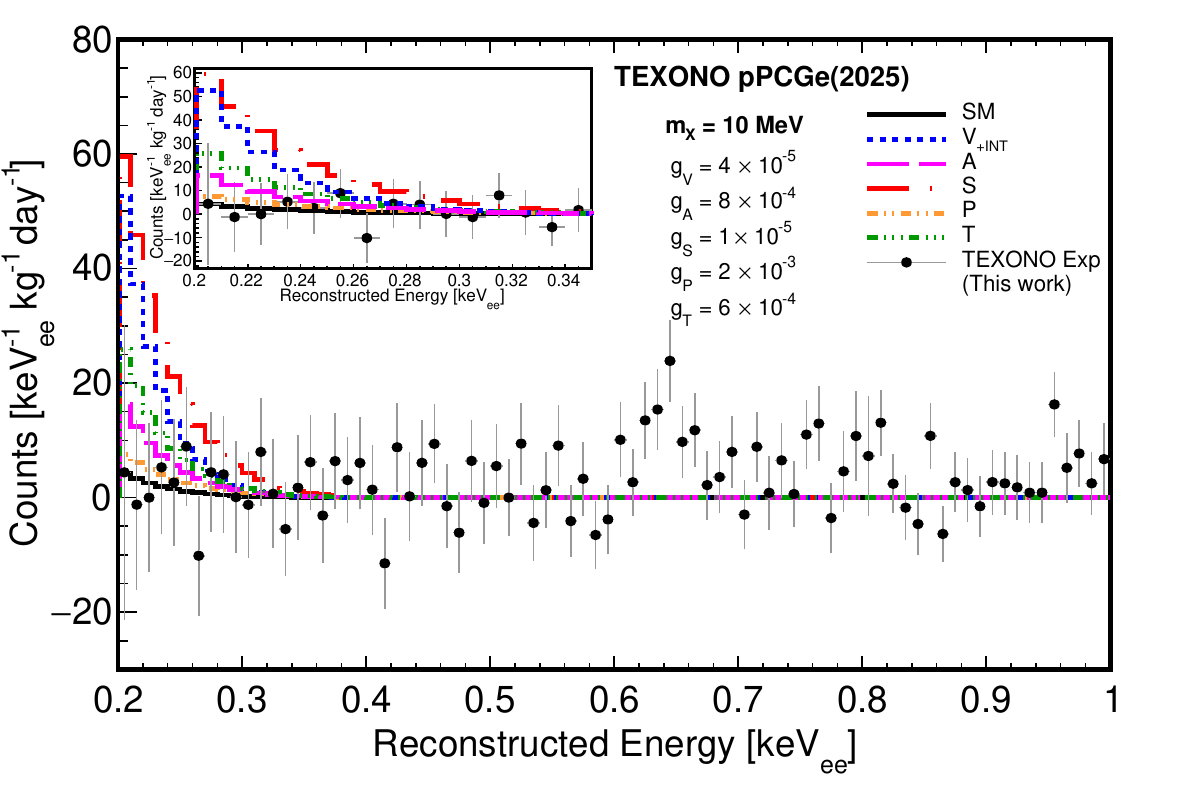}}
	\subfigure[]{\includegraphics[scale=0.4]{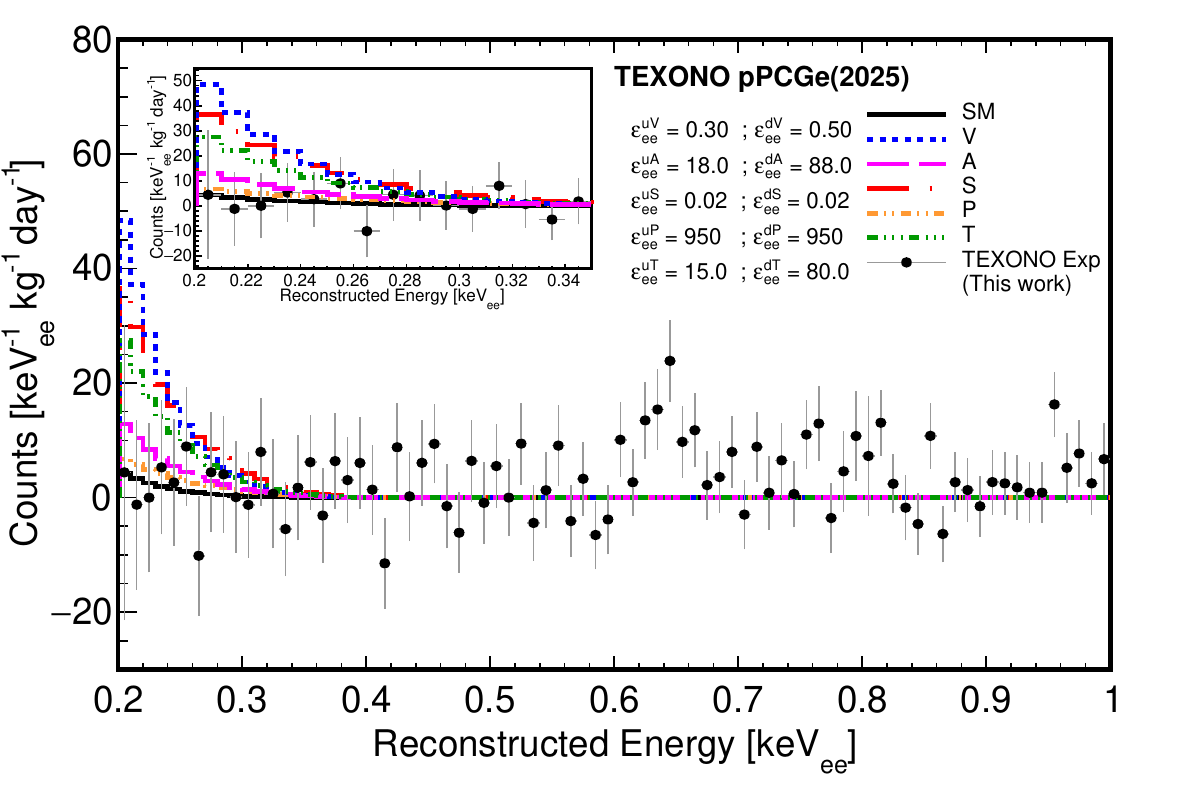}}
\caption{Residual and expected event rate for (a) new light mediators with specific values of mass and coupling, and (b) model independent NSI with specific values of $\varepsilon_{ee}^{uX}-\varepsilon_{ee}^{dX}$ parameters for the TEXONO pPCGe (2025) experiment. The SM prediction is superimposed. These predictions incorporate quenching factor and response function provided by the TEXONO Collaboration~\cite{Texono:2024}. The dotted, dashed, long-dashed, long-dash-dotted, dash-double-dotted, and dash-triple-dotted histograms represent the expected event rates for various NSI scenarios.}
	\label{fig:HIST_COH_TEX}
\end{figure*}

\begin{table} [!hbt]
\caption{Summary of $90\%$ CL upper limits on the coupling constant parameter for V, A, S, P, and T type NSI interactions, obtained from TEXONO nPCGe (2016) and pPCGe (2025) experimental data, for both $m_X^2 \ll 2m_N T_N$ and $m_X^2 \gg 2m_N T_N$ scenarios.}
\label{tab:simp_model_limits}
\begin{ruledtabular}
\begin{tabular}{lcccclcccc}
\multicolumn{3}{c}{($m_X^2 \ll 2m_N T_N$)}  &
\multicolumn{3}{c}{($m_X^2 \gg 2m_N T_N$)} \smallskip \\ \hline \\ 
\multicolumn{1}{l}{Fitting} &
\multicolumn{2}{c}{TEXONO($\times10^{-5}$)} &
\multicolumn{1}{l}{Fitting} &
\multicolumn{2}{c}{TEXONO($\times10^{-6}$)} \smallskip \\
\multicolumn{1}{l}{Par.} &
\multicolumn{1}{c}{nPCGe} &
\multicolumn{1}{c}{pPCGe} &
\multicolumn{1}{l}{Par} &
\multicolumn{1}{c}{nPCGe} &
\multicolumn{1}{c}{pPCGe} \\ \hline \\

$g_V$ & $< 5.54$ & $< 2.57$ & $g_V/m_V$ & $< 4.24$ & $< 2.18$  \\
$g_A$ & $< 148.5$ & $< 61.3$ & $g_A/m_A$ & $< 108.7$ & $< 50.39$  \\
$g_S$ & $< 1.33$ & $< 0.56$ & $g_S/m_S$ & $< 0.97$ & $< 0.46$  \\
$g_P$ & $< 412.3$ & $< 183.0$ & $g_P/m_P$ & $< 296.6$ & $< 147.1$  \\ 
$g_T$ & $< 100.8$ & $< 41.2$ & $g_T/m_T$ & $< 74.2$ & $< 34.02$  \\

\end{tabular}
\end{ruledtabular}
\end{table}

To constrain the parameters, we adapt the analysis method based on the minimum $\chi^2$ method. The TEXONO nPCGe data from 2016 and upgraded pPCGe from 2025 data are analyzed using a one-bin analysis by minimizing
\begin{equation}
  \chi^2 = \sum_{i}\Bigg[\frac{R_{\text{expt}}(i)-R_{\text{SM}}(i)-R_X(i)}{\Delta(i)}\Bigg]^2,
\end{equation}
where $R_{\text{expt}}$ is the measured residual event rate; $R_{\text{SM}}$ and $R_X$ are the expected event rates for the SM; $X$ represents V, A, S ,P, and T, respectively; and $\Delta$ is the $i^{th}$ bin statistical uncertainty published by the experiments.

\begin{figure*}[htb!]
	\centering
	\subfigure[]{\includegraphics[scale=0.4]{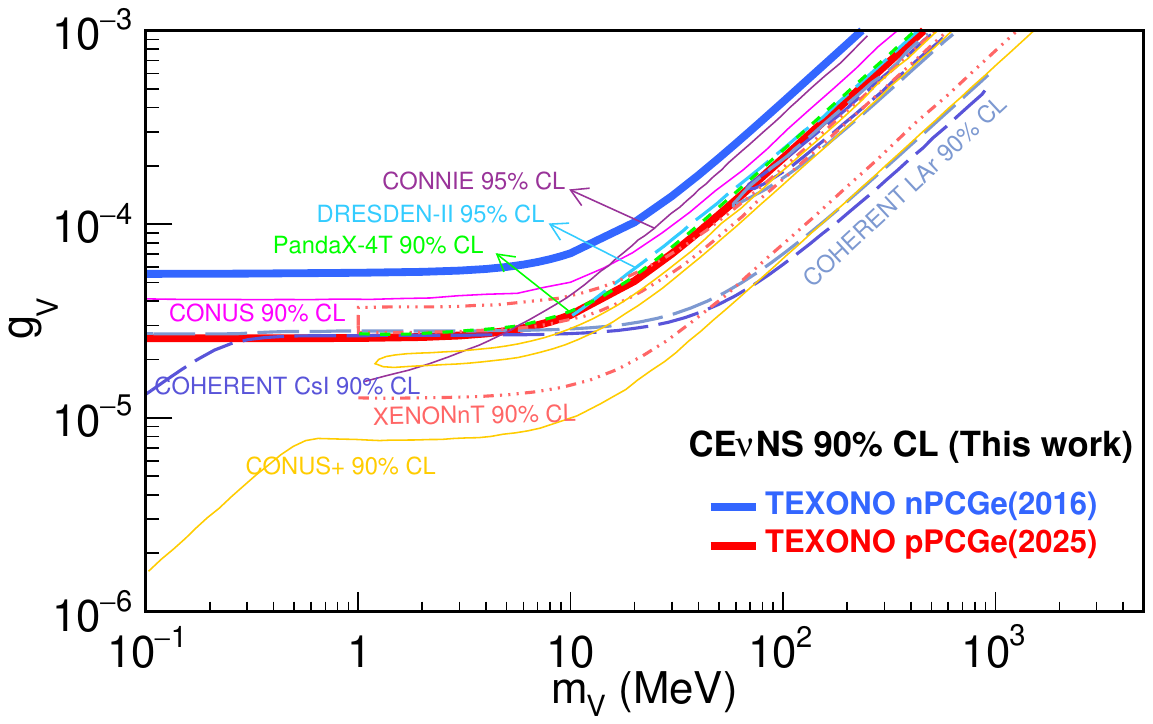}}
	\subfigure[]{\includegraphics[scale=0.4]{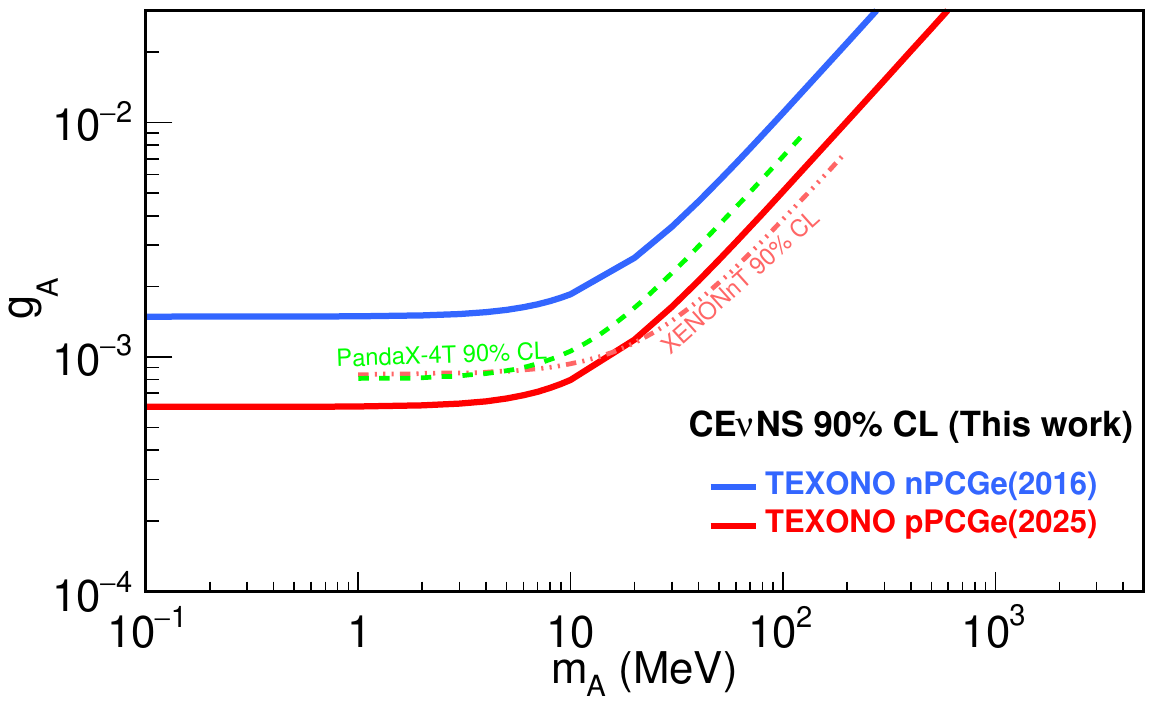}}
         \subfigure[]{\includegraphics[scale=0.4]{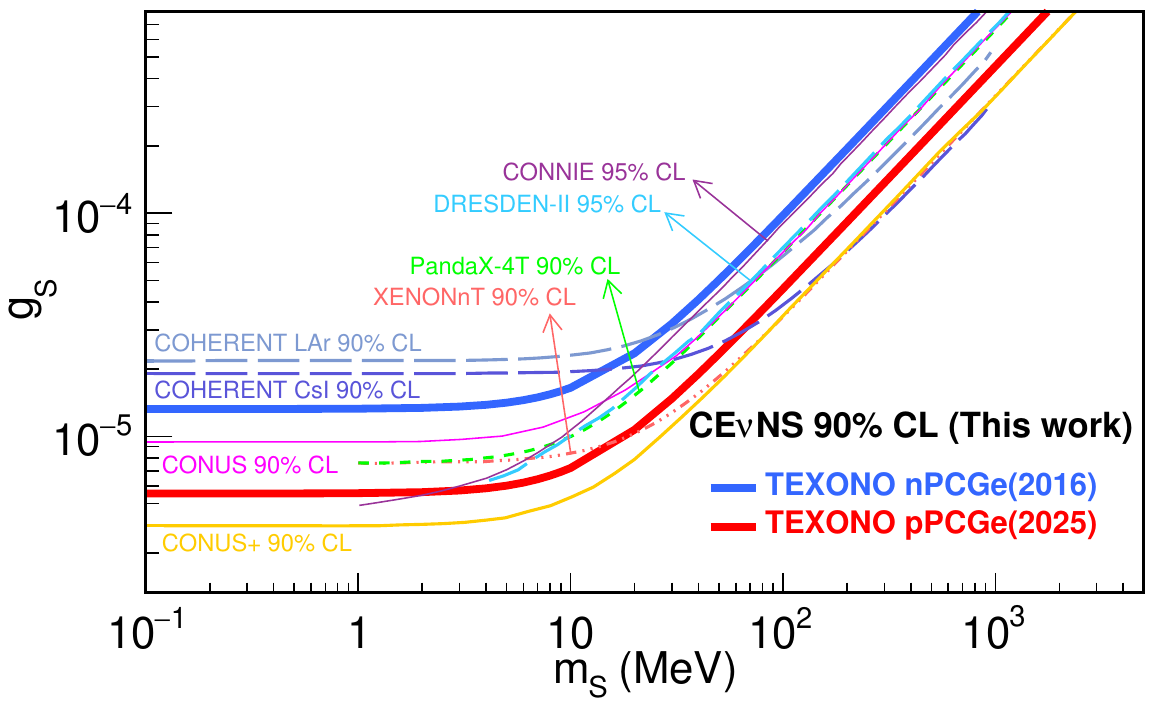}}
	\subfigure[]{\includegraphics[scale=0.4]{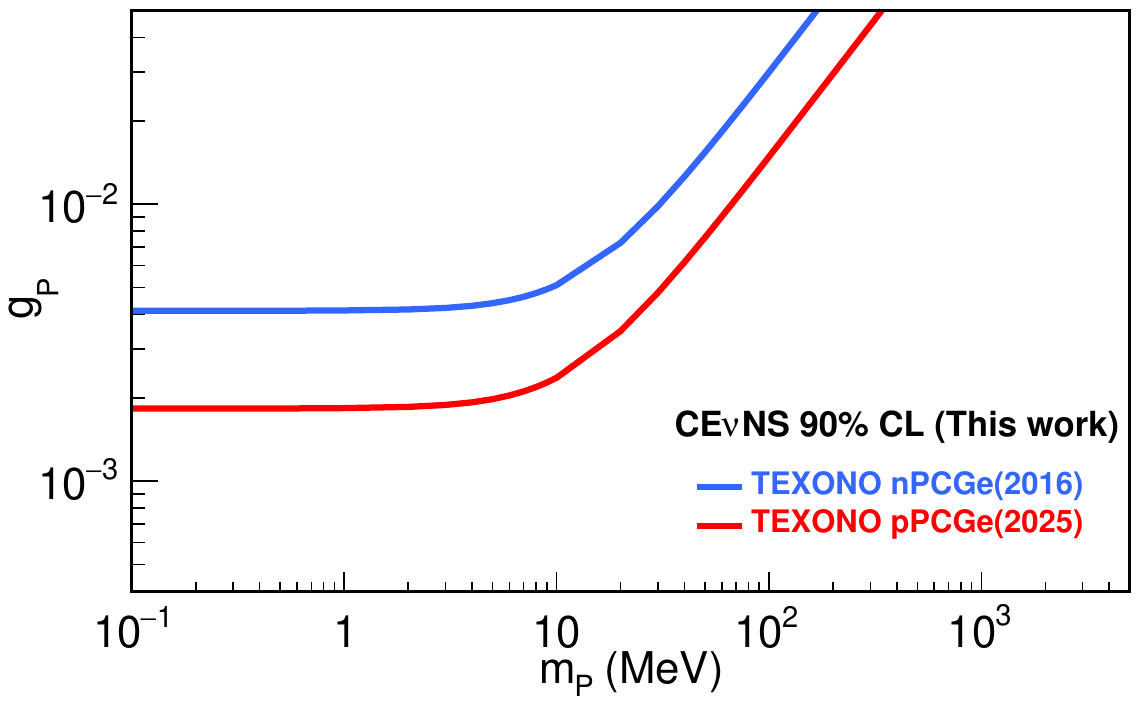}}
	\subfigure[]{\includegraphics[scale=0.4]{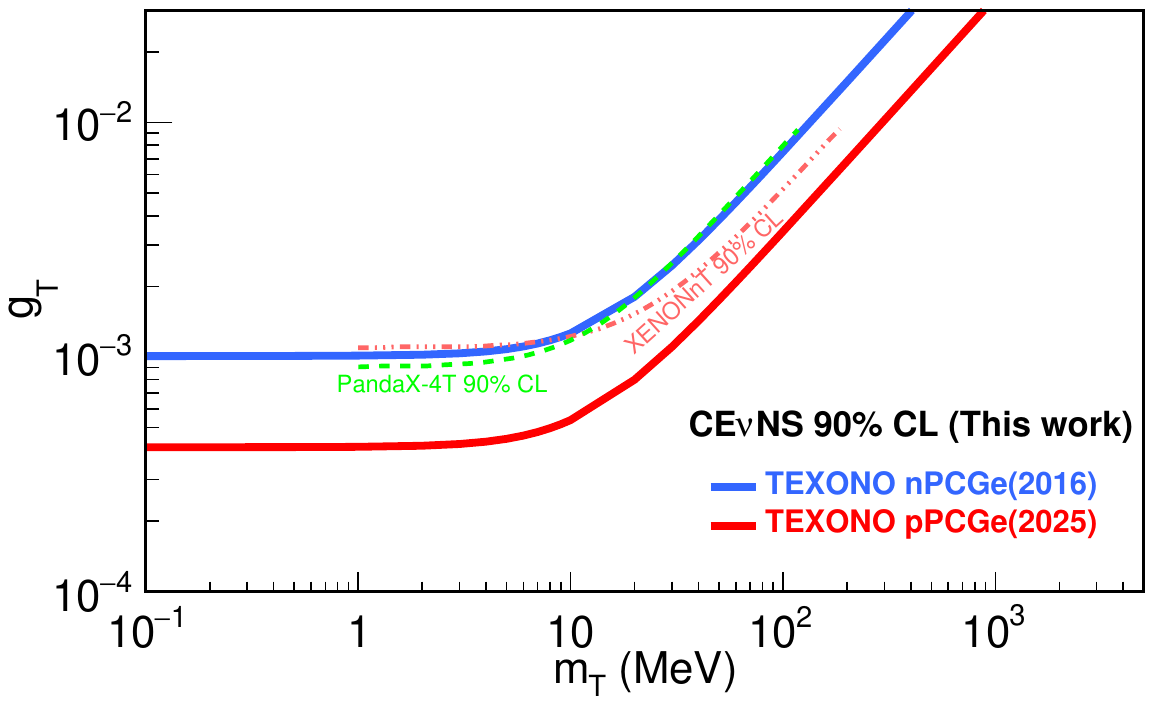}}
\caption{Bounds at $90\%$ CL are presented for (a) vector, (b) axialvector, (c) scalar, (d) pseudoscalar, and (e) tensor mediator masses and couplings, obtained from the TEXONO nPCGe (2016) and TEXONO pPCGe (2025) experiments comparison with the existing global bounds from various reactor-based experiments, CONUS+~\cite{CONUSpl:2025-BSM,Chattaraj:CONUSpl2025}, CONUS~\cite{CONUS:2022}, CONNIE~\cite{CONNIE:2020}, and DRESDEN-II~\cite{DresdenII:2022}; accelerator-based experiments, COHERENT CsI(Na) and LAr~\cite{Romeri-exo3:jhep2023}; as well as dark matter experiments sensitive to CE$\nu$NS, XENONnT, and PandaX-4T~\cite{Romeri:DM-LM2024}.}
\label{fig:gXmX}
\end{figure*}

Event rates as a function of recoil energy $T_N$ for V, A, S, P, and T type NSI are shown in Fig.~\ref{fig:dRdT}. These are calculated using the data from the TEXONO pPCGe (2025) experiment, based on specific values of mass and coupling constant parameters for the simplified model and specific values of $\varepsilon_{ee}^{uX}- \varepsilon_{ee}^{dX}$ parameters for the model-independent approach. In Fig.~\ref{fig:dRdT}, we present the predicted differential event rates for the SM CE$\nu$NS process (solid black) alongside possible new interactions considered in both the simplified model (left panel) and the model-independent NSI framework (right panel). The event rates are categorized by interaction type: V (dotted-light blue), V$_{\text{+INT}}$ (dashed-dark blue), A (long dashed-magenta), S (long dash dotted-red),  P (dash double dotted-orange), and T (dash triple dotted-green).  The SM differential event rates are calculated using Eq.~\eqref{eq:event-rate}, while the corresponding differential cross sections are derived from Eq.~\eqref{eq:smcevns}. For the NSI differential cross sections, the equations from Eq.~\eqref{eq:dSdT_V} to Eq.~\eqref{eq:dSdT_T} are employed for simplified model and the equations from Eq.~\eqref{eq:dSdT_V_eps} to Eq.~\eqref{eq:dSdT_T_eps} are used for the model-independent approach. In this analysis, we consider a mediator mass parameter of $m_X = 10$ MeV and coupling constants varying in the range $g_X = 0.1-20 \times 10^{-4}$ for each interaction type. The representative values chosen for the coupling constants are as follows: $g_V= 4 \times 10^{-5} $, $g_A= 8 \times 10^{-4}$, $g_S= 1 \times 10^{-5}$, $g_P= 2 \times 10^{-3}$, and $g_T= 6 \times 10^{-4}$.

We observe that when the recoil energy, $T_N$, is below 1~keV for the TEXONO pPCGe (2025) experiment,  the new interactions deviate significantly from the SM and also differ notably from each other. However, as the nuclear recoil energy increases, the new interactions tend to converge toward the SM prediction and, in some cases, overlap with them. This behavior can primarily be attributed to the dependence of the cross sections on the square of the $2m_N T_N$ term in the denominator. As the nuclear recoil energy decreases, this term becomes dominant, leading to a rapid increase in the cross sections for the new interactions. Consequently, the spectra of these interactions behave differently from the SM prediction,  with the divergence becoming more pronounced at lower nuclear recoil energies.

For model-independent NSI,  we identify values near the best-fit points for all possible new interaction types in scenarios where two epsilon parameters are present simultaneously. As a reference, we present the $\varepsilon_{ee}^{uX}-\varepsilon_{ee}^{dX}$ case in Fig.~\ref{fig:dRdT} (right panel). The values adopted to generate this reference plot are as follows: 
$\varepsilon_{ee}^{uV} = 0.30$, $\varepsilon_{ee}^{dV} = 0.50$, 
$\varepsilon_{ee}^{uA} = 18.0$, $\varepsilon_{ee}^{dA} = 88.0$, 
 $\varepsilon_{ee}^{uS} = 0.02$, $\varepsilon_{ee}^{dS} = 0.02$, 
 $\varepsilon_{ee}^{uP} = 950$, $\varepsilon_{ee}^{dP} = 950$, 
 $\varepsilon_{ee}^{uT} = 15.0$, and $\varepsilon_{ee}^{dT} = 80.0$.

We present the expected number of events for the TEXONO pPCGe (2025)~\cite{Texono:2024} experiment in Fig.~\ref{fig:HIST_COH_TEX}, plotted against reconstructed energy (keV$_{ee}$), with experimental data represented by black dots where the SM event predictions are shown as solid black lines.  To explore the effects of NSI on CE$\nu$NS, we also provide the estimated number of events with specific values in both the mass-coupling (left panel) and $\varepsilon_{ee}^{uX}-\varepsilon_{ee}^{dX}$(right panel) parameter spaces.

\subsection{Bounds from simplified model} \label{subsec:simpUL}
Equations ~\eqref{eq:dSdT_V} - \eqref{eq:dSdT_T} indicate that the cross section for all interaction types depends on the mass and coupling constant in the form of $g_X^2/(m_X^2 + 2 M_N T_N)$.  Based on this dependence,  the resulting limits exhibit distinct trends: in the low-mass regime ($m_X \lesssim 10$ MeV),  the upper limits are determined solely by the coupling constant $g_X$,  while in the high-mass regime ($m_X \gtrsim 10$ MeV), the limits scale with $g_X / m_X$ (as shown in Fig.~\ref{fig:gXmX}). Taking this behavior into account, Table~\ref{tab:simp_model_limits} provides a summary of the corresponding upper limits at $90\%$ CL for the coupling constant parameters associated with V, A, S, P, and T type NSI within the simplified model for two distinct mass regimes, $m_X^2 \ll 2m_N T_N$ and $m_X^2 \gg 2m_N T_N$, across the TEXONO nPCGe (2016) and pPCGe (2025) experiments. The first two columns of Table~\ref{tab:simp_model_limits} correspond to the $m_X^2 \ll 2m_N T_N$ case, and list the $90\%$ CL upper limits for the coupling constants. The latter two columns represent the $m_X^2 \gg 2m_N T_N$ case, showing the $90\%$ CL upper limits for the slope of $g_X/m_X$. Table~\ref{tab:simp_model_limits} highlights the varying sensitivity of each experiment and interaction type, with some coupling constants being more tightly constrained than others. In particular, the upper bounds obtained from the TEXONO pPCGe (2025) data represent a significant improvement over those obtained from the TEXONO nPCGe (2016) data, with the scalar interaction providing the most stringent constraint ($g_S <0.56\times10^{-5}$) and the pseudoscalar interaction providing the least stringent constraint ($g_P <18.3\times10^{-4}$).

Building on the results summarized in Table~\ref{tab:simp_model_limits}, Fig.~\ref{fig:gXmX} offers a detailed visualization of the $90\%$ CL  exclusion plots for all possible interaction types (V, A, S, P, and T) in the mass-coupling $m_X-g_X$ parameter space. These plots illustrate how the upper limits on the coupling constants evolve across different mediator masses.  For comparison, we also include the constraints from other reactor-based experiments, such as CONUS+~\cite{CONUSpl:2025,Chattaraj:CONUSpl2025}, CONUS~\cite{CONUS:2022}, CONNIE~\cite{CONNIE:2020}, and Dresden-II~\cite{DresdenII:2022}, and from accelerator-based COHERENT CsI and LAr experiments~\cite{Romeri-exo3:jhep2023}, as well as from dark matter experiments XENONnT and PAndaX-4T~\cite{Romeri:DM-LM2024}, which derive CE$\nu$NS constraints from solar neutrino induced nuclear recoils.

In Fig.~\ref{fig:gXmX}(a),  we see that CONUS+ achieves the most stringent upper limits for vector interactions by combining both CE$\nu$NS and neutrino-electron scattering ($\nu$ES) channels, with a significant improvement in the low mediator mass region. Notably,  COHERENT CsI provides the next strongest constraints for mediator masses below $\sim$0.2 MeV, also benefiting from the inclusion of $\nu$ES. 
At higher mediator masses (above $1$ MeV), XENONnT and CONNIE provide the second and third most stringent bounds, respectively. They are followed by TEXONO pPCGe(2025), COHERENT CsI, PandaX-4T, COHERENT LAr, Dresden-II, CONUS, and TEXONO nPCGe(2016). In addition, CONUS+, XENONnT, COHERENT CsI, and LAr exhibit allowed regions resulting from parameter degeneracies, which originate from the interference between new vector interaction and the SM. 
Such degeneracies arise when specific parameter combinations produce indistinguishable effects on the measured event rates. These effects can be mitigated by combining CE$\nu$NS data from detectors with different nuclear targets, as variations in weak charge dependencies help break these degeneracies more effectively, leading to stronger and more robust constraints.
These allowed regions emerge above 1 MeV of mediator mass for CONUS+ and XENONnT, while they appear at masses exceeding 30 MeV for COHERENT CsI and LAr.  In the case of TEXONO experiments, although destructive interference is present in the theoretical formulations given in Eq.~\eqref{eq:dSdT_Vint} (as visible in the difference between V and V$_{\text{+INT}}$ curves in Fig.~\ref{fig:dRdT}), the current experimental sensitivities are not sufficient to resolve this as a distinct allowed region in the exclusion plots.

Similarly, in Fig.~\ref{fig:gXmX}(c), the most stringent constraint on new scalar interaction is set by CONUS+, followed by CONNIE, TEXONO pPCGe(2025), and Dresden-II, respectively. The subsequent limits are given by XENONnT, PandaX-4T, CONUS, TEXONO nPCGe(2016), COHERENT CsI, and COHERENT LAr. As in the vector case, the CONUS+ and COHERENT limits also include contributions from both CE$\nu$NS and $\nu$ES channels. However, the impact of the $\nu$ES contribution is considerably reduced for scalar interactions. This is because the scalar cross section scales as $\sim 1/T_N$, making the analysis more sensitive to low nuclear recoil energy region, where CE$\nu$NS dominates. This behavior makes reactor-based CE$\nu$NS experiments, such as CONUS+, CONNIE, and TEXONO, particularly suitable for probing scalar interactions, due to their low energy thresholds. Compared to the vector interaction, whose cross section scales as $\sim 1/T_N^2$ and is thus more strongly suppressed at higher nuclear recoil energies, the scalar interaction, with a cross section scaling as $\sim 1/T_N$, generally provides stronger constraints, particularly in the low nuclear recoil energy region.

For the axialvector and tensor interactions, shown in Figs.~\ref{fig:gXmX}(b) and \ref{fig:gXmX}(e) respectively, the constraints are significantly weaker compared to the vector and scalar cases, as these interactions depend on the spin of individual nucleons rather than depending on the total number of nucleons. In most nuclei, these spin-dependent interactions are primarily dominated by unpaired nucleons, limiting the coherent enhancement observed in spin-independent interactions, where the cross section scales with the square of the neutron number. 
In both axialvector and tensor cases, TEXONO pPCGe(2025) provides strongest constraints. For the axialvector interaction, PandaX-4T and XENONnT yield bounds that are closer to TEXONO pPCGe(2025), while TEXONO nPCGe(2016) gives the weakest constraint. In the tensor case, the subsequent limits after TEXONO pPCGe(2025) are set by PandaX-4T, TEXONO nPCGe(2016), and XENONnT, which exhibit closely aligned sensitivities.
Between these two cases, the tensor interaction produces slightly stronger limits compared to the axialvector due to taking advantage of the distinct recoil energy and spin dependence of the tensor interaction. For the pseudoscalar interaction, shown in Fig.~\ref{fig:gXmX}(d), the constraints are the least stringent among all cases, as its cross section decreases rapidly at low recoil energies, further weakening the sensitivity.
As for the vector and scalar interactions,  the scalar interaction, with a cross section scaling as $\sim 1/T_N$, generally provides stronger constraints than the vector interaction, whose cross section scales as $\sim 1/T_N^2$ and is thus more strongly suppressed at higher recoil energies. Therefore, the nuclear recoil data at lower energy threshold from TEXONO significantly enhance the sensitivities to the scalar interaction compared to those for vector interaction. Importantly, across all interaction types considered, TEXONO pPCGe (2025) consistently provides stronger limits than TEXONO nPCGe (2016), highlighting the improved performance of the pPC detector setup.

\begin{table*} [!hbt]
\caption {Summary of the best-fit parameter values (BFPVs) and corresponding limits at $1\sigma$ and $90\%$ CL for V, A, S, P, and T type NSI measurements, based on one-parameter-at-a-time analysis.  The results are presented in comparison with two datasets from the TEXONO Collaboration: nPCGe (2016) data and pPCGe (2025) datasets. Here only one parameter is assumed to be nonzero for each bound.}
\label{tab:Texono_nsi}
\begin{ruledtabular}
\begin{tabular}{lccclccc}
\multicolumn{1}{l}{Fitting}&
\multicolumn{4}{c}{TEXONO nPCGe (2016)} &
\multicolumn{3}{c}{TEXONO pPCGe (2025)} \\ 
Par & BFPV$\ \pm \ 1\sigma$  & 90\% CL  & $\chi^2$/dof & \ &BFPV$\ \pm \ 1\sigma$  & 90\% CL  & $\chi^2$/dof  \\ \hline \\

$ \varepsilon_{ee}^{uV}$ & $0.18 \pm 1.421$ & $-1.71 < \varepsilon_{ee}^{uV} <2.084$ 
& 81.36/120 & \ & $0.0004^{+0.527}_{-0.153}$ & $-0.223 < \varepsilon_{ee}^{uV} < 0.597$
& 87.87/79 \smallskip \\ 

$ \varepsilon_{ee}^{dV}$ & $0.17 \pm 1.313$ & $-1.58 < \varepsilon_{ee}^{dV} < 1.926$ 
& 81.36/120 & \ & $0.0004^{+0.487}_{-0.142}$ & $-0.206 < \varepsilon_{ee}^{dV} < 0.551$ 
& 87.87/79 \smallskip \\ 

$\big ( \varepsilon_{e\mu(\tau)}^{uV} \big )^{2}$ & $-0.004 \pm 1.412 $ & 
$ \varepsilon_{e\mu(\tau)}^{uV} < 1.375$ & 81.37/120 & \  & $(-0.04 \pm 8.1)$$\times 10^{-2}$ & 
$ \varepsilon_{e\mu(\tau)}^{uV} < 0.364$ & 87.87/79 \smallskip \\ 

$\big ( \varepsilon_{e\mu(\tau)}^{dV} \big )^{2}$ & $-0.004 \pm 1.304 $ & 
$\varepsilon_{e\mu(\tau)}^{dV} < 1.322$ & 81.37/120 & \ & $(-0.04 \pm 6.9)$$\times 10^{-2}$ & 
$\varepsilon_{e\mu(\tau)}^{dV} < 0.337$ & 87.87/79 \smallskip \\ \hline  \\

$\big ( \varepsilon_{ee(\mu;\tau)}^{uA} \big )^{2}$ & $(-0.93 \pm 4.35)$$\times  10^{5}$ & 
$ \varepsilon_{ee(\mu;\tau)}^{uA} < 788.9$ & 81.3/120 & \ & $(0.12 \pm 2.08)$$\times  10^{4}$ & 
$ \varepsilon_{ee(\mu;\tau)}^{uA} < 188.4$ & 87.86/79 \smallskip \\ 

$\big ( \varepsilon_{ee(\mu;\tau)}^{dA} \big )^{2}$ & $(-1.88 \pm 8.84)$$\times 10^{4}$ & 
$\varepsilon_{ee(\mu;\tau)}^{dA} < 355.8$ & 81.3/120 & \  & $(0.23 \pm 4.24)$$\times 10^{3}$ & 
$\varepsilon_{ee(\mu;\tau)}^{dA} < 84.92$ & 87.86/79 \smallskip \\ \hline  \\

$\big ( \varepsilon_{ee(\mu;\tau)}^{uS}\big )^{2}$ & $(-0.03 \pm 7.8)$$\times 10^{-2}$ & 
$ \varepsilon_{ee(\mu;\tau)}^{uS} < 0.323 $ & 81.37/120 & \ & $(0.24 \pm 3.71)$$\times 10^{-4}$ & 
$ \varepsilon_{ee(\mu;\tau)}^{uS} < 0.025$ & 87.86/79 \smallskip \\ 

$\big ( \varepsilon_{ee(\mu;\tau)}^{dS}\big )^{2}$ & $(-0.03 \pm 7.7)$$\times 10^{-2}$ & 
$\varepsilon_{ee(\mu;\tau)}^{dS} < 0.321 $ & 81.37/120 & \ & $(0.22 \pm 3.65)$$\times 10^{-4}$ & 
$\varepsilon_{ee(\mu;\tau)}^{dS} < 0.024 $ & 87.86/79 \smallskip \\ \hline  \\

$\big ( \varepsilon_{ee(\mu;\tau)}^{uP}\big )^{2}$ & $(-1.40 \pm 6.53)$$\times 10^{7}$ & 
$ \varepsilon_{ee(\mu;\tau)}^{uP} < 9674.7 $ & 81.32/120 & \ & $(0.40 \pm 3.88)$$\times 10^{6}$ & 
$ \varepsilon_{ee(\mu;\tau)}^{uP} < 2605.0$ & 87.86/79 \smallskip \\ 

$\big ( \varepsilon_{ee(\mu;\tau)}^{dP}\big )^{2}$ & $(-1.44 \pm 6.71)$$\times 10^{7}$ & 
$\varepsilon_{ee(\mu;\tau)}^{dP} < 9801.0 $ & 81.32/120  & \ & $(0.40 \pm 3.98)$$\times 10^{6}$ & 
$\varepsilon_{ee(\mu;\tau)}^{dP} < 2638.2 $ & 87.86/79 \smallskip \\ \hline  \\

$\big ( \varepsilon_{ee(\mu;\tau)}^{uT}\big )^{2}$ & $(-0.25 \pm 1.15)$$\times  10^{6}$ & 
$ \varepsilon_{ee(\mu;\tau)}^{uT} < 1284.5$ & 81.32/120 & \ & $(0.42 \pm 11.2)$$\times  10^{4}$ & 
$ \varepsilon_{ee(\mu;\tau)}^{uT} < 433.1$ & 87.86/79 \smallskip \\ 

$\big ( \varepsilon_{ee(\mu;\tau)}^{dT}\big )^{2}$ & $(-0.9 \pm 4.22)$$\times  10^{4}$ & 
$\varepsilon_{ee(\mu;\tau)}^{dT} < 245.8$ & 81.32/120 & \  & $(0.86 \pm 19.1)$$\times  10^{2}$ & 
$\varepsilon_{ee(\mu;\tau)}^{dT} < 56.80$ & 87.86/79 \\ 
\end{tabular}
\end{ruledtabular}
\end{table*}

\subsection{Bounds from model-independent NSI} \label{subsec:epsUL}
Table~\ref{tab:Texono_nsi} presents the best-fit measurement results and corresponding limits at $90\%$ CL for individual epsilon parameters in the TEXONO nPCGe (2016) and pPCGe (2025) experiments. This analysis assumes one nonzero parameter at a time, with all other parameters set to zero,  and the table presents the resulting constraints for both FC and FV NSI scenarios. As in the simplified model, five types of interactions are considered: V. A, S, P, and T, with couplings denoted as $\varepsilon_{\alpha \beta}^{fX}$, where $f=u,d$; $\alpha, \beta = e,~ \mu,~ \tau$; and $X =$ V, A, S, P, and T. Because the composition of the reactor neutrino flux consists solely of $\bar{\nu}_e$, the TEXONO experiment is sensitive only to interactions involving electron neutrino flavor. Assuming real-valued couplings and applying the Hermiticity condition $\varepsilon_{\beta \alpha }^{fX} \equiv (\varepsilon_{\alpha \beta }^{fX})^*$,  the number of independent parameters per quark flavor reduces to six.  However, the reactor flux allows constraints on only two of these per flavor. 
Except for vector NSI, the FC and certain FV couplings are expected to be identical (i.e., $\varepsilon_{ee}^{fX}=\varepsilon_{e \mu}^{fX}=\varepsilon_{e \tau}^{fX}$). This equivalence arises because the strength of the FC coupling on nuclei corresponds to that of the FV parameter, due to the absence of interference between NSI and the SM and their dependency on the same flux term. Therefore, the relevant coupling parameters in the TEXONO experiment are $\varepsilon_{ee}^{fX}$ for the FC case and $\varepsilon_{e\mu(\tau)}^{fX}$ for the FV case.

As summarized in Table~\ref{tab:Texono_nsi}, FV vector couplings provide tighter constraints, while FC couplings remain less constrained due to the destructive nature SM interference term. However, while the presence of SM interference allows well-defined regions to be determined for FC vector couplings, only upper bounds can be determined for FV couplings, as they appear only in quadratic form in the cross section. NSI couplings associated with the other interaction types -- A, S, P, and T -- are similarly constrained as FV vector couplings, as they do not exhibit interference with the SM. As we see in the Table~\ref{tab:Texono_nsi},  since TEXONO operates with a single-flavor neutrino flux, both FC and FV couplings yield the same constraints in this scenario, except for the vector interactions due to the SM interference. Although the pPCGe (2025) setup offers clearly improved bounds over nPCGe (2016) configuration, reactor neutrino experiments remain limited in its ability to probe various NSI parameters due to its reliance on a single-flavor neutrino flux.

\begin{figure*}[htb!]
	\centering
	\subfigure[]{\includegraphics[scale=0.35]{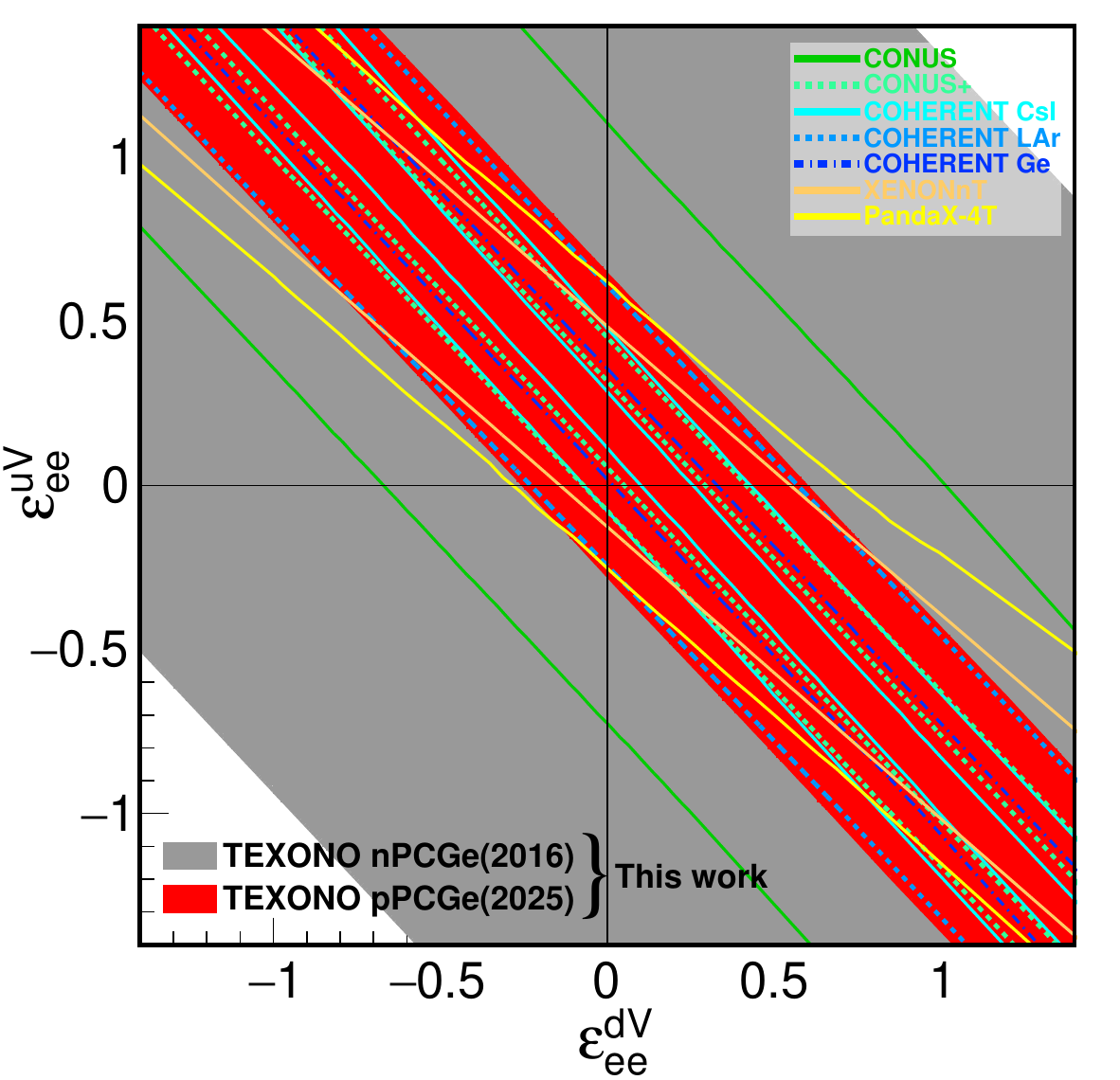}}
 	\subfigure[]{\includegraphics[scale=0.35]{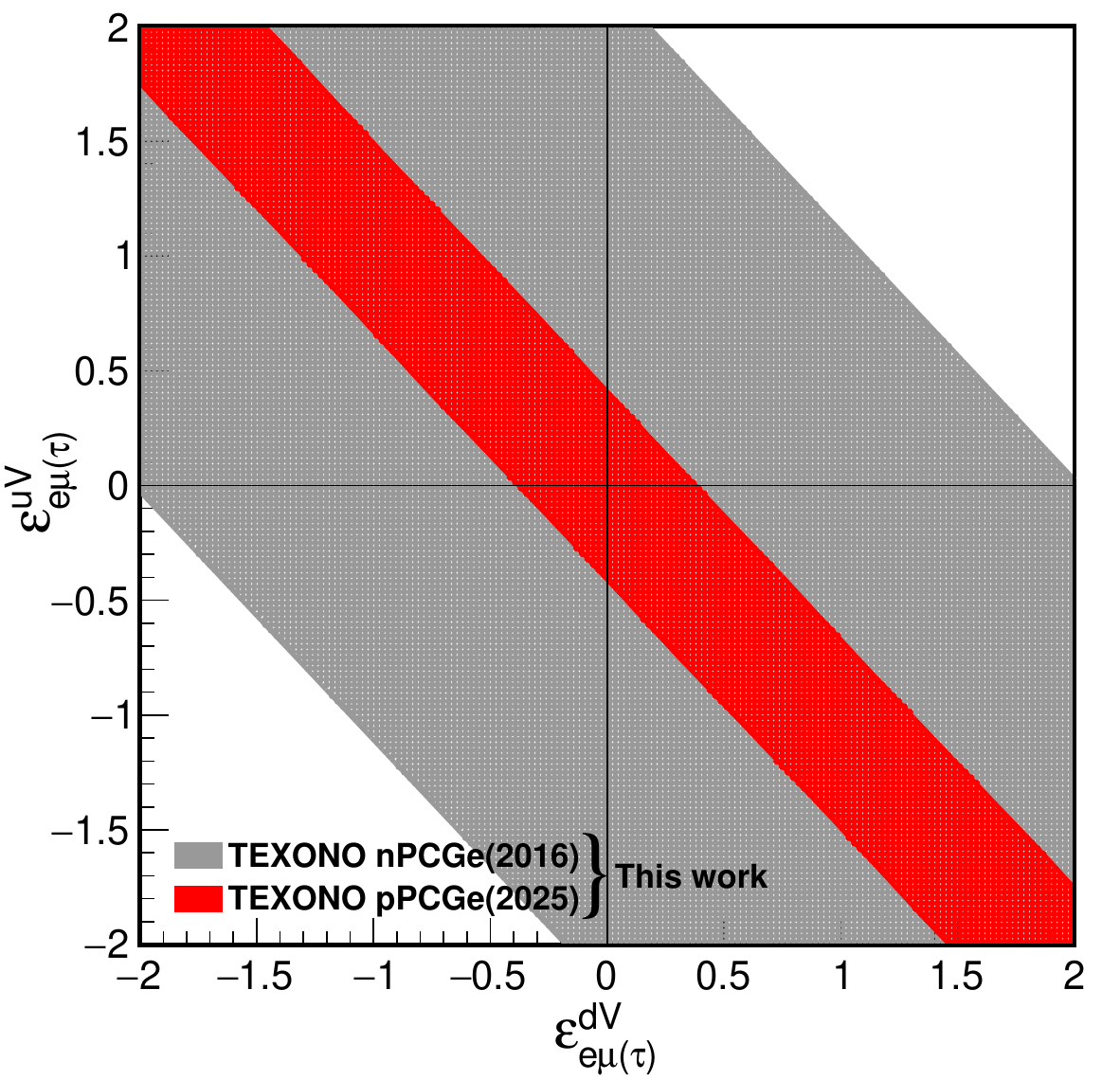}}
	\caption{$90\%$ CL allowed region from TEXONO on (a) single-type FC-FC parameter space for vector-type NSI: $\varepsilon_{ee}^{dV}-\varepsilon_{ee}^{uV}$ for comparison with the bounds from other reactor-based experiments (CONUS+~\cite{CONUSpl:2025-BSM}, CONUS~\cite{CONUS:2022}), accelerator based-experiments (COHERENT CsI and COHERENT LAr~\cite{Romeri-exo3:jhep2023}, COHERENT Ge~\cite{Coherent-Ge:PRD2024}), as well as dark matter experiments (XENONnT and PandaX-4T) sensitive to CE$\nu$NS are included~\cite{Sierra:DM2025}, and (b) single-type FV-FV parameter space for vector-type NSI: $\varepsilon_{e\mu(\tau)}^{dV}-\varepsilon_{e\mu(\tau)}^{uV}$.}
\label{fig:eps_FC_V}
\end{figure*}

We further explore the sensitivities to pairs of NSI parameters. While a wide variety of NSI parameter pairs can be analyzed, our study is limited to a few illustrative cases. These selected examples effectively capture the key features and trends observed across the parameter space. The allowed regions for other parameter combinations generally exhibit similar behavior. By focusing on these representative cases, we provide a clear and concise overview of the parameter space, minimizing redundancy and ensuring the clarity and consistency in this work.

\begin{figure*}[htb!]
\centering
  \subfigure[]{\includegraphics[scale=0.29]{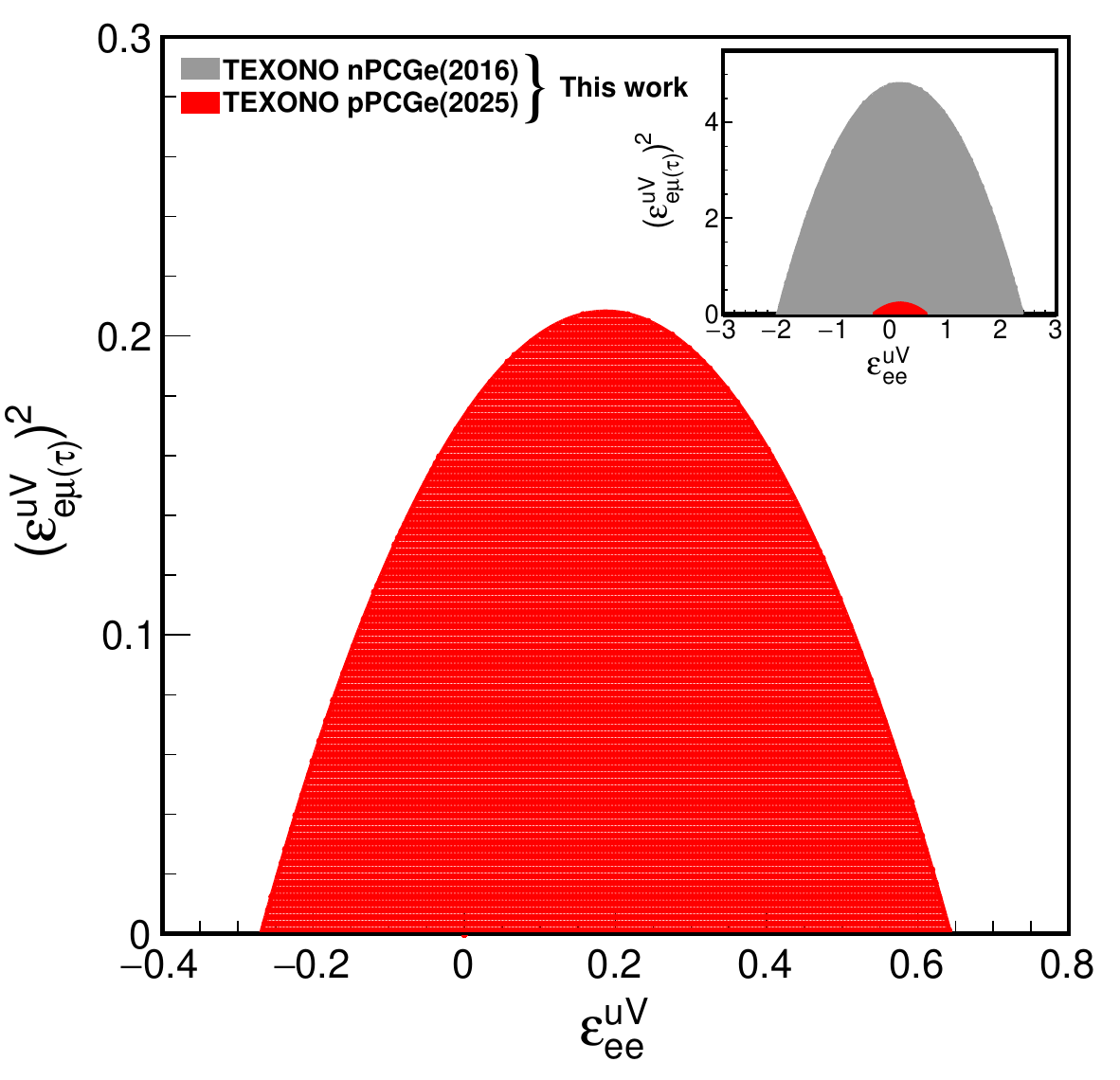}}
  \subfigure[]{\includegraphics[scale=0.29]{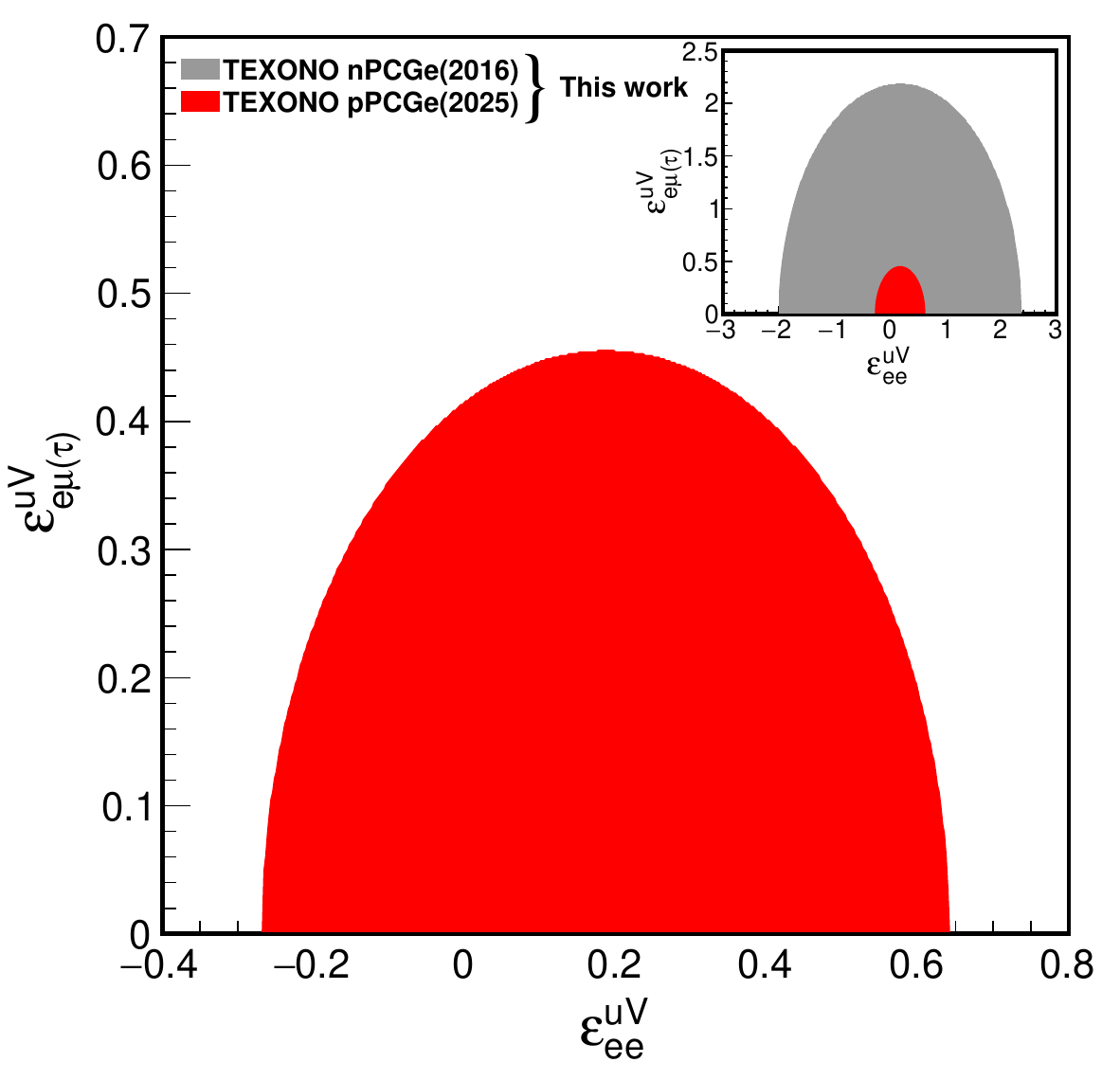}}
  \subfigure[]{\includegraphics[scale=0.29]{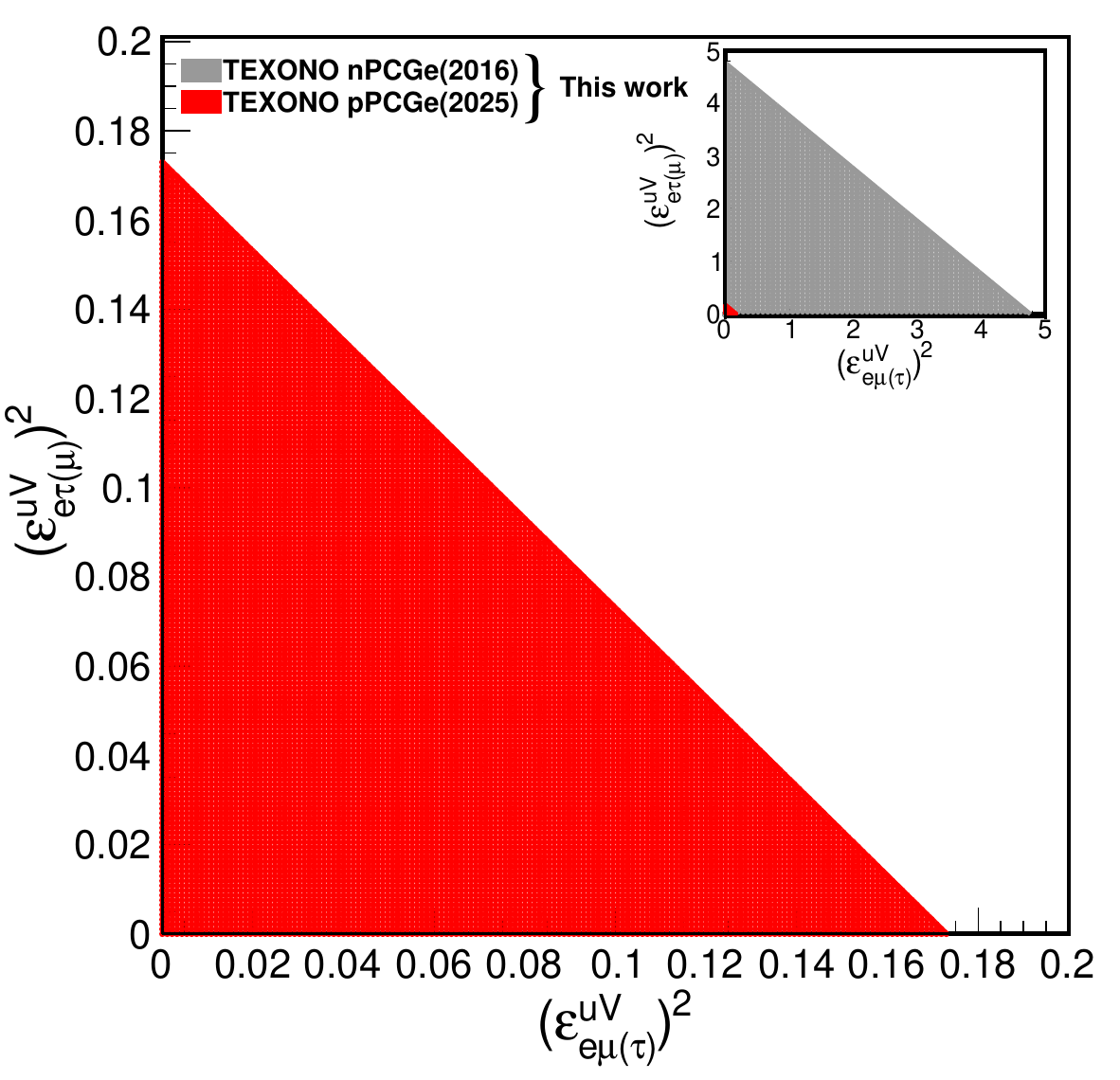}}
  \subfigure[]{\includegraphics[scale=0.29]{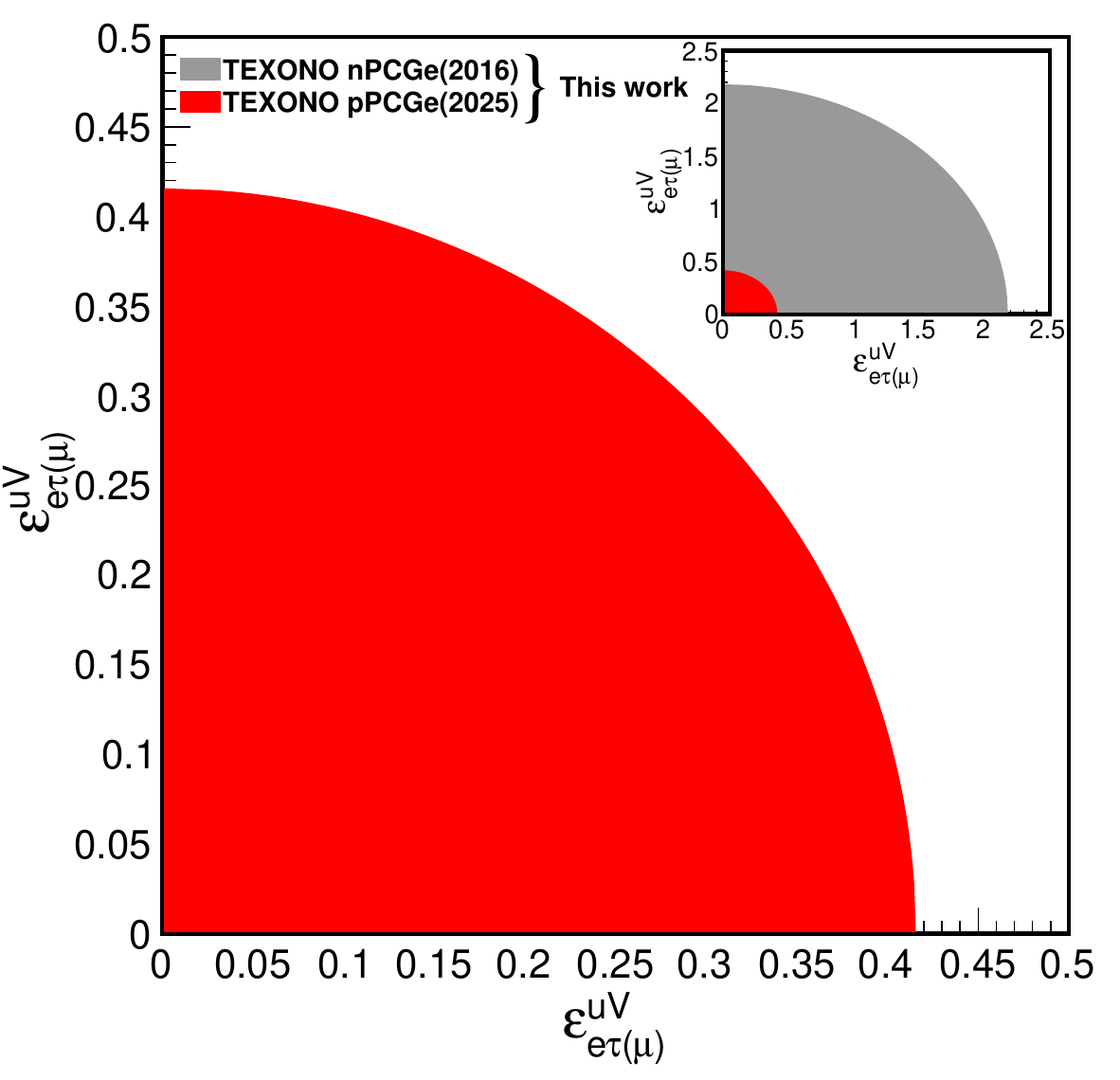}}
  \subfigure[]{\includegraphics[scale=0.29]{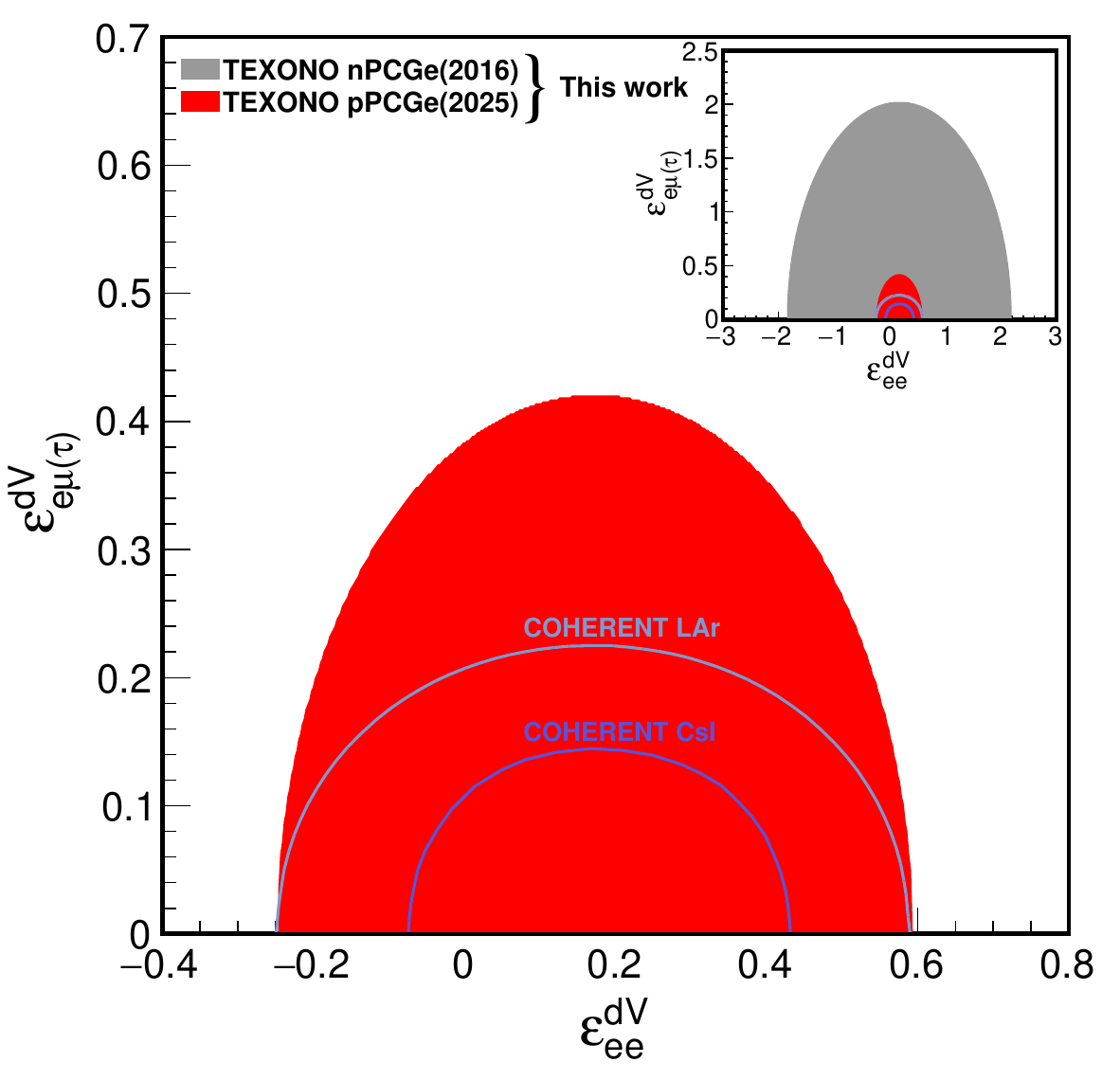}}
  \subfigure[]{\includegraphics[scale=0.29]{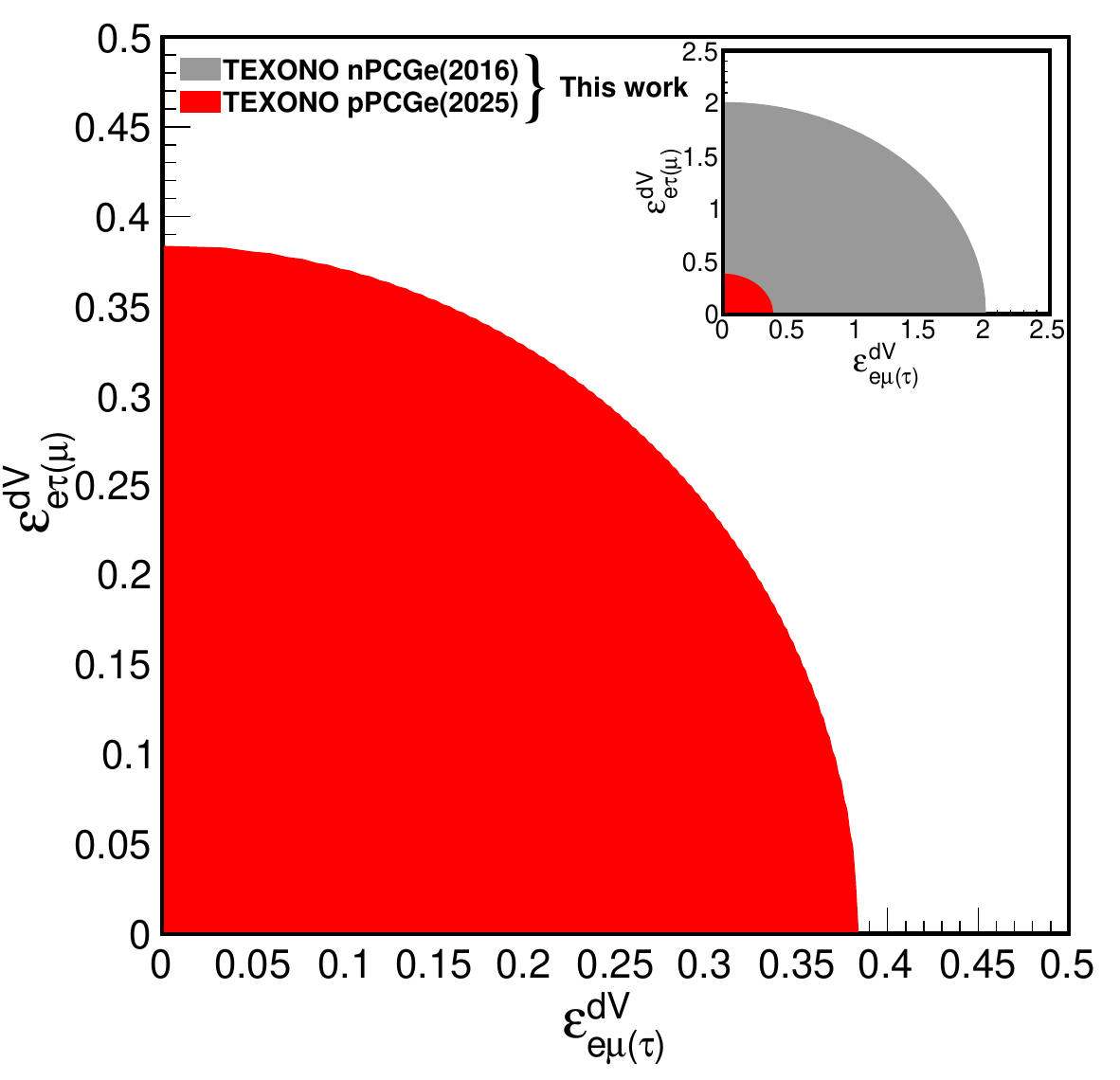}}
\caption{$90\%$ CL allowed region for mixed FC-FV and FV-FV parameter spaces for vector-type NSI:
(a) $\varepsilon_{ee}^{uV}-(\varepsilon_{e\mu(\tau)}^{uV})^{2}$,  
(b) $\varepsilon_{ee}^{uV}-\varepsilon_{e\mu(\tau)}^{uV}$,
(c) $(\varepsilon_{e\mu(\tau)}^{uV})^{2}-(\varepsilon_{e\tau(\mu)}^{uV})^{2}$, 
(d) $\varepsilon_{e\mu(\tau)}^{uV}-\varepsilon_{e\tau(\mu)}^{uV}$, 
(e) $\varepsilon_{ee}^{dV}-\varepsilon_{e\mu(\tau)}^{dV}$, and
(f) $\varepsilon_{e\mu(\tau)}^{dV}-\varepsilon_{e\tau(\mu)}^{dV}$. For comparison, constraints from COHERENT CsI and COHERENT LAr for the (e) $\varepsilon_{ee}^{dV}-\varepsilon_{e\mu}^{dV}$ scenario are also superimposed~\cite{Romeri-exo3:jhep2023}.}
\label{fig:eps_mixedFV_V}
\end{figure*}

In Figs.~\ref{fig:eps_FC_V} and \ref{fig:eps_mixedFV_V}, we present the $90\%$ CL bounds for all possible flavor-diagonal (FC) and flavor-off-diagonal (FV) cases of vector NSI interactions. Figure~\ref{fig:eps_FC_V} focuses on single type FC-FC and single type FV-FV parameter spaces, with the left panel showing limits for a scenario involving a single flavor-diagonal term (e.g., $\varepsilon_{ee}^{uV}-\varepsilon_{ee}^{dV}$), and the right panel showing bounds for a scenario involving a single flavor-off-diagonal term (e.g., $\varepsilon_{e\mu}^{uV}-\varepsilon_{e\mu}^{dV}$). In both parameter spaces, the constraints appear as diagonal parallel bands; the FV-FV case provide slightly tighter bounds compared to the FC-FC case as it does not interfere with the SM which gives only destructive contribution to the total cross section. For comparison, we adopt existing bounds from various reactor-based experiments,  CONUS~\cite{CONUS:2022} and CONUS+~\cite{CONUSpl:2025-BSM}, from accelerator-based experiments, COHERENT CsI, LAr~\cite{Romeri-exo3:jhep2023}, and Ge~\cite{Coherent-Ge:PRD2024}, as well as dark matter experiments,  XENONnT and PandaX-4T~\cite{Sierra:DM2025} in Fig.~\ref{fig:eps_FC_V}(a). Notably, the COHERENT Ge provides the most stringent limits, with CONUS+ offering the second strongest bound. While accelerator-based experiments continue to be leading contributors, improved constraints from the latest CONUS+ and TEXONO pPCGe (2025) experiments underscore the growing importance of reactor neutrino experiments. This is driven by increased sensitivities and advancements in detection technologies,  which are crucial for studies beyond the SM.

For certain cases, particularly in CONUS+ and COHERENT CsI, in Fig.~\ref{fig:eps_FC_V}(a),
degenerate regions emerge due to interference between the SM and NSI differential cross sections. 
These degeneracies are exclusive to vector interactions. The degeneracy region appearing in Fig.~\ref{fig:eps_FC_V}(a) is a consequence of the dependence of the cross section on specific linear combinations of the parameters and interference of these parameters with the SM, as derived from Eqs.\eqref{eq:dSdT_V_eps}, and Eq.~\eqref{eq:dSdT_Vint_eps}. Specifically, the cross section depends on the condition $([\varepsilon_{ee}^{uV} (2Z+N) + \varepsilon_{ee}^{dV}(\mathcal{Z}+2\mathcal{N})]+Q_{SM_V})^2 = $ constant, resulting in two allowed bands with slopes determined by the ratio $ (2\mathcal{Z}+\mathcal{N})/(\mathcal{Z}+2\mathcal{N})$~\cite{NSItensor1:prd2024,Barranco:2005}. Even though these degeneracy regions are not visible in current TEXONO data due to the sensitivity limitations,  their appearance in other reactor experiments like CONUS+ highlights the potential of reactor experiments to probe such features.

In the case of single flavor-off-diagonal (FV-FV) parameter spaces,  as shown in Fig.~\ref{fig:eps_FC_V}(b), the cross section takes the form 
$\varepsilon_{\alpha \beta}^{uV} (2\mathcal{Z}+\mathcal{N}) + \varepsilon_{\alpha \beta}^{dV}(\mathcal{Z}+2\mathcal{N}) = constant$, with a center at (0,0), and the upper limits form parallel bands. However, unlike the flavor-diagonal FC-FC case in Fig.~\ref{fig:eps_FC_V}(a), in this scenario, with no contribution from SM interference, the band becomes narrower, and results in more stringent bound. Additionally, due to the presence of only a single-type neutrino flux in TEXONO,  bounds from $\varepsilon_{e\mu}^{uV}-\varepsilon_{e\mu}^{dV}$ and $\varepsilon_{e\tau}^{uV}-\varepsilon_{e\tau}^{dV}$parameter spaces are identical. We represent such cases as $\varepsilon_{e\mu}=\varepsilon_{e\tau} \equiv \varepsilon_{e\mu(\tau)}$.

In Fig.~\ref{fig:eps_mixedFV_V}, we present the $90\%$ CL upper limits for mixed FC-FV and two-type FV-FV parameter spaces. For these scenarios, we explored two different parameter spaces: combinations of  
$\varepsilon_{\alpha \alpha}^{fV}-\varepsilon_{\alpha \beta}^{fV}$, and 
$\varepsilon_{\alpha \beta}^{fV}-\varepsilon_{ij}^{fV}$, where $\alpha,~\beta \neq i,~j$. 
In certain FV cases, where the parameter space involves either two flavor-off-diagonal parameters or a combination of one flavor-diagonal and one flavor-off-diagonal parameter space, the absence of SM interference and the lack of first-order terms of the relevant epsilon parameters in the cross section, the upper limits cannot be directly set on these parameters. Instead, as shown in Figs.~\ref{fig:eps_mixedFV_V}a and \ref{fig:eps_mixedFV_V}c, the $\chi^{2}$ analysis is performed on the epsilon-squared parameters, where $(\varepsilon_{\alpha \beta}^{fV})^{2}>0$,  and then subsequently converted into the $\varepsilon_{\alpha \beta}^{fV}$ parameter space. This method allows us to eliminate certain regions of the full elliptical bounds, resulting in (semielliptical [Fig.~\ref{fig:eps_mixedFV_V}(b)] or quarter-elliptical [Fig.~\ref{fig:eps_mixedFV_V}(d)] allowed regions. This difference arises because $\varepsilon_{ee}^{fV}$ parameters interfere with the SM and it appears linearly in the cross section. This linear dependence allows us to place both upper and lower bounds on the parameter. In contrast, for parameters like $\varepsilon_{e\mu}^{fV}$, which do not interfere with the SM, both terms appear only quadratically in the cross section.  As a result, the sensitivity is limited to the absolute values of the parameters, and only upper limits can be set. Finally, the analyses of other FV parameter spaces in Figs.~\ref{fig:eps_mixedFV_V}(e) and ~\ref{fig:eps_mixedFV_V}(f) follow a similar methodology. To ensure clarity and provide a more concise representation,  plots related to the epsilon-squared, $(\varepsilon_{\alpha \beta}^{dV})^{2}$,  are excluded and only the corresponding $\varepsilon_{\alpha \beta}^{dV}$ parameter spaces are displayed. Also, we adopted existing COHERENT CsI and LAr~\cite{Romeri-exo3:jhep2023} bounds for $\varepsilon_{ee}^{dV}-\varepsilon_{e\mu}^{dV}$ parameter space in Fig. ~\ref{fig:eps_mixedFV_V}(e), for comparison.

\begin{figure*}[htb!]
	\centering
\subfigure[]{\includegraphics[scale=0.35]{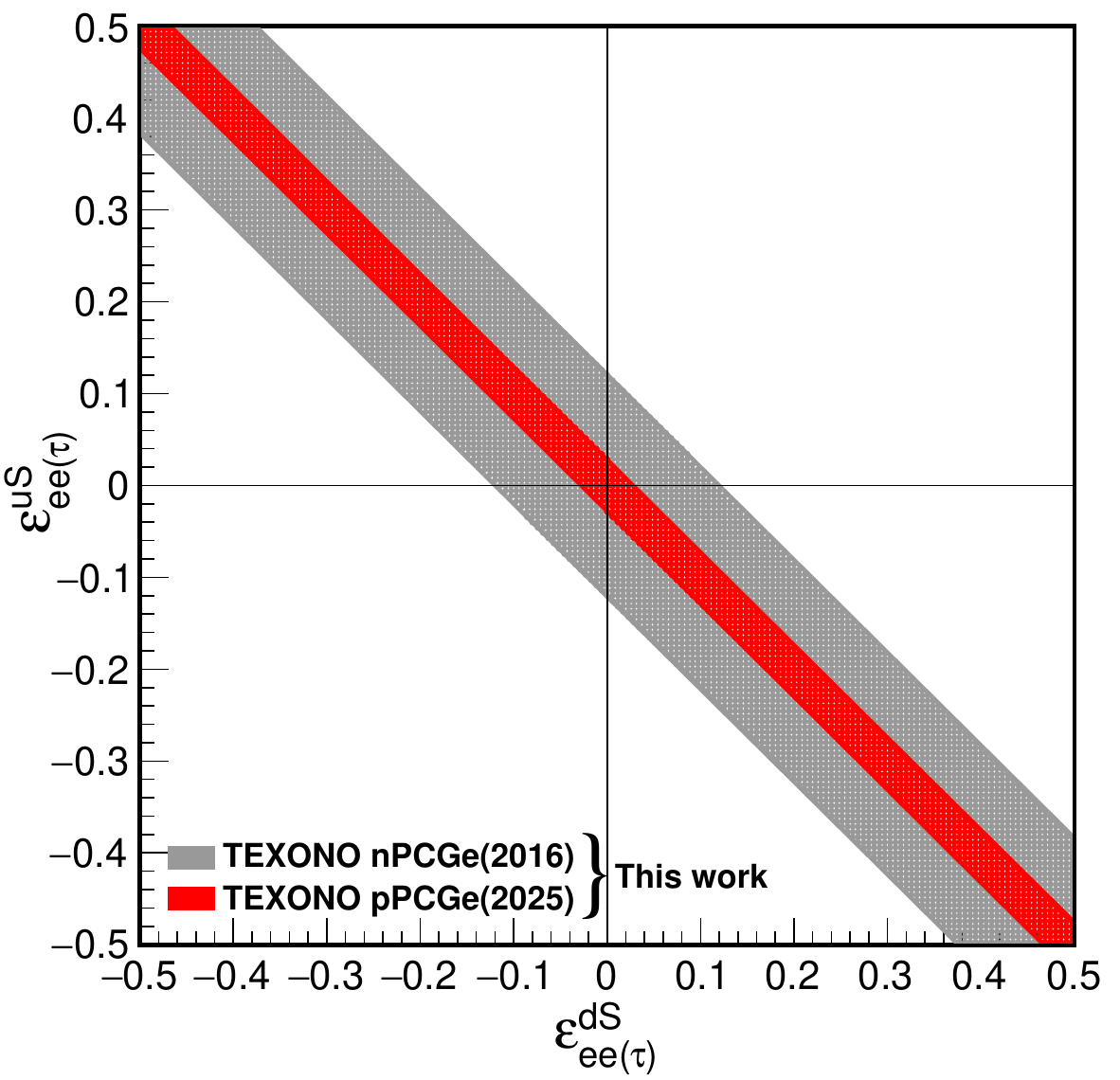}}
\subfigure[]{\includegraphics[scale=0.35]{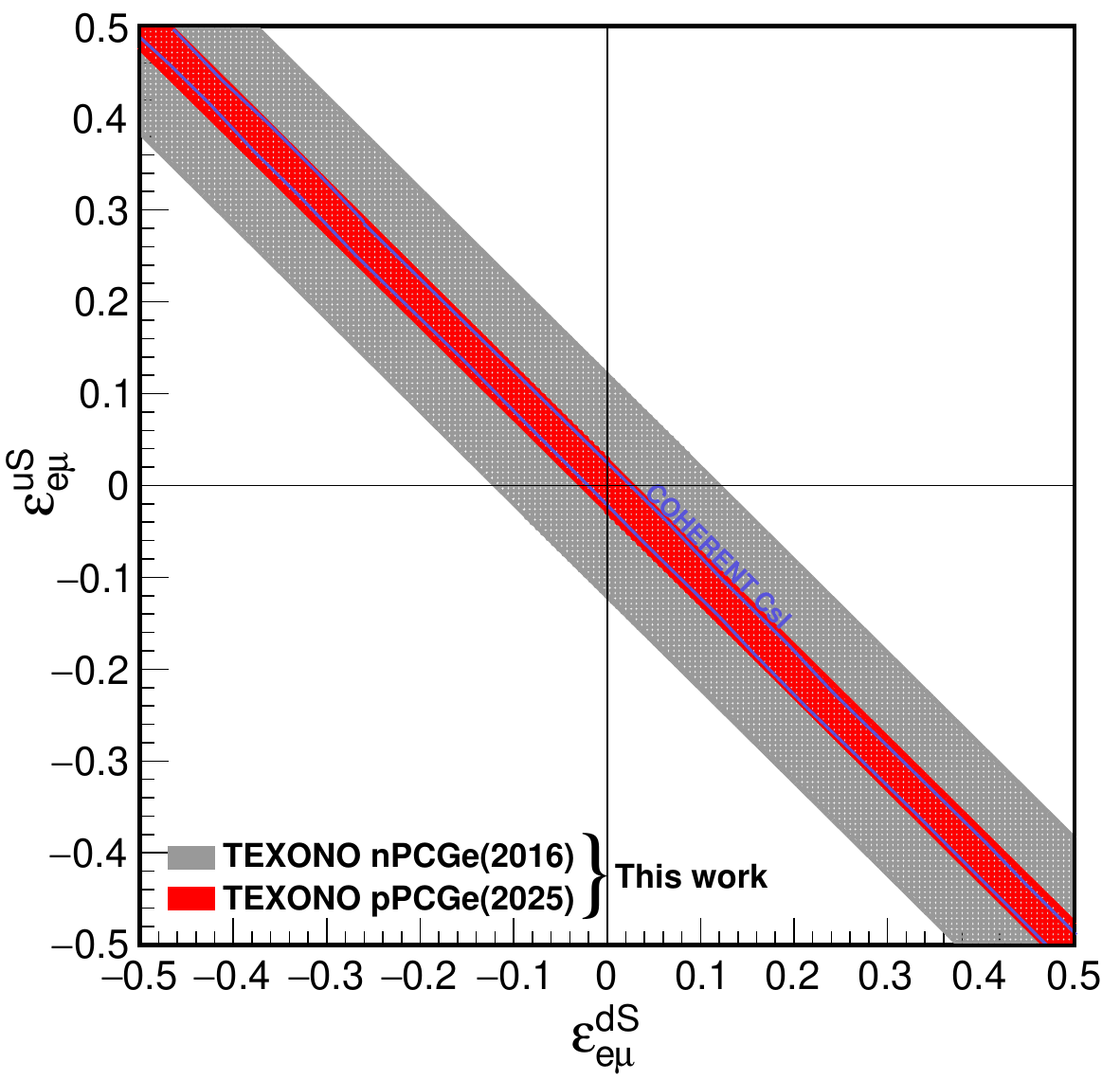}}
  \subfigure[]{\includegraphics[scale=0.35]{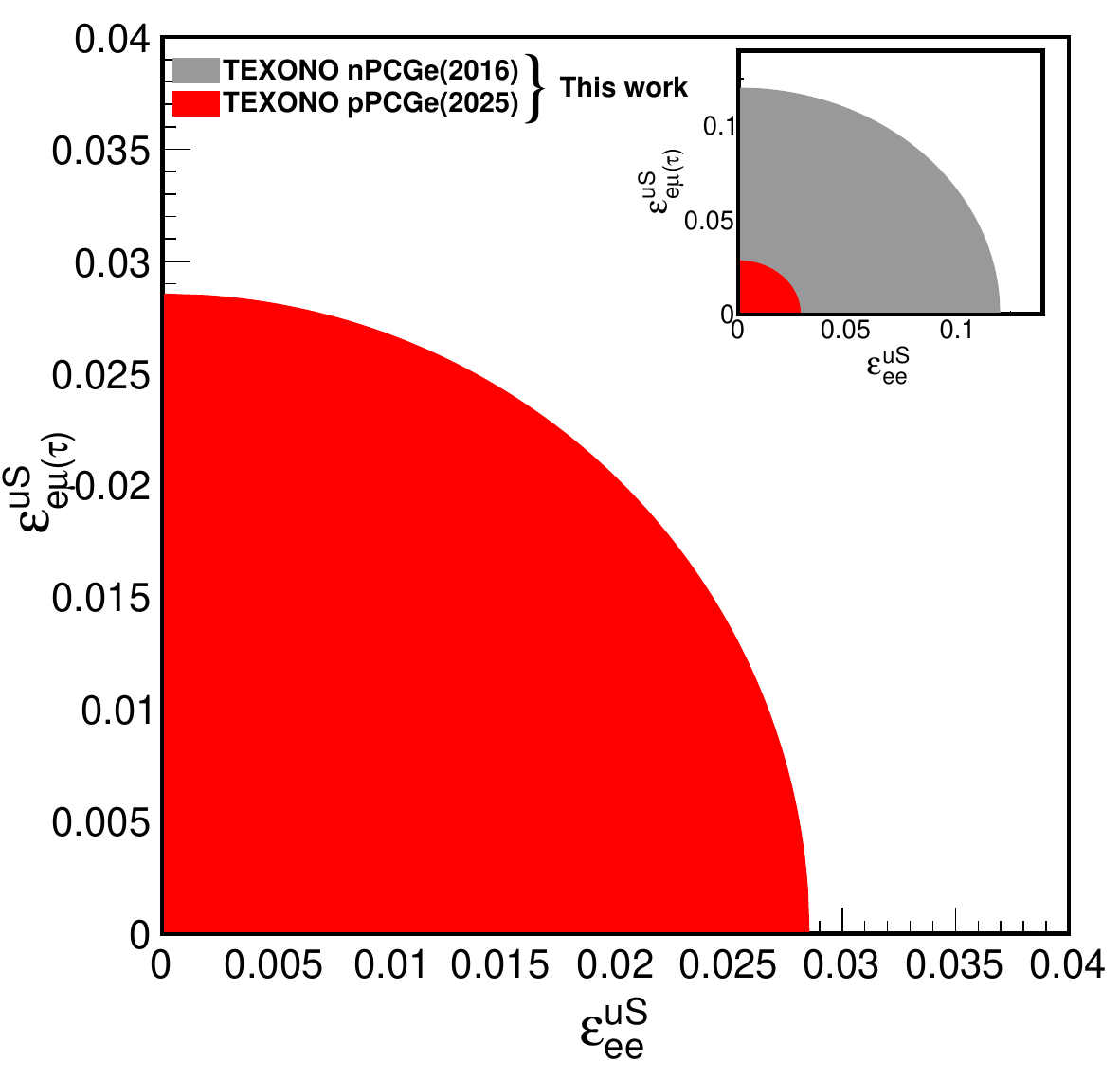}}
  \subfigure[]{\includegraphics[scale=0.35]{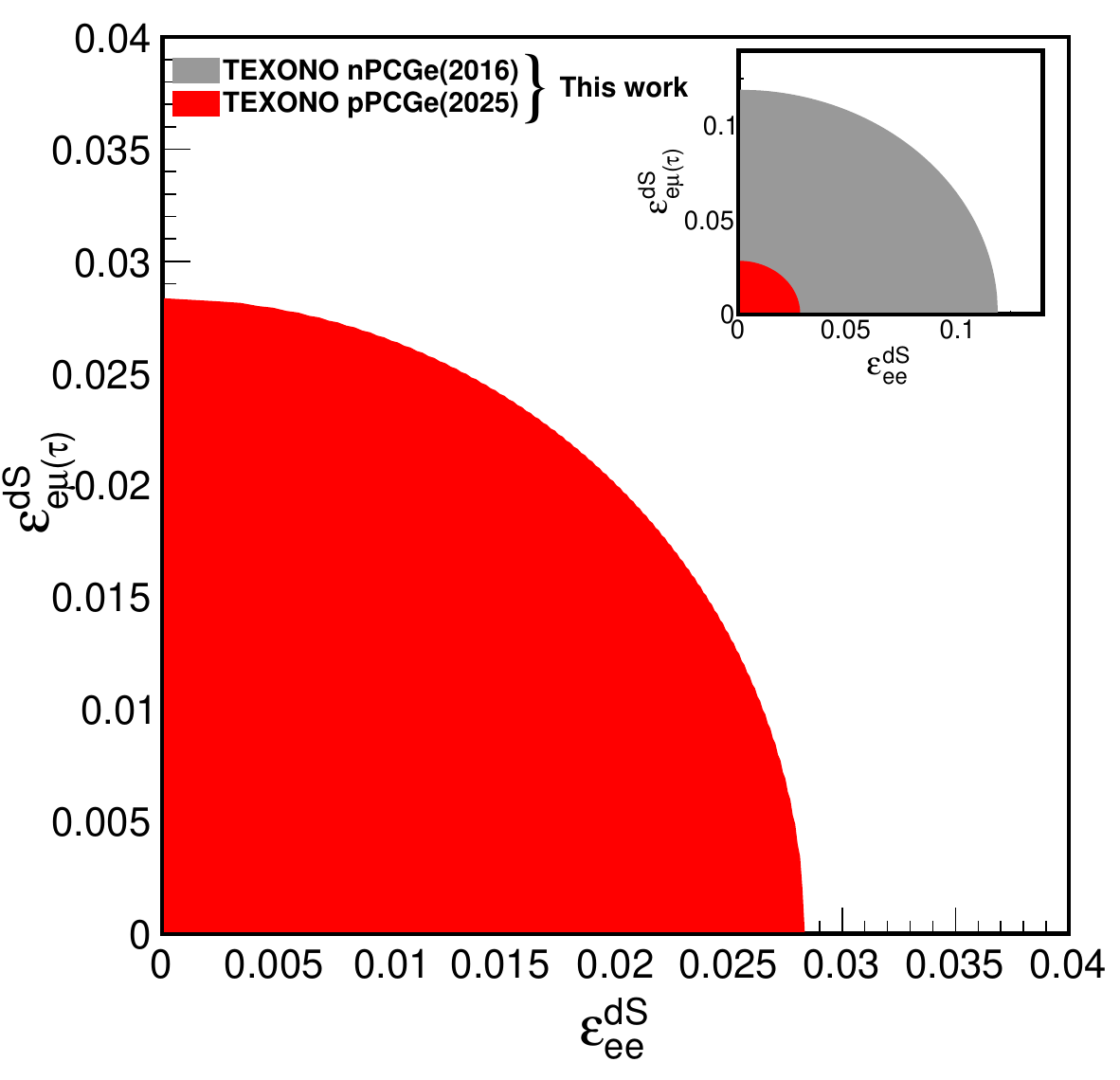}}
\caption{$90\%$ CL allowed region for FC and FV parameter spaces for scalar-type NSI:
(a) $\varepsilon_{ee(\tau)}^{dS}-\varepsilon_{ee(\tau)}^{uS}$, 
(b) $\varepsilon_{e\mu}^{dS}-\varepsilon_{e\mu}^{uS}$, 
(c) $\varepsilon_{ee}^{uS}-\varepsilon_{e\mu(\tau)}^{uS}$, and  
(d) $\varepsilon_{ee}^{dS}-\varepsilon_{e\mu(\tau)}^{dS}$.
For comparison, the constraint of the (b) $\varepsilon_{e\mu}^{dS}-\varepsilon_{e\mu}^{uS}$ parameter space from COHERENT CsI is also superimposed~\cite{NSItensor1:prd2024}.}
\label{fig:eps_FC-FV_S}
\end{figure*}

Figure~\ref{fig:eps_FC-FV_S} shows the $90\%$ CL bounds for scalar interactions in all possible FC and FV scenarios. The cross section for scalar interactions is expressed as a linear combination of the relevant epsilon parameters:
$\varepsilon_{\alpha \beta}^{uS} 
[\mathcal{N} (m_n / m_u) f_{Tu}^n + \mathcal{Z} (m_p / m_u) f_{Tu}^p]
+ \varepsilon_{\alpha \beta}^{dS}[\mathcal{N} (m_n / m_d) f_{Td}^n + \mathcal{Z} 
(m_p / m_d) f_{Td}^p] = $ constant, as derived in Eq.\eqref{eq:dSdT_S_eps}. 
This linear dependence leads to exclusion bounds in the form of linear bands in the $\varepsilon_{\alpha \beta}^{uS} - \varepsilon_{\alpha \beta}^{dS}$ parameter space. The slope of the linear bands is influenced by the relative contributions of the up- and down-quark parameters, $f_{Tq}^p$ and $f_{Tq}^n$, as well as the nuclear masses of the target material.
This results in stronger bounds for heavier nuclei, as larger nuclear masses lead to greater cross sections. This behavior is evident in Fig.~\ref{fig:eps_FC-FV_S}(b), where we show, for comparison,  COHERENT CsI bounds~\cite{NSItensor1:prd2024} in the $\varepsilon_{e\mu}^{dS}-\varepsilon_{e\mu}^{uS}$ parameter space, which appears slightly tighter than TEXONO pPCGe (2025).

The bounds shown in Figs.\ref{fig:eps_FC-FV_S}(c) and \ref{fig:eps_FC-FV_S}(d) include parameter spaces for two different FC-FV parameters, resulting in quarter-elliptical regions. To derive these limits, we begin with the epsilon-squared parameter space, $(\varepsilon_{\alpha \beta}^{fS})^{2}$, and then transition to the epsilon parameter space, $\varepsilon_{\alpha \beta}^{fS}$, as explained above. In contrast to the vector case, scalar interactions do not interfere with the SM. This leads to identical bounds in some parameter spaces, thus reducing the number of distinct cases for which independent constraints can be set. For instance, in Fig.\ref{fig:eps_FC-FV_S}(a), it can be seen that the contributions from the flavor-diagonal $\varepsilon_{ee}^{fS}$ term and the flavor-off-diagonal $\varepsilon_{e \mu(\tau)}^{fS}$ terms yield the same bounds, as both are related to the electron neutrino flux. Similarly, the limits provided by the parameter spaces $\varepsilon_{ee}^{fS}-\varepsilon_{e\tau(\mu)}^{fS}$ in Figs.~\ref{fig:eps_FC-FV_S}(b) and \ref{fig:eps_FC-FV_S}(c) are identical to those in the $\varepsilon_{e\mu(\tau)}^{fS}-\varepsilon_{e\tau(\mu)}^{fS}$ parameter spaces. Since these parameter spaces result in the same bounds, only one representative case is shown for clarity.

\begin{figure*}[htb!]
	\centering
\subfigure[]{\includegraphics[scale=0.29]{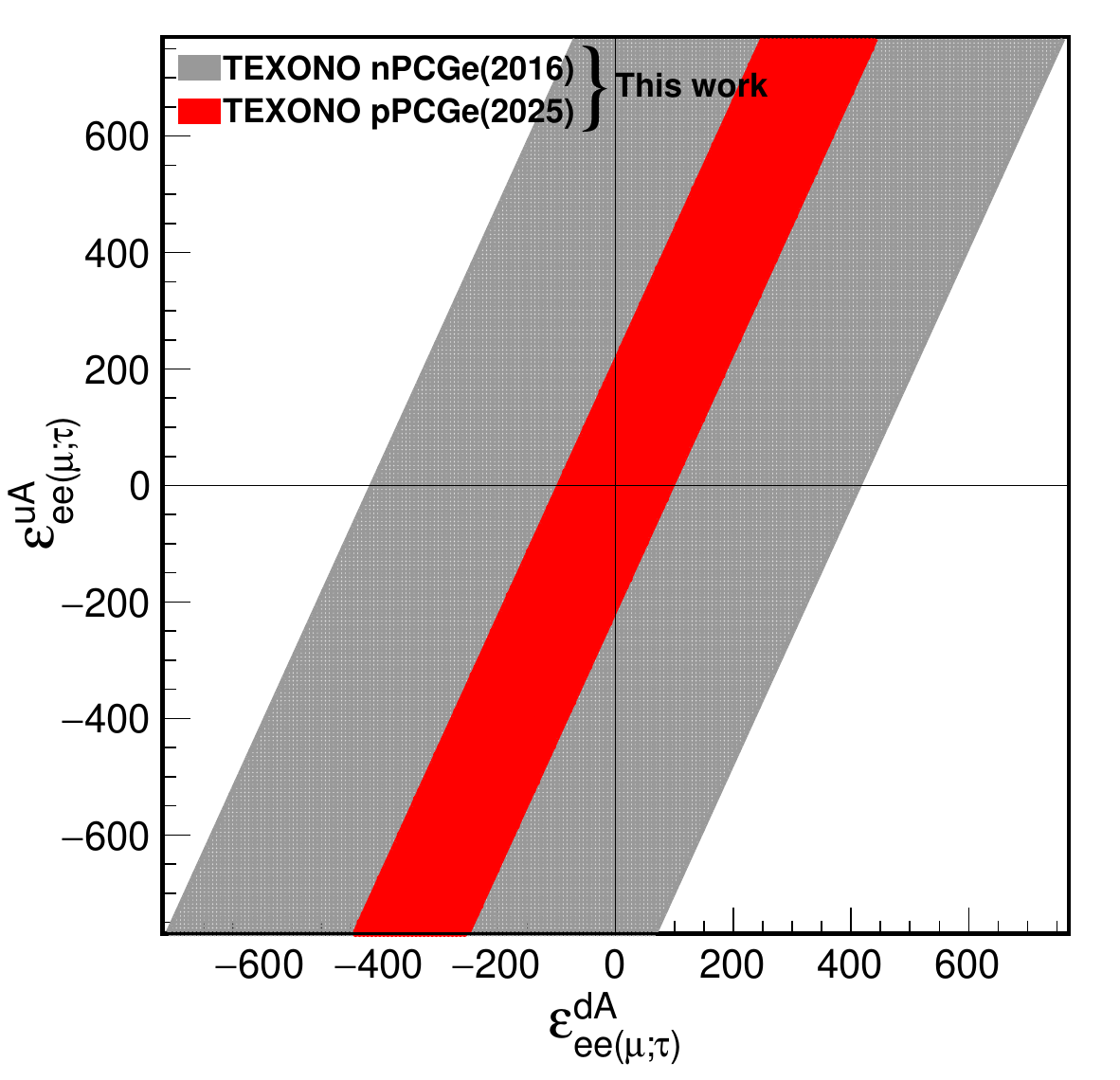}}
\subfigure[]{\includegraphics[scale=0.29]{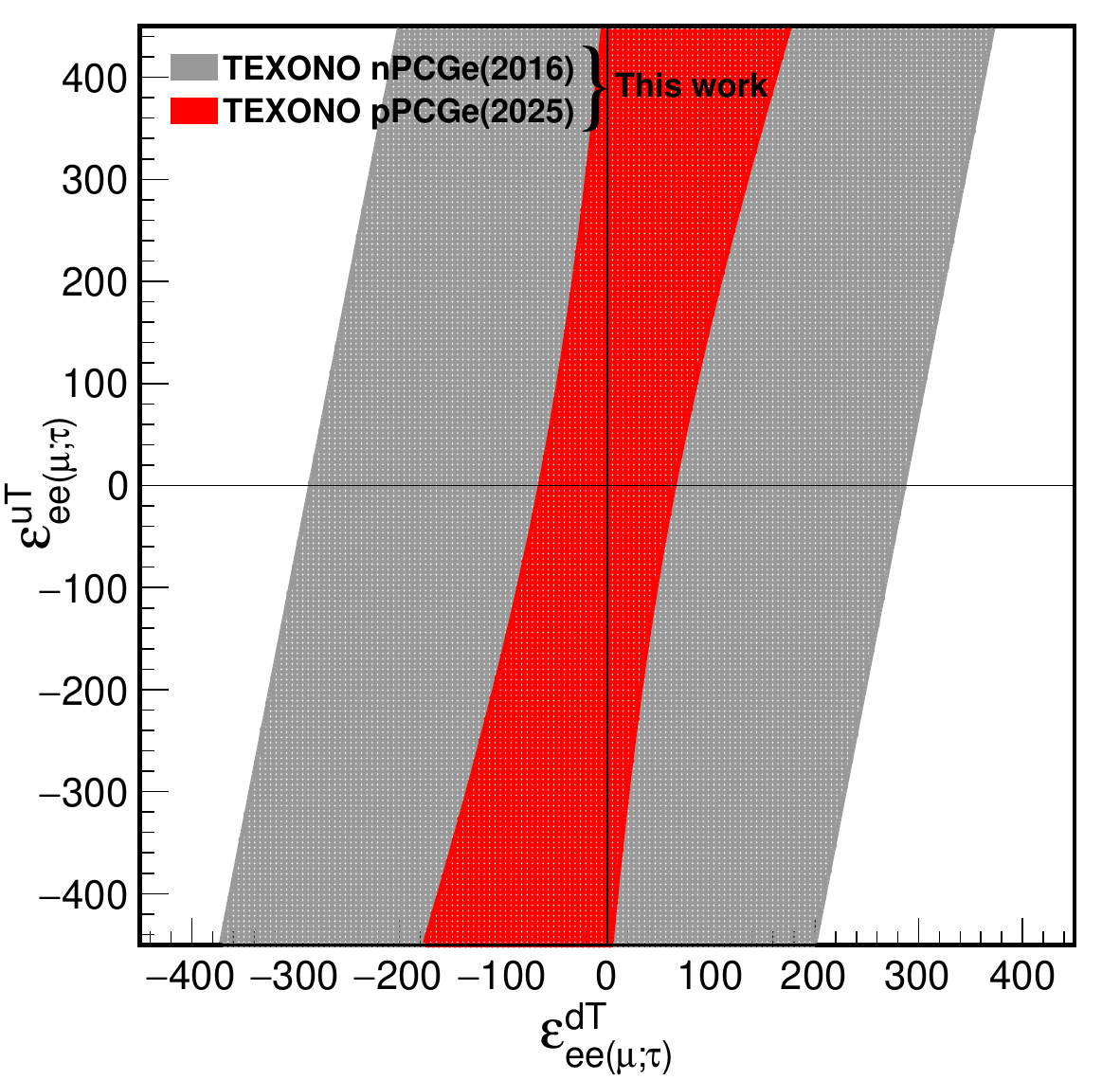}}
\subfigure[]{\includegraphics[scale=0.29]{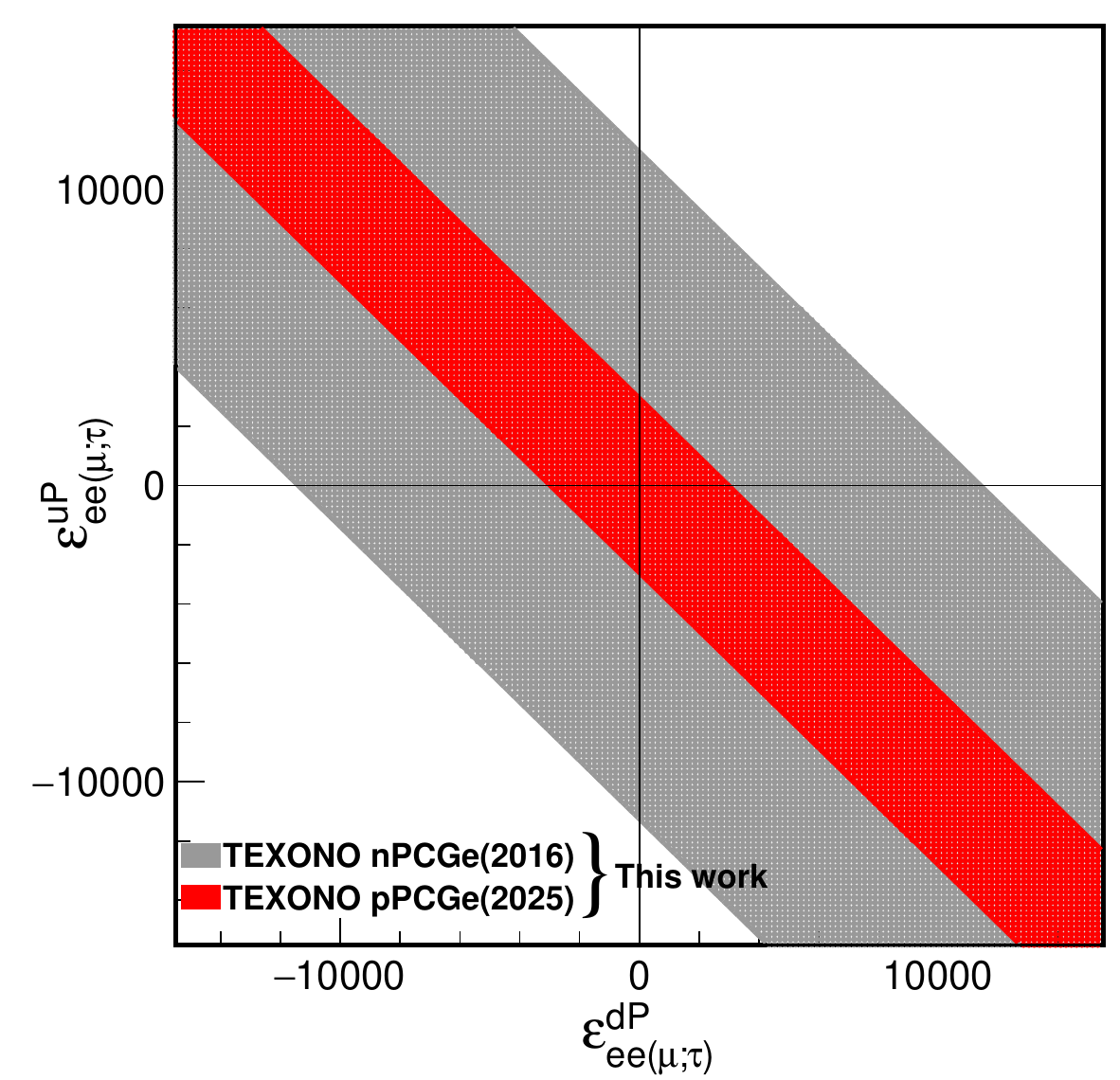}}
\caption{$90\%$ CL allowed region for FC-FC 
parameter spaces for A, P, and T type NSI:
(a) $\varepsilon_{ee(\mu;\tau)}^{dA}-\varepsilon_{ee(\mu;\tau)}^{uA}$,
(b) $\varepsilon_{ee(\mu;\tau)}^{dT}-\varepsilon_{ee(\mu;\tau)}^{uT}$, and
(c) $\varepsilon_{ee(\mu;\tau)}^{dP}-\varepsilon_{ee(\mu;\tau)}^{uP}$.}
\label{fig:eps_FC-FC_ATP}
\end{figure*}

\begin{figure*}[htb!]
	\centering
	\subfigure[]{\includegraphics[scale=0.29]{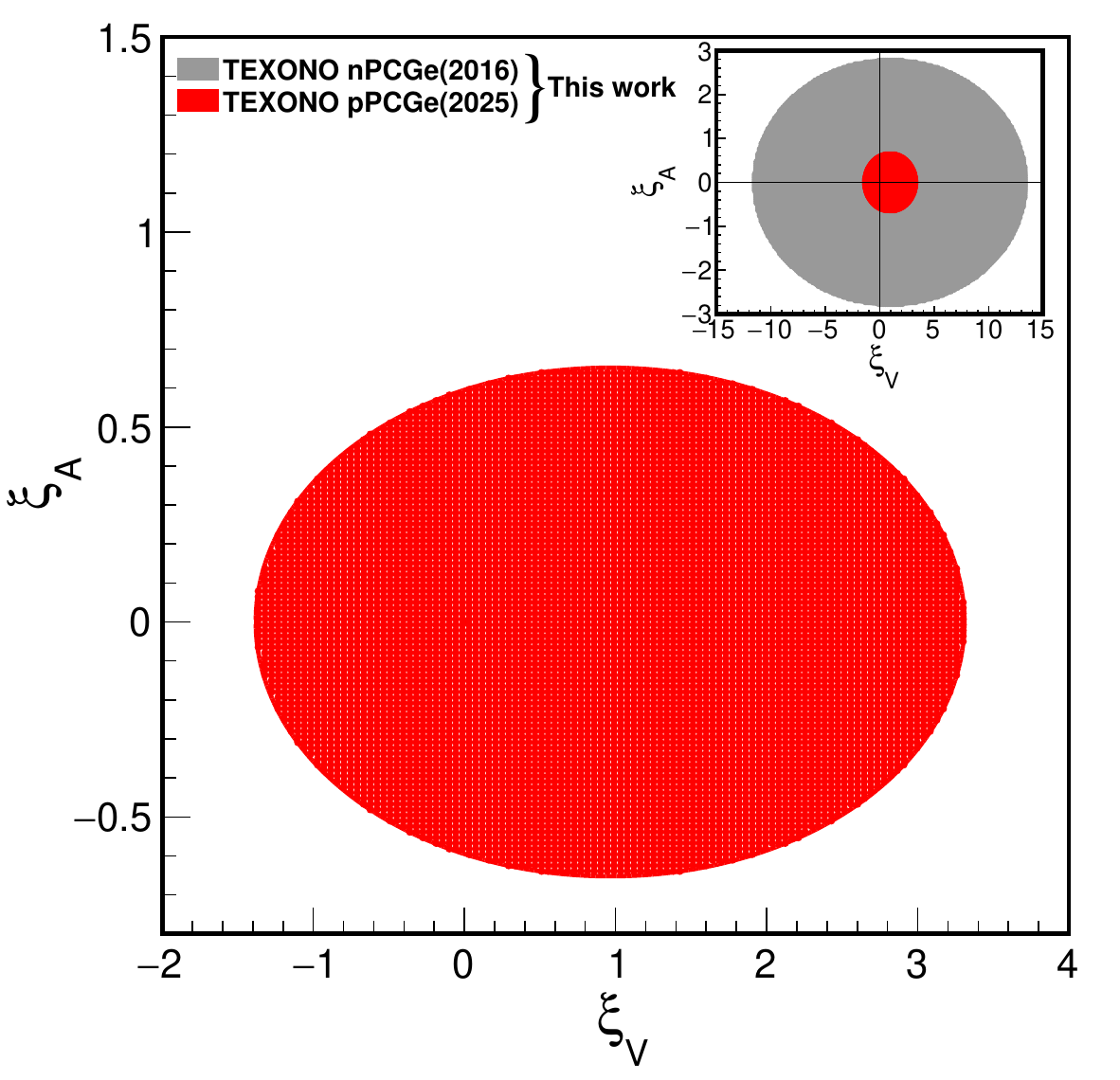}}
	\subfigure[]{\includegraphics[scale=0.29]{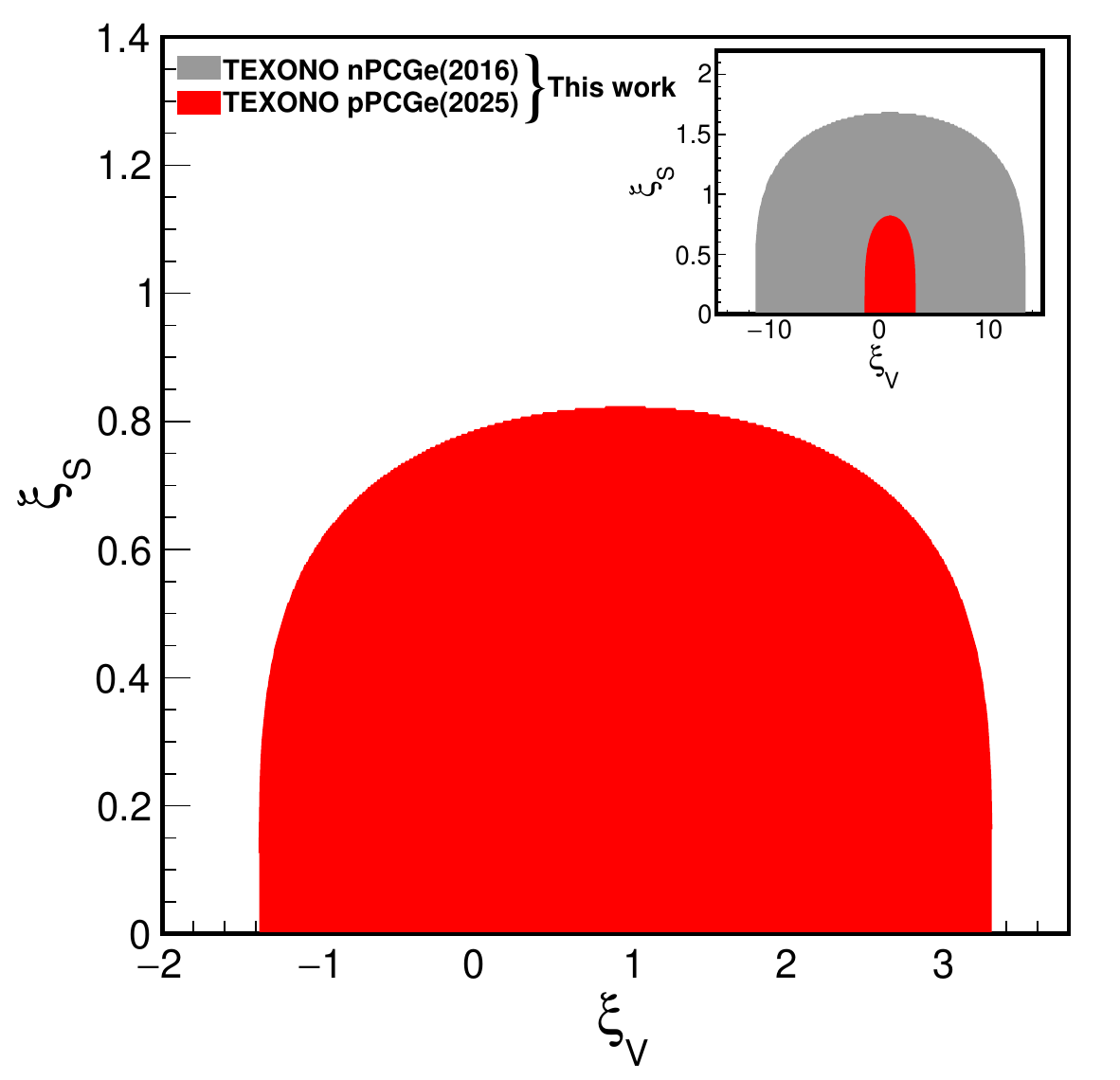}}
	\subfigure[]{\includegraphics[scale=0.29]{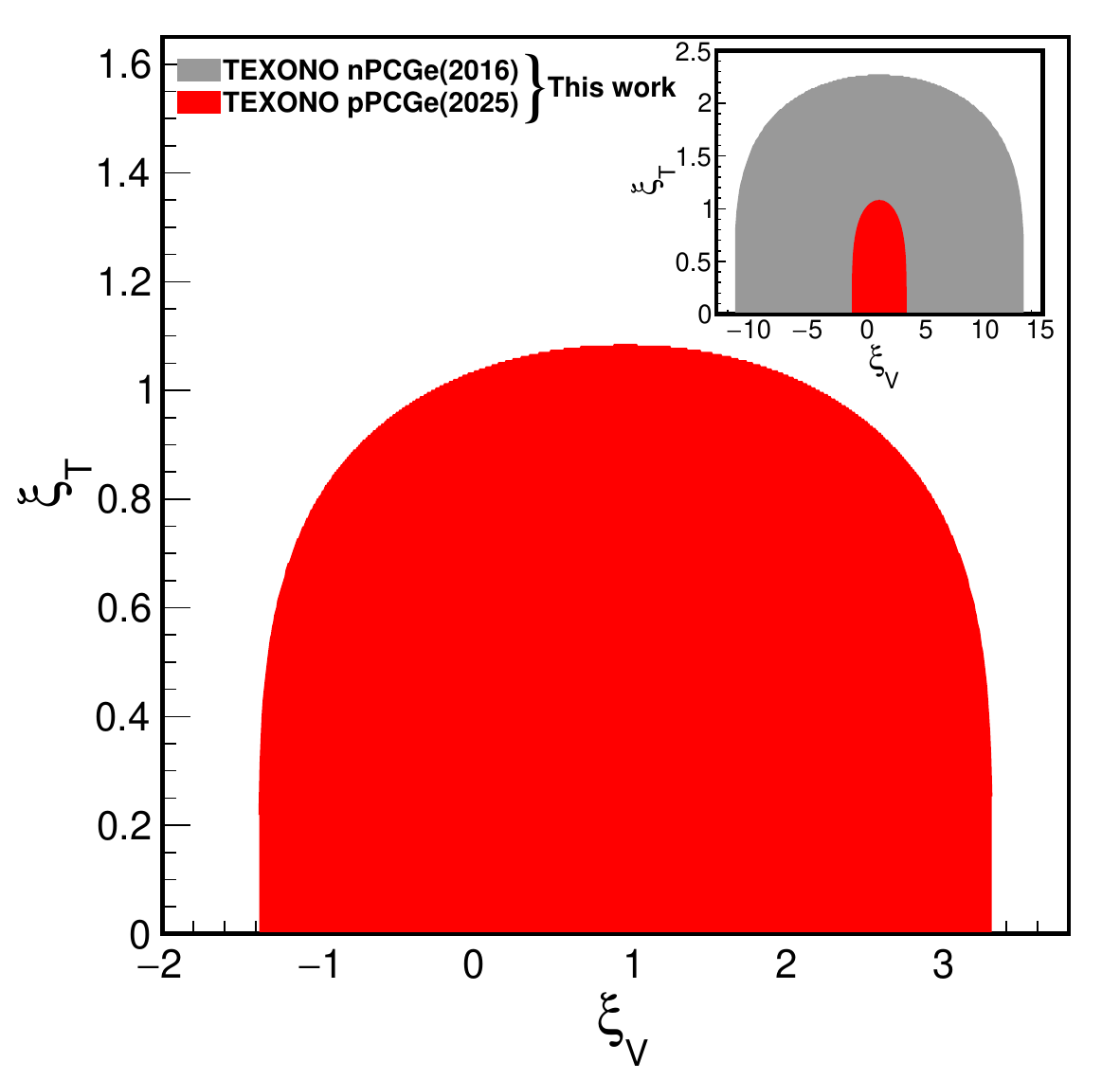}}

	\subfigure[]{\includegraphics[scale=0.29]{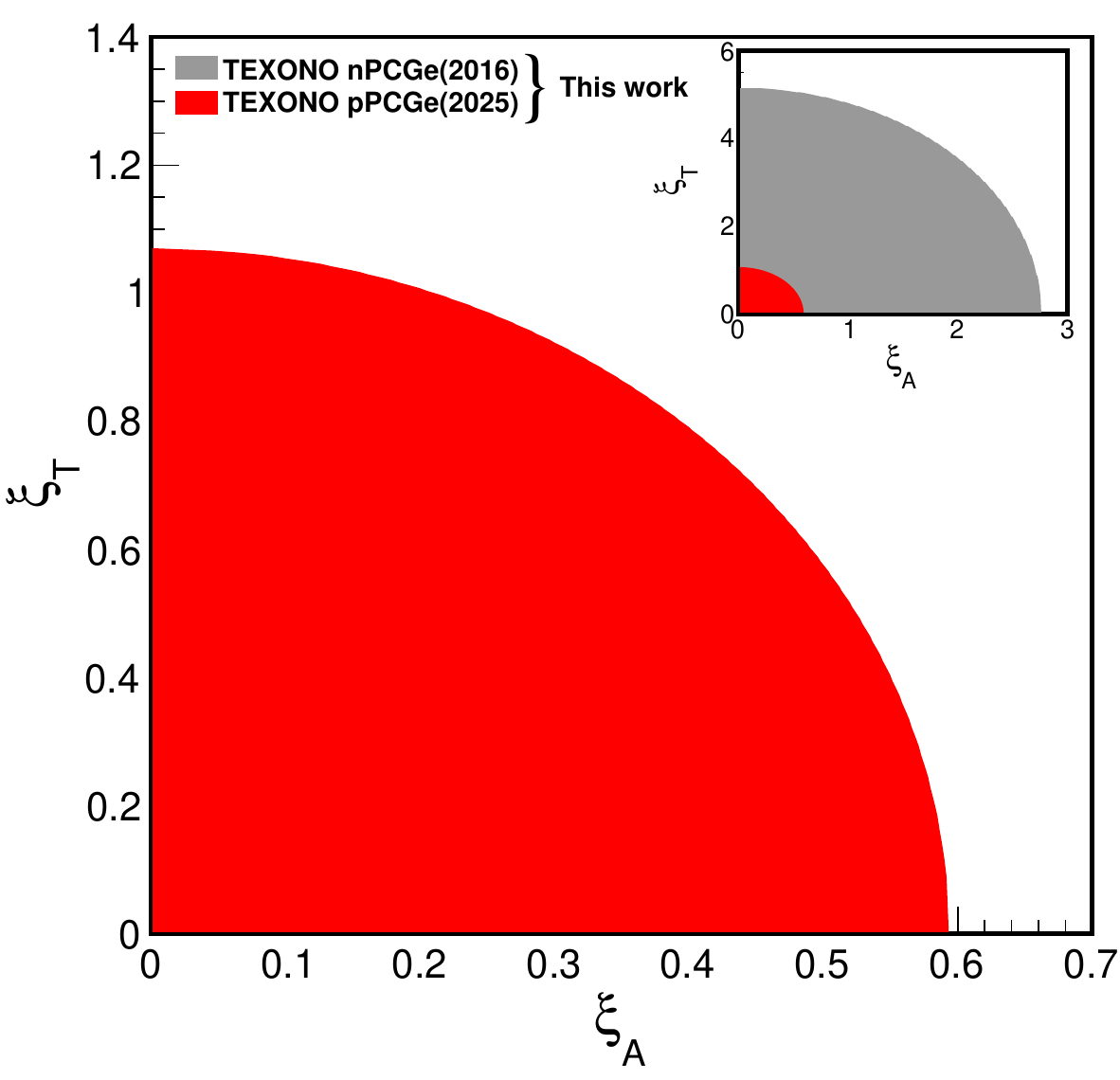}}
	\subfigure[]{\includegraphics[scale=0.29]{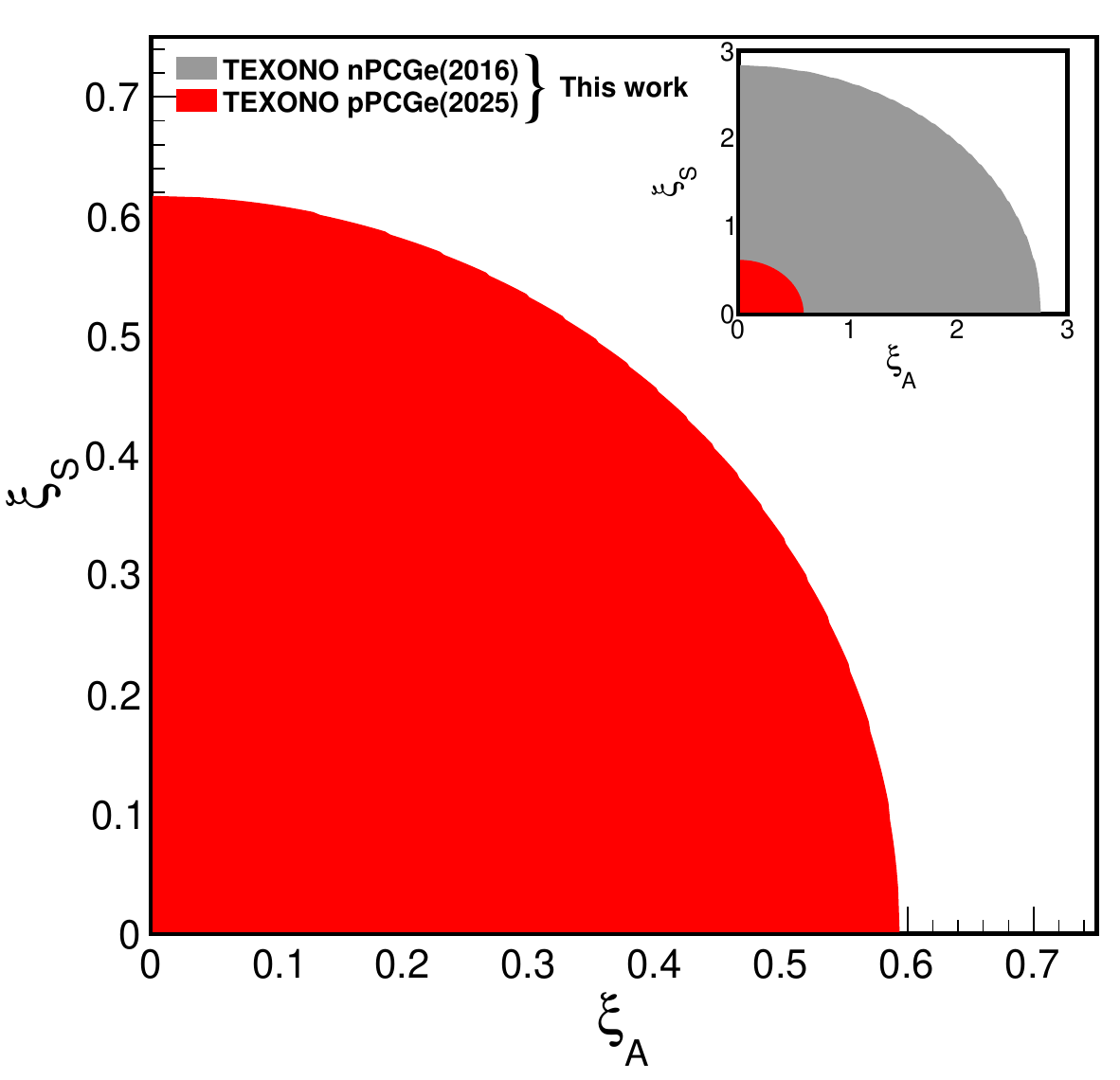}}
	\subfigure[]{\includegraphics[scale=0.29]{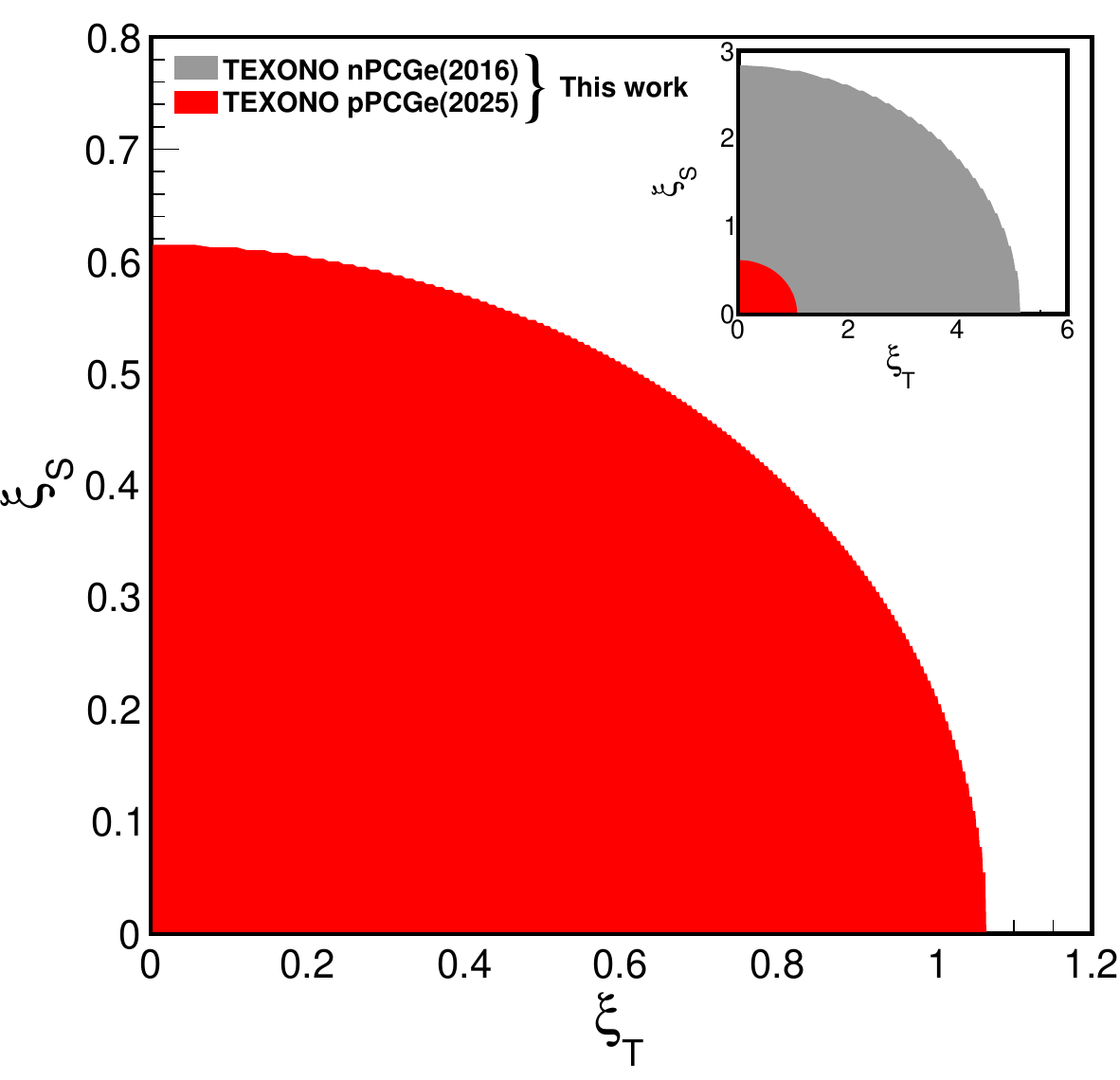}}

	\caption{$90\%$ CL allowed region of
(a) $\xi_{V}-\xi_{A}$, (b) $\xi_{V}-\xi_{S}$, (c) $\xi_{V}-\xi_{T}$,
(d) $\xi_{A}-\xi_{T}$, (e) $\xi_{A}-\xi_{S}$, and (f) $\xi_{T}-\xi_{S}$ parameter spaces for exotic neutral current interactions.}
        \label{fig:exo_xi}
\end{figure*}

Following the scalar interaction, we now present a comparative summary of A, T, and P type NSI constraints. Figure~\ref{fig:eps_FC-FC_ATP} shows the $90\%$ CL upper limits for these three interactions in $\varepsilon_{e\alpha}^{fX}-\varepsilon_{e\alpha}^{fX}$ parameter space, where $X=$ A, T, and P. To focus on the most representative and interpretable cases, we omit other parameter combinations. All three interactions produce linear exclusion bands, but with differing slopes due to their distinct dependencies on nuclear recoil energy $T_N$ and nuclear spin. As described in Eqs.~\eqref{eq:dSdT_A_eps} and ~\eqref{eq:dSdT_T_eps}, both axialvector and tensor interactions are spin dependent; however, their sensitivities differ depending on how spin-structure functions enter the cross section.  As a result, the axialvector interaction [Fig.~\ref{fig:eps_FC-FC_ATP}(a)] yields comparatively weaker constraints than the tensor interaction [Fig.~\ref{fig:eps_FC-FC_ATP}(b)]. In contrast, pseudoscalar interactions in Fig.~\ref{fig:eps_FC-FC_ATP}(c), while formally spin-dependent, are highly suppressed by the nuclear recoil energy.  As in the simplified model case, this suppression makes the spin contribution negligible, resulting in the weakest bounds among all five types of interactions.

\begin{table*} [!hbt]
\caption {Summary of the best-fit results and corresponding upper bounds at $90\%$ CL for exotic coupling constant parameters. The results are presented in comparison with two datasets from the TEXONO Collaboration: nPCGe (2016) and pPCGe (2025). Here only one coupling parameter is assumed to be nonzero for each bound.}
\label{tab:Texono_exo}
\begin{ruledtabular}
\begin{tabular}{lccclccc}
\multicolumn{1}{l}{Fitting}&
\multicolumn{4}{c}{TEXONO nPCGe (2016)} &
\multicolumn{3}{c}{TEXONO pPCGe (2025)} \\ 
Par & BFPV$\ \pm \ 1\sigma$  & 90\% CL  & $\chi^2$/dof & \ &BFPV$\ \pm \ 1\sigma$  & 90\% CL  & $\chi^2$/dof  \\ \hline \\

$\xi_V$   & $0.96 \pm 8.04$ & $[-9.79, 11.7]$ & $81.36/120$ & & $0.004^{+2.72}_{-0.79}$ & $[-1.147,3.075]$ & $87.87/79$ \smallskip \\

$\xi_A^2$ & $-0.87 \pm 3.97$ & $\xi_A<2.38$ & $81.32/120$ & & $0.009 \pm 0.162$ & $\xi_A<0.525$ &  $87.86/79$ \smallskip \\

$\xi_S^2$ & $-0.91 \pm 4.18$ & $\xi_S<2.44$ & $81.32/120$ & &  $0.011 \pm 0.175$ & $\xi_S<0.547$ &  $87.86/79$ \smallskip \\

$\xi_T^2$ & $-3.06 \pm 13.8$ & $\xi_T<4.44$ & $81.32/120$ & &  $0.024 \pm 0.529$ &  $\xi_T<0.946$ &  $87.86/79$\\
\end{tabular}
\end{ruledtabular}
\end{table*}

\subsection{Bounds from  exotic neutral-current interactions} \label{subsec:exoUL}
We now turn to the final aspect of our study, presenting results on exotic neutral current interactions of CE$\nu$NS.  Based on Eq.~\eqref{eq:exo}, there are ten independent parameters dependent on $C_X$ and $D_X$. However, for simplicity,  we may choose to set upper limits on the $\xi_X$ parameters, where $X =$ V, A, S, and T. We present our best-fit measurement results at $1\sigma$ and upper limits at $90\%$ CL in Table~\ref{tab:Texono_exo}, assuming only one parameter remains while setting others to zero for both TEXONO nPCGe (2016) and and pPCGe (2025) experiments. Similar to the model-independent NSI, the vector term, $\xi_V$, in this analysis includes interference with the SM.  Additionally,  both vector and axialvector terms also exhibit interference.  Therefore, when performing a $\chi^2$ analysis based on a single-parameter scenario, it is possible to set limits directly on the $\xi_V$ term. As shown in the first row of Table \ref{tab:Texono_exo}, the bounds at $90\%$ CL are presented as allowed regions rather than upper limits. For the parameters $\xi _A$, $\xi_S$, and $\xi_T$, since these parameters appear in the $\chi^2$ function as the square of the parameters, we turn the parameter-square space into the parameter space, similar to the approach used for model-independent NSI. This method allows us to extract upper limits for the parameters at $90\%$ CL.  

In Fig.~\ref{fig:exo_xi}, we present the $90\%$ CL   upper bounds for the two-parameter scenario across six possible parameter spaces: $\xi_V-\xi_A$, $\xi_V-\xi_S$, $\xi_V-\xi_T$, $\xi_A-\xi_T$,  $\xi_A-\xi_S$, and $\xi_T-\xi_S$. Unlike the simpler one-parameter case, the two-parameter scenario involves more complex interactions between the parameters, leading to changes in the structure of the allowed regions. The interference between the vector term and both the SM and axialvector terms in Eq.~\eqref{eq:exo} results in a different analysis procedure for Fig.~\ref{fig:exo_xi}(a) compared to Figs.~\ref{fig:exo_xi}(b) and ~\ref{fig:exo_xi}(c). Similar behavior is observed when analyzing the vector parameter, $\xi_V$,  and other parameters, $\xi_A,~\xi_S$, and $\xi_T$, in the one-parameter scenario.  These interference terms influence how the constraints are applied, which in turn affects the resulting bounds and the shape of the allowed regions in each scenario.

In Fig.~\ref{fig:exo_xi}(a), we display the allowed regions for TEXONO pPCGe (2025) and nPCGe (2016) in the $\xi_V-\xi_A$ parameter space, while in Figs.~\ref{fig:exo_xi}(b) and ~\ref{fig:exo_xi}(c), we present the corresponding bounds in the $\xi_V-\xi_S$ and $\xi_V-\xi_T$ parameter spaces, respectively. Due to the quadratic dependence of $\xi_S$ and $\xi_T$ in the cross section, only the positive regions of these parameters are shown,  resulting in semielliptical allowed regions, with the $\xi_V$–$\xi_S$ parameter space providing more stringent constraints than $\xi_V$–$\xi_T$. In contrast,  the $\xi_V$–$\xi_A$ parameter space, where both $\xi_V$ and $\xi_A$ appear linearly in the cross section,  leads to a symmetric and fully shown parameter space. In the bottom panel of Fig.~\ref{fig:exo_xi}, we present the bounds for the $\xi_A-\xi_T$, $\xi_A-\xi_S$, and $\xi_T-\xi_S$ parameter spaces. Unlike the figures in the upper panel, these scenarios do not include SM interference or mutual interference between the parameters.  Since the parameters enter the $\chi^2$ function in quadratic form, only the positive regions are shown, resulting in quarter-elliptical upper bounds. Notably, TEXONO pPCGe (2025) exhibits a remarkable improvement over TEXONO nPCGe (2016) across all exotic neutral current interaction parameter spaces.

\subsection{Overview of our results} \label{subsec:SNS-vs-Reactor}
In this work, we present a comprehensive analysis of NSI by examining the reactor-based TEXONO CE$\nu$NS experiment, comparing our results with existing bounds from both reactor-based and accelerator-based experiments, and highlighting the role of detector advancements in improving parameter constraints. We focus on five different NSI interaction types: vector (V), axialvector (A), scalar (S), pseudoscalar (P) and tensor (T), and present the $90\%$ CL exclusion bounds for key NSI parameters using a one-bin analysis of the data from TEXONO nPCGe (2016) and pPCGe (2025).

Besides, one of the key features of this work is to present the first NSI analysis of TEXONO pPCGe (2025) data with a detector threshold of 200~eV. This advancement improves NSI constraints compared to the previous TEXONO nPCGe (2016) data and achieves results competitive with other reactor-based experiments. Furthermore, the low-energy nature of reactor neutrinos increases the sensitivity to the NSI model with light mediators, strengthening the constraints, especially in the low momentum transfer regime, where deviations from the SM are more pronounced. Specifically,  in certain scenarios, TEXONO experiments yield stronger constraints than accelerator-based results, COHERENT CsI and LAr. For instance,  both TEXONO nPCGe (2016) and pPCGe (2025) provides more stringent upper limits compared to COHERENT in the light scalar model interaction shown in Fig.~\ref{fig:gXmX}(c).  Also, in the light vector mediator case [Fig. ~\ref{fig:gXmX}(a)] TEXONO pPCGe (2025) achieves bounds comparable to COHERENT CsI.
Moreover,  TEXONO pPCGe (2025) demonstrates a significant improvement over the earlier TEXONO nPCGe (2016) results and remains competitive with other reactor-based experiments,  including CONUS, CONNIE, and Dresden-II [Figs.~\ref{fig:gXmX}(a) and ~\ref{fig:gXmX}(c)].  These results highlight the improvements in detector capabilities and its strong potential for investigating NSI.

\section{Conclusions}\label{sec:conc}
Different experimental setups with varying detection thresholds and neutrino energy spectra contribute to the exploration of different regions of the NSI parameter space in CE$\nu$NS.
High-energy neutrinos from the SNS are better suited to exploring new physics associated with heavier mediators, while reactor neutrinos emitted at lower energies provide a more powerful search for light mediators models. This study provides a comprehensive analysis of NSI within the framework of CE$\nu$NS, utilizing both the simplified model and model-independent approach, as well as exploring exotic neutral interactions, with a particular focus on the TEXONO experiment.

We note that the studies of different NSI scenarios have been extensively explored in previous works~
\cite{Farzan:2018gtr, Barranco:2005, Barranco:2011wx, CONUSpl:2025-BSM,
Lindner:2017,Miranda:2015, Venegas:CONUSpl2025, 
Chattaraj:CONUSpl2025,
Kosmas:2017tsq,CONUS:2022, CONNIE:2020, 
DresdenII:2022, Corona:2022, Miranda:2020nsi,
Coloma:2022nsi,Lindner:2024nsi,Cadeddu:2021nsi, 
DemirciM:2024nsi,Majumdar:prd2022,
Romeri-exo3:jhep2023,Romeri:DM-LM2024,
Giunti:2019xpr, Mustamin:2021mtq, NSItensor1:prd2024, 
Chatterjee:2023nsi,Canas:2020nsi,Coherent-Ge:PRD2024,
Flores-exo2:prd2022, AristizabalSierra:2018eqm,Majumdar:DresdenExo,Romeri:CONUSpl2025}. 
In this study, we present the first dedicated NSI analysis based solely on TEXONO reactor neutrino experiments, using both the nPCGe data from 2016 and the recently released pPCGe data from 2025.  This study provides a systematic assessment of how advances in detector technology have strengthened constraints on NSI parameters and provides new insights into the improving sensitivity of reactor-based CE$\nu$NS searches. We compare our results with existing bounds in the literature from both accelerator-based experiments, such as COHERENT CsI, LAr, and Ge, and other reactor-based experiments including CONUS+, CONUS, CONNIE, and Dresden-II, as well as the DM experiments like XENONnT, and PandaX-4T.

The abundance of neutrino sources that meet the detection criteria, including terrestrial sources, solar and atmospheric neutrinos as well as those from accelerators and nuclear power plants, is expected to drive increased scientific activity in the near future. This work underscores the potential for future CE$\nu$NS experiments to further probe NSI  and enhance our understanding of new physics BSM. Moreover, we highlight that future improvements in detector sensitivities could resolve degeneracies and enable tighter constraints, particularly in two-parameter scenarios. Our results may also offer clues for the search for new mediators in the low-energy realm, likely driven by advancements in neutrino or DM experiments.

\begin{acknowledgments}
This work is supported by the Investigator Award No. ASIA-106-M02 and Thematic Project No. AS-TP-112-M01 from the Academia Sinica, Taiwan, and Contracts No. 106-2923-M-001-006-MY5, No. 107-2119-M-001-028-MY3, No. 110-2112-M-001-029-MY3, and No. 113-2112-M-001-053-MY3 from the National Science and Technology Council, Taiwan.
\end{acknowledgments}

\section{Data Availability}\label{sec:data}

The data that support the findings of this article are openly available~\cite{Texono:2024}.

\end{document}